\documentclass[twocolumn,times,tighten]{aastex61}

\usepackage[all]{hypcap} 

\usepackage{amsmath,amsfonts, amssymb}

\usepackage{mathrsfs}

\usepackage{dsfont}

\usepackage{url}
\usepackage{paralist}

\usepackage{float}
\restylefloat{deluxtable}

\usepackage{subfigure}

\shorttitle{The Profile of the Galactic Halo from Pan-STARRS1 3$\pi$ RR Lyrae}
\shortauthors{Hernitschek et~al.}

\begin{document}


\title{The Profile of the Galactic Halo from Pan-STARRS1 3$\pi$ RR Lyrae}


\author{Nina Hernitschek}
\affiliation{Division of Physics, Mathematics and Astronomy, Caltech, Pasadena, CA 91125}

\author{Judith G. Cohen}
\affiliation{Division of Physics, Mathematics and Astronomy, Caltech, Pasadena, CA 91125}

\author{Hans-Walter Rix}
\affiliation{Max-Planck-Institut f{\"u}r Astronomie, K{\"o}nigstuhl 17, 69117 Heidelberg, Germany}

\author{Branimir Sesar}
\affiliation{Max-Planck-Institut f{\"u}r Astronomie, K{\"o}nigstuhl 17, 69117 Heidelberg, Germany}

\author{Nicolas F. Martin}
\affiliation{Max-Planck-Institut f{\"u}r Astronomie, K{\"o}nigstuhl 17, 69117 Heidelberg, Germany}    
\affiliation{Observatoire astronomique de Strasbourg, Universit\'{e} de Strasbourg, CNRS, UMR 7550, 11 rue de l'Universit\'{e}, F-6700 Strasbourg, France}

\author{Eugene Magnier}
\affiliation{Institute for Astronomy, University of Hawai’i at Manoa, Honolulu, HI 96822, USA}

\author{Richard Wainscoat}
\affiliation{Institute for Astronomy, University of Hawai’i at Manoa, Honolulu, HI 96822, USA}

\author{Nick Kaiser}
\affiliation{Institute for Astronomy, University of Hawai’i at Manoa, Honolulu, HI 96822, USA}

\author{John L. Tonry}
\affiliation{Institute for Astronomy, University of Hawai’i at Manoa, Honolulu, HI 96822, USA}

\author{Rolf-Peter Kudritzki}
\affiliation{Institute for Astronomy, University of Hawai’i at Manoa, Honolulu, HI 96822, USA}

\author{Klaus Hodapp}
\affiliation{Institute for Astronomy, University of Hawai’i at Manoa, Honolulu, HI 96822, USA}

\author{Ken Chambers}
\affiliation{Institute for Astronomy, University of Hawai’i at Manoa, Honolulu, HI 96822, USA}

\author{Heather Flewelling}
\affiliation{Institute for Astronomy, University of Hawai’i at Manoa, Honolulu, HI 96822, USA}

\author{William Burgett}
\affiliation{GMTO Corporation (Pasadena), 465 N. Halstead Street, Suite 250, Pasadena, CA 91107, USA}

\correspondingauthor{Nina Hernitschek}
\email{ninah@astro.caltech.edu}





\begin{abstract}
We characterize the spatial density of the Pan-STARRS1 (PS1) sample of Rrab stars to study the properties of the old Galactic stellar halo. This sample, containing 44,403 sources, spans Galactocentric radii of $0.55\;\mathrm{kpc}\leq R_{\mathrm{gc}} \leq 141 \;\mathrm{kpc}$ with a distance 
precision of 3\% and thus is able to trace the halo out to larger distances than most previous studies. After excising stars that are attributed to dense regions such as stellar streams, the Galactic disc and bulge as well as halo globular clusters, the sample contains ${\sim}11,000$ sources within $20\;\mathrm{kpc}\leq R_{\mathrm{gc}}\leq 131\;\mathrm{kpc}$.\newline
We then apply forward modeling using Galactic halo profile models with a sample selection function. Specifically, we use ellipsoidal stellar density 
models $\rho(l,b,R_{\mathrm{gc}})$ with a constant and a radius-dependent halo flattening 
$q(R_{\mathrm{gc}})$. Assuming constant flattening $q$, the distribution of the sources is reasonably well fit by a single power law with $n=4.40^{+0.05}_{-0.04}$ and $q=0.918^{+0.016}_{-0.014}$, and comparably well by an Einasto profile with $n=9.53^{+0.27}_{-0.28}$, an effective radius $r_{\mathrm{eff}}=1.07\pm 0.10\; \mathrm{kpc}$ and a halo flattening of $q=0.923\pm 0.007$. If we allow for a radius-dependent flattening $q(R_{\mathrm{gc}})$, we find evidence for a distinct flattening of $q{\sim}0.8$ of the inner halo at ${\sim}25\;\mathrm{kpc}$. Additionally, we find that the south Galactic hemisphere is more flattened than the north Galactic hemisphere.\newline
The results of our work are largely consistent with many earlier results, e.g. \cite{Watkins2009}, \cite{Iorio2017}.
We find that the stellar halo, as traced in RR Lyrae stars, exhibits a substantial number of further significant over- and underdensities, even after masking all known overdensities.
\end{abstract}



\section{Introduction}
\label{sec:Introduction}

The Milky Way's extended stellar halo contains only a small fraction 
(${\sim}1\%$) of the
Galaxy's stars but is an important diagnostic of the Milky Way's formation, dark 
matter distribution and mass.

The stellar halo shows great complexity in its spatial structure, with abundant 
globular clusters, dwarf galaxies and stellar streams. This makes it difficult 
to dissect with local spectroscopic or photometric data.
While the radial density profile can be derived from data of a limited number of 
sight-lines through the Galaxy, a sensible description of the overall stellar 
halo shape requires nearly complete coverage of the sky.

As stellar haloes formed from disrupted satellites and still show signs of their accretion history 
in the form of overdensities such as streams, they are central to studies on galaxy formation such as the 
hierarchical galaxy formation in the $\Lambda \mathrm{CMD}$ model. The spatial distribution, as well as kinematics and metallicities and thus ages of halo stars enable us to
get information on those merger processes as well as to compare them to simulations from theoretical models.

Many studies were carried out within the last 50 years to map the Galactic halo, 
and those studies often took advantage of RR Lyrae stars as reliable halo 
tracers. These old and metal-poor pulsators are ideal for this task as they can 
be selected with a high purity, thus showing only very little contamination from 
other populations of the Milky Way. Furthermore, RRab are luminous 
variable stars pulsating in the fundamental mode which obey a well defined period - luminosity relation, albeit with a small 
dependence on metallicity.  Thus the mean luminosity of a RRab variable, and hence its 
distance, can be determined with knowledge of the light curve only.
RRab stars were used by many previous studies, including those of
\cite{Hawkins1984}, \cite{Saha1984}, \cite{WettererMcGraw1996}, 
\cite{Ivezic2000},
\cite{VivasZinn2006}, \cite{Juric2008}, \cite{Catelan2009}, \cite{Watkins2009}, 
\cite{deJong2010},  \cite{Sesar2010}, \cite{Deason2011} \cite{Sesar2011}, 
\cite{Akhter2012}, \cite{Drake2014}, \cite{Torrealba2015}, \cite{Cohen2015}, \cite{Xue2015}, 
\cite{Soszynski2016}, \cite{Bland2016}, \cite{Iorio2017}, \cite{Cohen2017}.

The key to using RRab to explore the Galactic halo is having a reliable list of 
RRab variables selected from a suitable multi-epoch imaging survey covering a wide distance
range and as much of the sky as possible. Recently the inner halo out to $\sim$30~kpc was explored by a sample of 
${\sim} 5000$ RRab generated from a recalibration of the LINEAR catalog by \cite{Sesar2013b},
and most recently by \cite{Iorio2017} using a sample selected from the 
combination of the Gaia Data Release 1 \citep[GDR1,][]{gaiaDR1} and 2MASS \citep{Skrutskie2006}.

\cite{Drake2014} used the
Catalina Real-Time Transient Surveys (CRTS) DR1 to select a sample of 47,000 periodic variables
of which 16,797 are RR Lyrae, the bulk of them are at $R_{\mathrm{gc}} < 40$~kpc.
In total, the Catalina Surveys
RR Lyrae Data Release 1\footnote{\url{http://nesssi.cacr.caltech.edu/DataRelease/RRL.html}} \citep{Drake2013a,Drake2013b,Drake2014,Torrealba2015,Drake2017} contains 43,599 RR Lyrae of which 32,980 are RRab stars.

To reach larger distances with larger samples was very difficult in the past.  
One approach was to use brighter tracer stars, usually K giants and usually selected from the 
SDSS, but with larger distance uncertainties and only modest sample sizes,
see e.g. \cite{Xue2015} who probe the Galactic halo out to 80 kpc using 1757 
stars from the SEGUE K-giant Survey. There have also been efforts to reach the outer halo using blue horizontal 
branch (BHB) stars, which to first order have a fixed luminosity similar to that of RRab, see, e.g. \cite{Deason2014}, but 
these run into problems of confusion with much more numerous blue stragglers at the same apparent magnitude and with quasars.  Prior to the 
present work, perhaps the most successful attempt to probe
the density distribution in the outer Galactic halo was by \cite{Cohen2017}, reaching out to above 100~kpc,
with a small (${\sim} 450$) sample of RRab stars selected from the Palomar Transient Facility (PTF) database.

Here, we overcome these difficulties by using a selection of RRab from the PS1 
survey which covers the entire northern sky to a limiting magnitude such that detection of RRab out 
to more than 100~kpc is not difficult. \cite{Hernitschek2016} and \cite{Sesar2017b}
exploited the PS1 survey to create a sample of RRab which reaches far into the outer halo,
which is very large (44,403 RRab), with known high purity and completeness.
The details of the machine learning techniques which were used to select this sample and the assessment of its purity and completeness as a 
function of distance are described in \cite{Hernitschek2016} and \cite{Sesar2017b}.

In this paper we exploit the PS1 RRab sample to study the Milky Way halo out 
to distances in excess of 100~kpc.

We develop and apply a rigorous density modeling approach for Galactic 
photometric surveys that enables investigation of the
structure of the Galactic halo as traced by RR Lyrae stars from 20 kpc to more than 100 kpc.
We fit models that characterize the radial density and flattening of the Milky 
Way's stellar halo, while accounting for the complex selection function 
resulting from both the survey itself as well as the selection of sources within 
the survey data.

In Section \ref{sec:PS1}, we lay out the properties of the PS1 RRab stars. In Section \ref{sec:DensityFitting} we present the method of fitting a series of parameterized models to the RRab stars while considering a selection function. This step is key to obtaining accurate radial profiles. In the following, first two types of parameterized models for the radial stellar density are shown in \ref{sec:StellarDensityModels}, followed by the description of the selection function in Section \ref{sec:SelectionFunction}, and our approach to constrain the model parameters in Section \ref{sec:ConstrainingModelParameters}. A test method, relying on mock data, is shown in Section \ref{sec:TestsOnMockData}. In Section \ref{sec:Results}, we present the results for the profile and flattening of the Milky Way's stellar halo, as well as findings of previously unknown halo overdensitis.
In Section \ref{sec:Discussion}, we discuss results and methodology, and compare them to work by others. Finally, Section \ref{sec:Summary} summarizes the paper.

\section{RR Lyrae Stars from the PS1 Survey}
\label{sec:PS1}

Our analysis is based on a sample of highly likely RRab stars, as selected by \cite{Sesar2017b} from the Pan-STARRS1 3$\pi$ survey. 
In this section, we describe the pertinent properties of the PS1 3$\pi$ survey and the RR Lyrae light curves obtained, and recapitulate briefly the process of selecting the likely RRab, as laid out in \cite{Sesar2017b}. We also briefly characterize the obtained candidate sample.

The Pan-STARRS 1 (PS1) survey \citep{Kaiser2010} collected multi-epoch, multi-color observations undertaking a number of surveys, among which the PS1 3$\pi$ survey \citep{Chambers2016} is currently the largest. It has observed the entire sky north of declination $-30^{\circ}$ in five filter bands ($g_{\rm P1},r_{\rm P1},i_{\rm P1},z_{\rm P1},y_{\rm P1}$) 
with a 5$\sigma$ single epoch depth of about 22.0, 22.0, 21.9, 21.0 and 19.8 magnitudes in $g_{\rm P1},r_{\rm P1},i_{\rm P1},z_{\rm P1}$, and $y_{\rm P1}$, respectively  \citep{Stubbs2010, Tonry2012}. 

Starting with a sample of more than  $1.1\times 10^9$ PS1 3$\pi$ sources, 
\cite{Hernitschek2016} and \cite{Sesar2017b} subsequently selected a sample of 
44,403 likely RRab stars, of which $\sim$17,500
are at $R_{\mathrm{gc}} \geq 20$~kpc, by applying machine-learning techniques based on 
light-curve characteristics. RRab stars are the most common type of RR Lyrae, 
making up ${\sim}91\%$ of all observed RR Lyrae \citep{Smith2004}, and 
displaying the steep rises in brightness typical of RR Lyrae.

The identification of the RRab stars is highly effective, and the sample of RRab stars is pure ($90\%$), and complete ($\geq80\%$ at 80~kpc) at high galactic latitudes. The distance estimates are precise to $3\%$, based on newly derived period-luminosity relations for the optical/near-infrared PS1 bands \citep{Sesar2017b}. Overall, this results in the widest (3/4 of the sky) and deepest (reaching $> 120$~kpc) sample of RR Lyrae stars to date, allowing us to observe them globally across the Milky Way. Out of these sources, 1093 exist beyond a Galactocentric distance of 80 kpc, and 238 beyond 100 kpc.

In the subsequent analysis, we refer to this sample \citep{Sesar2017b} as ``RRab stars''.

\begin{figure*}
\begin{center}  
\includegraphics[trim=0.0inch 0.0inch 0.0inch 0.0inch, clip=true]{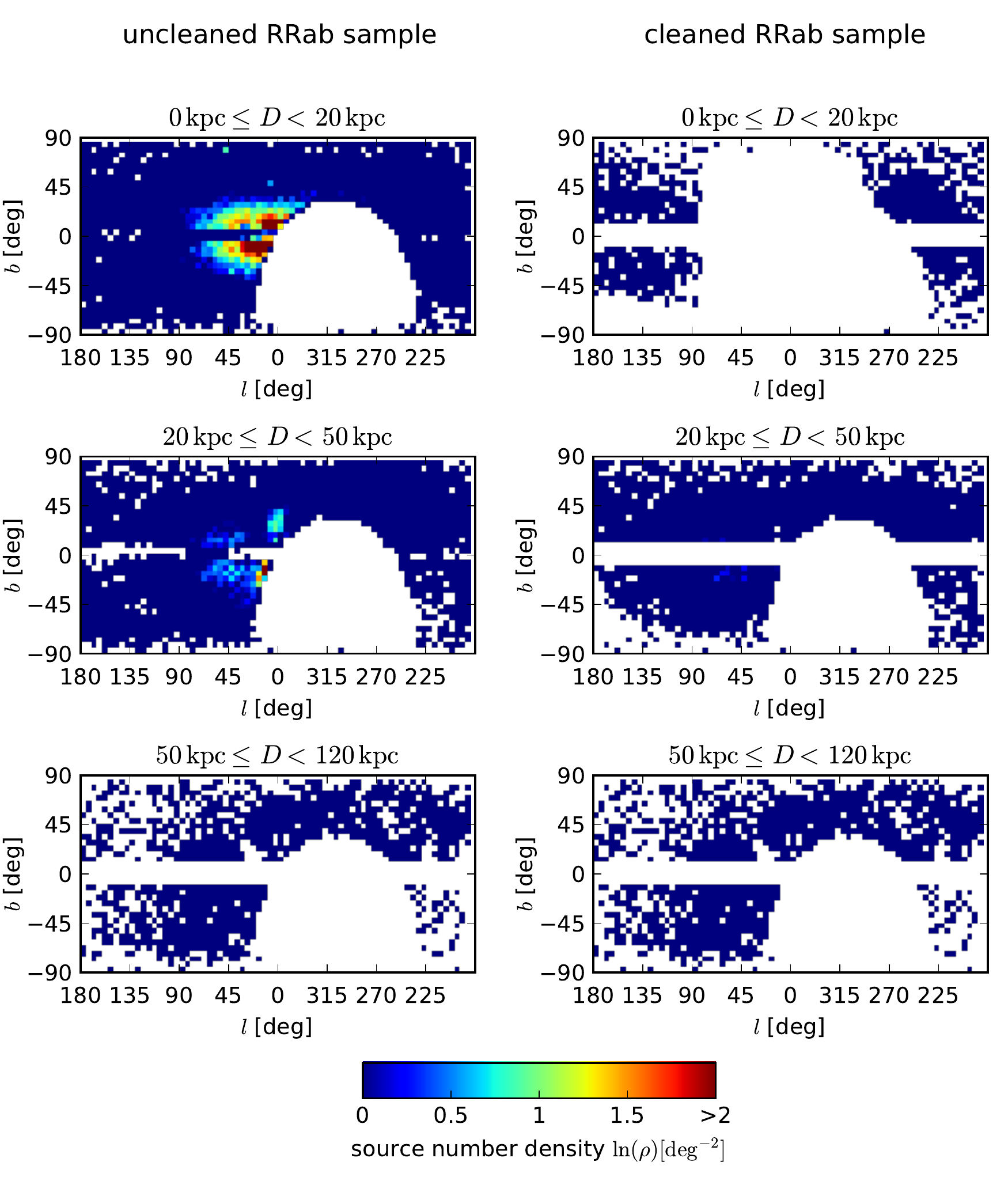}
\caption[The density of the uncleaned and cleaned PS1 3$\pi$ sample of RRab stars, shown in Galactic coordinates $(l,b)$ for different heliocentric distance 
bins.]
{{The density of the uncleaned and cleaned PS1 3$\pi$ sample of RRab stars, shown in Galactic coordinates $(l,b)$ for different heliocentric distance 
bins. The logarithmic source number density is given within 5 deg$^2$ wide bins, in units of deg$^{-2}$. This bin size was chosen to reduce Poisson noise. White cells are empty, and dark blue cells have 1 source per deg$^2$.\newline
Starting from a sample of 44,403 sources \citep{Sesar2017b}, containing overdensities like globular clusters and streams and affected by sample incompleteness near the Galactic plane and apocenter (here shown in the left column as ``uncleaned sample"), we construct a sample of 11,025 sources outside of such known overdensities. To do so, we apply the selection cuts described in Sec. \ref{sec:SelectionFunction}, to geometrically excise such overdensities. The largest overdensities removed are the Sagittarius stream (we remove sources associated with the Sgr stream according to \cite{Hernitschek2017}), as well as the thick disc (we remove sources within $\vert b \vert <10 \arcdeg$) and close to the Galactic center and the bulge (we remove sources within $R_{\mathrm{gc}} \leq 20~{\mathrm{kpc}}$).\newline
Showing the source density in three different distance bins shows major overdensities as well as how excising such overdense regions affects the cleaned sample.
}
\label{fig:sample_map_gl}}
\end{center}  
\end{figure*}

The left panels of Fig. \ref{fig:sample_map_gl} show the source density of the PS1 sample of RRab stars for different distance bins $0 \; \mathrm{kpc} < D \leq 20 \; \mathrm{kpc}$, $20 \; \mathrm{kpc} < D \leq 50 \; \mathrm{kpc}$, $50 \; \mathrm{kpc} < D \leq 120 \; \mathrm{kpc}$. The right panels of the same figure show the sample after a cleaning to remove overdensities was applied; the details of this cleaning are descried later.

Fig. \ref{fig:sample_map} is based on the same data but shown in the Cartesian reference frame $(X,Y,Z)$ for an easier comparison with subsequent plots of halo models, as well as to highlight the individual effects of removing certain overdensities.

While the sample covers the entire sky above a declination $\delta > -30^\circ$, which enables a view of halo substructure like the Sagittarius stream \citep{Hernitschek2017}, in this paper we focus on stars away from the Galactic plane and center, and also away from known large overdensities like the Sagittarius stream. 
Details of the process of removing these overdensities are given in Sec. \ref{sec:SelectionFunction}.

\section{Density Fitting}
\label{sec:DensityFitting}

In this section we lay out a forward-modeling approach to describe the spatial distribution of the stellar halo using a set of flexible but ultimately smooth and
symmetric functions. 

We presume that the stellar halo distribution can be sensibly approximated by a spheroidal distribution with a parameterized radial profile. 
Similar approaches were carried out by e.g. \cite{Sesar2013b}, \cite{Xue2015}, \cite{Cohen2015}, \cite{Iorio2017}, but all with either a smaller sample size than in our analysis, or probing a smaller distance range.

The number of halo parameters depends on the complexity of the model assumed for the stellar halo distribution. The mathematics of this approach essentially follows \cite{Bovy2012} and \cite{Rix2013}.

A number of very different models have been proposed for the density profile of the stellar halo.
We denote the spatial number density here as $\rho_{\mathrm{RRL}}(l,b,D)$ and the general form of the models as $\rho_{\mathrm{RRL}} (\mathcal{D} \vert \boldsymbol{\theta})$, where $\boldsymbol{\theta}$ are the model parameters (see Sec. \ref{sec:StellarDensityModels}) and $\mathcal{D}=(l,b,D)$ are the observables with Galactic coordinates and the heliocentric distance $D$.

An approach for fitting the spatial-density profiles of the RRab sample must account for the fact that the observed star counts do not reflect the underlying stellar distribution, but are strongly shaped by selection effects both from the survey itself as well as from selection cuts we chose while preparing the sample. We denote the spatial selection function as $\mathcal{S}(l, b,D)$ (see Sec. \ref{sec:SelectionFunction}).

To properly take all of these effects into account, we need to use forward modeling: In what follows we fit stellar-density models to the data
by generating the expected observed distribution of stars in the RRab sample, based on our model for the selection function and the halo density models. This predicted distribution is then automatically compared to the observed star counts to calculate the likelihood of the observed RRab star counts.

\subsection{Stellar Density Models}
\label{sec:StellarDensityModels}

Stellar density models can take various functional forms. We first describe what the stellar density models we use for evaluating the RRab sample have in common.

In what follows we will assume that our models are characterized by a set of parameters denoted as $\boldsymbol{\theta}$, and that the density $\rho_{\mathrm{RRL}}$ is ellipsoidal, allowing for a halo flattening $q$ along the $Z$ direction. Oblate density distributions have $q < 1$, spherical have $q = 1$ and prolate have $q > 1$.

The density is a function of right-handed Cartesian coordinates $(X,Y,Z)$, that we evaluate through the Galactic longitude, Galactic latitude and heliocentric distance $(l,b,D)$, so its dimension is $\mathrm{kpc}^{-3}$:

\begin{align}
X &= R_{\odot} - D \cos l \cos b \label{eq:cartesian} \\
Y &= -D \sin l \cos b \nonumber\\
Z &= D \sin b. \nonumber
\end{align}
This reference frame is centered at the Galactic center. The Galactic disc is in the $(X,Y)$ plane, with the $X$ axis pointing to the Sun and the $Z$ axis to the North Galactic Pole.
$R_{\odot}$ denotes the distance of the Galactic center from the Sun, in this work assumed to be 8~kpc, and main results of our work should not change for other values of 
$R_{\odot}$ within the assumed observational uncertainties.

The vertical position of the Sun with respect to the Galactic disc is uncertain, but it is estimated to be smaller than 50 pc \citep{Karim2017, Iorio2017}, and thus negligible for the purpose of this work. 

Caution must be taken when comparing our work to others: Some papers use a left-handed system instead, e.g. \cite{Iorio2017}, where the $Y$-axis is flipped with respect to our definition.

With Equ. \eqref{eq:cartesian}, the Galactocentric distance $R_{\mathrm{gc}}$ is then defined as $R_{\mathrm{gc}} = \sqrt{X^2+Y^2+Z^2}$, and the flattening-corrected radius defined as $r_q = \sqrt{X^2+Y^2+(Z/q)^2}$ where $q$ gives the halo flattening along the $Z$ direction as a minor-to-major-axis ratio.
This describes an oblate stellar halo that is stratified on concentric ellipsoids, where $X$, $Y$, $Z$ are the ellipsoid principal axes.

Following a number of previous studies, we presume that the overall radial density profile
of the halo can be described by a power law or an Einasto profile, with
the density stratified on concentric ellipsoidal surfaces of constant $r_q$ in all cases.

\subsubsection{Power-Law Profile}

A simple power-law halo model $\rho_{\mathrm{halo}}$ is widely used \citep[e.g.][]{Sesar2013b} to describe the distribution of the halo stars:
\begin{equation}
\rho_{\mathrm{halo}}(X,Y,Z)=\rho_{\odot \mathrm{RRL}}\left(R_{\odot}/r_q \right)^{n}.
\label{eq:halomodel}
\end{equation}

For a power-law profile, the shape of the density profile is described by the parameter $n$. Larger values of $n$ indicate a steeper profile.

The free parameters are $\boldsymbol{\theta}=(n,q)$. Here, $\rho_{\odot \mathrm{RRL}}$ is the number density of RR Lyrae at the position of the Sun, $R_{\odot}$ the distance of the Sun from the Galactic center, and $r_q$ is the flattening-corrected radius. As we are not interested in absolute numbers, we are not fitting for $\rho_{\odot \mathrm{RRL}}$.

Others presume a broken power-law (BPL) \citep[e.g.][]{Xue2015}, where an inner and outer power law index are defined. The change in the power law index then occurs by a step function at the break radius. As our sample starts at a Galactocentric radius of 20 kpc, and the break radius is found to be around or below 20 kpc \citep[e.g.][]{Xue2015}, we cannot compare to the results by \cite{Xue2015}. However, in order to compare to the findings by \cite{Deason2014} who find a BPL with three ranges of subsequently steepening slope, where one of the breaks is occurring within the distance range present in our sample, we fit a BPL:

\begin{equation}
\rho_{\mathrm{halo}}(X,Y,Z)=\left\{\begin{array}{cl} \rho_{\odot \mathrm{RRL}}\left(R_{\odot}/r_q \right)^{n_{\mathrm{inner}}},  \mbox{if } r_q \leq r_{\mathrm{break}}       \\\rho_{\odot \mathrm{RRL}}  r_{\mathrm{break}}^{n_{\mathrm{outer}}-n_{\mathrm{inner}}}    \left(R_{\odot}/r_q \right)^{n_{\mathrm{outer}}},  \mbox{else. } \end{array}\right.
\label{eq:BPL_halomodel}
\end{equation}

\subsubsection{Einasto Profile}
The Einasto profile \citep{Einasto1965,Einasto1989} is the 3D analog to the S{\'e}rsic profile \citep{Sersic1963} for surface brightnesses and has been used to describe the halo density distribution \citep{Merritt2006,Deason2011, Sesar2011, Xue2015, Iorio2017} as well as dark matter haloes \citep{Merritt2006, Navarro2010}. It is given by

\begin{equation}
\gamma (r)\equiv -{\frac {d \ln \rho (r)}{d \ln r}}\propto r^{\alpha }
\end{equation}
where the steepness of the Einasto profile, $\alpha$, changes continuously as a function of the effective radius $r_{\mathrm{eff}}$,
\begin{equation}
\alpha = -\frac{d_n}{n}\left( \frac{r_q}{r_{\mathrm{eff}}} \right) ^{1/n}.
\label{eq:einastosteepness}
\end{equation}

This can be rearranged to
\begin{equation}
\rho_{\mathrm{halo}}(r_q) \equiv \rho_0 \exp\left\{-d_n\left[\left(r_q/r_{\mathrm{eff}}\right)^{1/n}-1\right ]\right\},
\label{eq:einasto}
\end{equation}
where $\rho_0$ is the (here irrelevant) normalization, $ r_{\mathrm{eff}}$ is the effective radius, $n$ is the concentration index. The parameter $d_n$ is a function of $n$, where for $n \ge 0.5$, a good approximation is given by $ \rm d_n \approx 3n-1/3+0.0079/n$ \citep{Graham2006}. 

For an Einasto profile, the shape of the density profile is described by the parameter $n$. This profile allows for a non-constant fall-off without the need
for imposing a discontinuous break radii:
Density distributions with steeper inner profiles and shallower outer
profiles are generated by large values of $n$, whereas small values of $n$ account for a shallower inner and steeper outer profile. The parameter $r_{\mathrm{eff}}$ describes the radius of the inner core of the profile.

The free parameters of an Einasto profile with a constant flattening $q$ are $\boldsymbol{\theta}=(r_{\mathrm{eff}}, n, q)$.

\subsubsection{Profiles with Varying Flattening}

The models described so far assume a constant flattening $q$. However, \cite{Preston1991}
found evidence for a decrease in the flattening with increasing radius. \cite{Carollo2007,Carollo2010} find evidence that at least the innermost part of the halo is quite flattened.

We thus increase the complexity of the model by allowing for a non-constant flattening of the halo, parameterized by the Galactocentric radius. To describe such a radial variations of the stellar halo's flattening, we consider the functional form for $q(R_{\mathrm{gc}})$ as:

\begin{equation}
q(R_{\mathrm{gc}}) = q_{\infty} - (q_{\infty} - q_0) \exp\left(1- \frac{\sqrt{R_{\mathrm{gc}}^2 + r_0^2}}{r_0}       \right)
\label{eq:q_r}
\end{equation}
with $q_0$ being the flattening at the center, $q_{\infty}$ being the flattening at large Galactocentric radii, and $r_0$ being the exponential scale radius over which the change of flattening occurs.

Thus, the flattening $q$ now varies from $q_0$ at the center to the asymptotic value $q_{\infty}$ at large radii and the variation is tuned by the exponential scale length $r_0$.

All other equations to describe the radial profile given above apply from the previously described Einasto and power law profile, replacing only $q$ with $q(R_{\mathrm{gc}})$, thus replacing the fitting parameter $q$ with three fitting parameters $q_0$, $q_{\infty}$ and $r_0$.

\subsection{Selection Function}
\label{sec:SelectionFunction}

In general, a selection function describes the fraction of stars that are targeted, as a function of e.g. position, distance, or magnitude.

We introduce the selection function for two reasons: to correct for the non-complete volume sampling naturally occurring during a survey, and to remove known overdensities
to build a ``clean" sample of RRab, eliminating all the stars belonging to the substructures from our original catalogue. 
Both cosmological models and observations imply that a good portion of halo stars, at least beyond 20 kpc, are in substructures. Especially the prominent
ones, such as the Sagittarius stream and the Virgo overdensity, can and will affect the fits of smooth models, as pointed out also by \cite{Deason2011}.

After we remove those known substructures it is, of course, still possible that there are previously unknown substructures, as well as the smooth component of the halo is also structured, but at a level that is below our resolution.

Our selection function $\mathcal{S}(l,b,D)$ is binary $\left[0,1\right]$ so that $\mathcal{S}$ is always equal to 1 except for the points $(l,b,D)$ that are excluded.
The predicted density of stars is then simply the product of the underlying density distribution with the selection function, suggesting that one constrains this underlying density by forward modeling of the observations.

The RRab candidates from \cite{Sesar2017b} were selected uniformly from the set of objects in the PS1 3$\pi$ survey in the area and apparent magnitude range available for this survey. The selection completeness and purity are uniform over a wide range of apparent magnitude up to a flux-averaged r-band
magnitude of 20 mag \citep{Sesar2017b}, which is described later on in Equ. \eqref{eq:selectionbrightness}.

Starting from the 44,403 RRab stars in the sample of \cite{Sesar2017b}, we exclude known overdensities in $(l,b,D)$. Among the largest overdensities are the Sagittarius stream, dwarf galaxies such as Draco dSph, and globular clusters. A complete list can be found in Tab. \ref{tab:overdensities}.
Also, we cut out sources too close to the Galactic plane ($\vert b \vert <10 \arcdeg$), or too close to the Galactic center ($R_{\mathrm{gc}} \leq 20~{\mathrm{kpc}}$), as we want to avoid regions with many overdensities such as streams as mostly found within 20 kpc, want to excise the Galactic bulge, and additionally the RRab sample is relatively sparse towards the Galactic disc.

From the 33,378 sources we exclude in total, 6,575 are within $\pm 10 \arcdeg$ of the Galactic plane, 26,951 are within 20 kpc of the Galactic center, 5,960 are in the Sgr stream and 578 are in other overdensities as listed in Tab. \ref{tab:overdensities}; as those regions partially overlap, the numbers stated here would add up to 35,484.

The selection function $\mathcal{S}(l,b,D)$ is thus composed of:
\begin{equation}
\mathcal{S}(l,b,D) = \mathcal{S}_{\mathrm{RRL}}(l,b,D) \times \mathcal{S}_{\mathrm{area}}(l,b,D)
\label{eq:selfunc}
\end{equation}
where $\mathcal{S}_{\mathrm{RRL}}(l,b,D)$ describes the selection cuts of the sample introduced by the survey and \cite{Sesar2017b} itself leading to the 44,403 RRab stars, and 
$\mathcal{S}_{\mathrm{area}}(l,b,D)$ describes area cuts to exclude overdensities.

The area and depth of the PS1 sample of RRab lead to

\begin{equation}
\mathcal{S}_{\mathrm{RRL}}(l,b,D)=\left\{\begin{array}{cl} 1, & \mbox{if } \mathrm{\delta}>-30\arcdeg \; \mathrm{and} \; D_{\mathrm{min}}<D<D_{\mathrm{max}}   \\ 0, & \mbox{else. } \end{array}\right.
\end{equation}

The spatial cuts to geometrically excise bulge and thick-disc stars beyond a Galactocentric distance of 20 kpc are

\begin{equation}
\mathcal{S}_{\mathrm{bulge,disc}}(l,b,D)=\left\{\begin{array}{cl} 1, & \mbox{if } \vert b \vert \geq 10 \arcdeg \; \mathrm{and} \; R_{\mathrm{gc}} \geq 20~{\mathrm{kpc}}\\0, & \mbox{else. } \end{array}\right.
\end{equation}

The spatial cuts to geometrically excise the Sagittarius (Sgr) stream are based on our previous work describing the Sgr stream's 3D geometry as traced by PS1 RRab stars \citep{Hernitschek2017}. To each star in the sample, we can assign a probability that it is associated with the Sgr stream, $p_{\mathrm{sgr}}$ \citep[][see Equ. (11) therein]{Hernitschek2017}. We excise sources with $p_{\mathrm{sgr}}>0.2$ as members of the Sgr stream, leading to a selection function of
\begin{equation}
\mathcal{S}_{\mathrm{sgr}}(l,b,D)=\left\{\begin{array}{cl} 1, & \mbox{if } p_{\mathrm{sgr}}(l,b,D)<0.2\\0, & \mbox{else. } \end{array}\right.
\label{eq:sgrselfunc}
\end{equation}

Additional spatial cuts are used to remove all stars in the boxes listed in Table \ref{tab:overdensities} in the Appendix, in order to excise known overdensities. This results into $\mathcal{S}_{\mathrm{other\;overdensities}}(l,b,D)$.

Taking into account that the RRab sample is not complete, with the completeness varying with magnitude, another term for the selection function needs to be introduced.

\cite{Sesar2017b} find that the RRab selection function is approximately constant at ${\sim} 90\%$ for a flux-averaged $r$-band magnitude $r_{\mathrm{F}} \lesssim 20$ mag, after which it steeply drops to zero at $r_{\mathrm{F}} \sim 21.5$  mag.
Writing $r_{\mathrm{F}}$ as $r_{\mathrm{F}}(D)$, the selection function characterizing the distance-dependent completeness is

\begin{equation}
\mathcal{S}_c(r_{\mathrm{F}}) = L - \frac{L}{1+\exp \left( -k \left( r_{\mathrm{F}}-x_0 \right) \right)   }
\label{eq:selectionbrightness}
\end{equation}

with \citep{Sesar2017b}:
\begin{align}
L&=0.91 \\
k&=4.0\\
x_0&=20.6\\
r_{\mathrm{F}}&=2.05\log(D)+11.
\end{align}

In addition, we estimated the distance-dependent purity to supplement the overall sample purity that was given as 90\% by \cite{Sesar2017b}.
Using the RRab sample within SDSS S82, as done by \cite{Sesar2017b} to estimate the distance-dependent completeness of our RRab sample, we find the purity staying stable at a level of 98\% to 95\% over a range from 15 to more than 20 mag in the $r$ band. In contrast, over the same magnitude range, the completeness drops from 91\% to 80\%.
The faintest RRab in S82 (which we use as validation set, see \citep{Sesar2017b}) is
found at $r_{\mathrm{F}}$=20.58 mag, and there are in total only two sources in this faintest 0.5 mag bin. The 10 faintest RRab stars within S82 span a distance range
from 85 to 102 kpc. This means that for sources fainter than 20.5 mag, the purity cannot be estimated in this way.
For sources beyond $D=90$ kpc, we adopted a purity of 94\%.
There is no SDSS source within S82 that was not picked up by PS1. The different distance dependency of purity and completeness reflects that it is easier to lose objects (i.e. not to classify them as RRab stars) than to get spurious sources into the catalog of PS1 RRab stars, given the rigorous definition adopted to consider a star as RRab \citep{Sesar2017b}.
Although the effect of the purity is negligible, as the effect of a dropping completeness at large distances dominates, and we cannot determine the purity beyond $D=90$ kpc, we included it as part of the selection function, $\mathcal{S}_{p}(D)$.

We end up with a selection function
\begin{align}
\label{eq:allselfunc}
\mathcal{S}(l,b,D) =& \mathcal{S}_{\mathrm{RRL}}(l,b,D) \times  \mathcal{S}_{c}(D)  \times  \mathcal{S}_{p}(D)\times \mathcal{S}_{\mathrm{area}}(l,b,D) \\ 
=& \mathcal{S}_{\mathrm{RRL}}(l,b,D) \times  \mathcal{S}_{c}(D) \nonumber \\ & \times \mathcal{S}_{\mathrm{bulge,disc}}(l,b,D) \nonumber \\ & \times \mathcal{S}_{\mathrm{Sgr}}(l,b,D) \times \mathcal{S}_{\mathrm{other\;overdensities}}(l,b,D). \nonumber
\end{align}

The overdensities listed in Table \ref{tab:overdensities} are chosen in the following way:
Based on a list of dwarf galaxies within 3 Mpc by \cite{McConnachie2012}, its update from 2014\footnote{\url{https://www.astrosci.ca/users/alan/Nearby_Dwarfs_Database.html}}
and a list of currently known halo streams by \cite{GrillmairBook}, we select overdensities that could show up in a survey that covers the position and distance cuts of PS1 3$\pi$. We check each overdensity to see if it appears in the RRab sample, and if so, cut it out by defining a selection box in $(l,b,D)$. We end up with the cuts described in Table \ref{tab:overdensities}. 

After excising stars using $\mathcal{S}_{\mathrm{area}}(l,b,D)$, the sample reduces to 11,025 RRab stars which we call the ``cleaned sample''. The original and the cleaned sample are shown in Fig. \ref{fig:sample_map}.

Out of these sources, 679 lie beyond a Galactocentric distance of 80 kpc, and 101 beyond 100 kpc, in contrast to 1093 sources beyond 80 kpc, and 
238 beyond 100 kpc in the original sample.

We now incorporate this selection function in fitting a parameterized model for the stellar density of the halo.

\subsection{Constraining Model Parameters}
\label{sec:ConstrainingModelParameters}

With the models $\rho_{\mathrm{RRL}} (\mathcal{D} \vert \boldsymbol{\theta})$ and the selection function $\mathcal{S}$ at hand, we can directly calculate the likelihood of the data $\mathcal{D}$ given the model $\rho_{\mathrm{RRL}}$, the fitting parameters $\boldsymbol{\theta}$ and the selection function $\mathcal{S}$ following \cite{Bovy2012}.

The normalized un-marginalized log likelihood for the $i$-th star with the observables $\mathcal{D}_i$ is then
\begin{equation}
\ln p(\mathcal{D}_i \vert \boldsymbol{\theta}) = \frac{\rho_{\mathrm{RRL}}(\mathcal{D}_i \vert \boldsymbol{\theta}) \vert \mathbf{J} \vert   \mathcal{S} (l_i,b_i,D_i) }{\int \int \int  \rho_{\mathrm{RRL}}(l,b,D \vert \boldsymbol{\theta})  \vert \mathbf{J}\vert  \mathcal{S}(l,b,D) \mathrm{d}l \mathrm{d}b \mathrm{d}D  }      
\label{eq:likelihood}
\end{equation}
where the normalization integral is over the observed volume.
The Jacobian term $\vert \mathbf{J}\vert=D^2 \cos b$ reflects the transformation from $(X,Y,Z)$ to $(l,b,D)$ coordinates.

We evaluate the logarithmic posterior probability of the parameters $\boldsymbol{\theta}$ of the halo model, given the full data $\mathcal{D}$ and a prior $p (\boldsymbol{\theta})$, $\ln p(\boldsymbol{\theta} \vert \mathcal{D}) = \ln p (\mathcal{D} \vert \boldsymbol{\theta}) + \ln p (\boldsymbol{\theta})$ with 
\begin{equation}
\ln p (\mathcal{D} \vert \boldsymbol{\theta}) = \sum_i \ln p(\mathcal{D}_i \vert \boldsymbol{\theta})
\label{eq:posterior}
\end{equation}
being the marginal log likelihood for the full data set.

To determine the best-fit parameters and their uncertainties, we sample the posterior probability over the parameters space with Goodman \& Weare's Affine
Invariant Markov Chain Monte Carlo \citep{Goodman2010}, making use of the Python module \texttt{emcee} \citep{Foreman2013}. 

The final best-fit values of the model parameters have been estimated using the median of the posterior distributions, the uncertainties have been estimated using the 15.87th and 84.13th percentiles. For a parameter whose \textit{pdf} can be well-described by a Gaussian distribution, the difference between the 15.87th and 84.13th percentile is equal to 1$\sigma$.

The calculation of the normalization integral in Equ. \eqref{eq:likelihood} is complicated by the presence of the selection function $\mathcal{S}$, leading to the fact that in some regions of the integrated space, the integrand function is not continuous and shows an abrupt decrease to 0. For this reason, the classical multi-dimensional quadrature methods in Python are not able to give robust results. We decided to calculate the integral instead on a fine regular grid that is $(\Delta l = 1\arcdeg) \times (\Delta b = 1\arcdeg) \times (\delta D = 1\;\mathrm{kpc})$ wide.

\subsubsection{Model Priors}

We now lay out the ``pertinent range'', across which the model priors are given. We set a different prior distribution $p (\boldsymbol{\theta})$ for each of the five following cases:

power law model:
\begin{align}
\ln p(\boldsymbol{\theta}) =& \mathrm{Uniform}( 1.0 < n < 6.0) \\&+ \mathrm{Uniform}( 0.1  < q < 1.0) \nonumber
\label{eqn:prior_powerlaw}
\end{align}

BPL model:
\begin{align}
\ln p(\boldsymbol{\theta}) =& \mathrm{Uniform}( 1.0 < n < 6.0) \\&+ \mathrm{Uniform}( 0.1  < q < 1.0) \nonumber\\&+ \mathrm{Uniform}( \log(R_{\mathrm{min}}) < \log(r_{\mathrm{break}})< \log (R_{\mathrm{max}})) \nonumber
\label{eqn:prior_BPL}
\end{align}
where $R_{\mathrm{min}}$, $R_{\mathrm{max}}$ give the Galactocentric distance range available in the sample.

Einasto profile:
\begin{align}
\ln p(\boldsymbol{\theta}) =& \mathrm{Uniform}( \log(0.01) < \log(r_{\mathrm{eff}})< \log (50)) \\&+ \mathrm{Uniform}( \log(0.01) < \log(r_0)< \log (50)) \nonumber \\
&+\mathrm{Uniform}( 0.5 < n < 20.0) \nonumber \\ &+ \mathrm{Uniform}( 0.1  < q < 1.0) \nonumber
\label{eqn:prior_einasto}
\end{align}

power law model with $q(R_{\mathrm{gc}})$:
\begin{align}
\ln p(\boldsymbol{\theta}) =& \mathrm{Uniform}( \log(0.01) < \log(r_0)< \log (50)) \\
&+\mathrm{Uniform}( 1.0 < n < 5.0) \nonumber \\&+ \mathrm{Uniform}( 0.1  < q_0 < 1.0) \nonumber \\&+ \mathrm{Uniform}( 0.1  < q_{\infty} < 1.0) \nonumber
\label{eqn:prior_powerlaw_qr}
\end{align}

Einasto profile with $q(R_{\mathrm{gc}})$:
\begin{align}
\ln p(\boldsymbol{\theta}) =& \mathrm{Uniform}( \log(0.01) < \log(r_{\mathrm{eff}})< \log (50)) \\
&+\mathrm{Uniform}( \log(0.01) < \log(r_0)< \log (50)) \nonumber \\
&+ \mathrm{Uniform}( \log(0.01) < \log(r_0)< \log (50)) \nonumber \\
&+\mathrm{Uniform}( 0.5 < n < 20.0) \nonumber \\&+ \mathrm{Uniform}( 0.1  < q_0 < 1.0) \nonumber \\&+ \mathrm{Uniform}( 0.1  < q_{\infty} < 1.0) \nonumber
\label{eqn:prior_einasto_qr}
\end{align}

\subsection{Fitting Tests on Mock Data}
\label{sec:TestsOnMockData}

In order to test the methodology for fitting the density as discussed in Section \ref{sec:DensityFitting}, we created mock data samples of RR Lyrae stars in the Galactic halo, which should have the same properties as the observed sample of RRab stars, using a combination of a density law and assumptions on the selection function imposed by both PS1 3$\pi$ and our selection cuts (Sec. \ref{sec:SelectionFunction}).
In detail, we first sampled ${\sim}$50,000 stars from mock halos generated with an underlying density given by a power law, Einasto profile, power law with $q(R_{\mathrm{gc}})$, or Einasto profile with $q(R_{\mathrm{gc}})$. We then applied a 3\% distance uncertainty, superimpose the sample with faint and far Gaussian blobs away from the regions excluded by the selection function to simulate unknown overdensities, added the RRab known as members of the Sgr stream, and then applied the selection function. After that, the sample has ${\sim}$12,000 sources, and we randomly sample 11,025 sources to match the cleaned observed sample.

An example of a simulated distribution of halo RR Lyrae is shown in the upper panel of Fig. \ref{fig:compare_predicted_observed}.

We then run the same analysis code on this sample as for the PS1 3$\pi$ RRab sample. This enables us to estimate which halo properties we are able to identify and constrain with our approach.

We find results that are consistent with the input model within reasonable uncertainties, which means that we are able to recover the input parameters for all models in their assumed parameter range, and compare well with results we got from the PS1 3$\pi$ data.

The one- and two-dimensional projections of the posterior probability distributions (\textit{pdf}) for fitting one of these mock halos, along with the parameters used to generate the mock halo, are given in Fig. \ref{fig:triangle_allsky_powerlaw_qr_mockhalo}.

\section{Results}
\label{sec:Results}

We now present the results of applying the modeling from Sec. \ref{sec:StellarDensityModels} to the cleaned sample of RRab stars as described in Sec. \ref{sec:SelectionFunction}.
We fitted the simplest model, a power law, to our data, as well as the Einasto model to allow easy comparison with density profiles obtained from
N-body simulations \citep{Navarro2004, Diemand2004, Merritt2006, Graham2006}. Both models are fitted with a constant halo flattening parameter $q$ as well as with a distance-dependent flattening $q(R_{\mathrm{gc}})$. We illustrate these results in three ways: first by showing the predicted distribution by the best-fit models, then by showing the joint posterior distribution functions of the halo model parameters of each model; and third, we compare the models using the Bayesian information criterion (BIC).

First, we discuss the result of fitting the complete cleaned 3$\pi$ sample, in order
to explore the broad trends in spatial structure. Subsequently, we split the sample into two hemispheres as well as into relatively broad  $\Delta l = 30 \arcdeg$, $\Delta b = 60 \arcdeg$ bins and map the local halo structure. Finally, we calculate and analyze the residuals of the best-fit model.

\subsection{Best-Fit Model Parameters}
\label{sec:BestFitModelParameters}

Based on the five models described above in Sec. \ref{sec:StellarDensityModels} and the selection function as described in Sec. \ref{sec:SelectionFunction}, we apply our likelihood approach (Sec. \ref{sec:ConstrainingModelParameters}) in order to constrain the best-fit model parameters.

We estimate those best-fit model parameters for the complete cleaned RRab sample which spans 3/4 of the sky and contains 8,917 sources. 
Fig. \ref{fig:halofit_allsky} compares the observed number density of RR Lyrae stars with the density predicted by best-fit models.

Table \ref{tab:bestfit} summarizes the best-fit parameters of our five halo density models. For each model, we give the type of the density model, its best-fitting parameters along with their 1$\sigma$ uncertainties estimated as the 15.87th and 84.13th percentiles, and the maximum log likelihood $\ln (\mathcal{L}_{\mathrm{max}})$. We also give the BIC, a measure for model comparison described in Sec. \ref{sec:ComparingModels}.

The one- and two-dimensional projections of the posterior probability distributions (\textit{pdf}) for each model are given in Fig. \ref{fig:triangle_powerlaw} to \ref{fig:triangle_einasto_qr}.

For the power-law and BPL model, the \textit{pdf} shows an almost Gaussian-like distribution with no covariance between the model parameters $q$ and $n$.\newline
For the Einasto profile, as for the power law, the concentration index $n$ and the flattening parameter $q$ show an almost Gaussian distribution with no covariance.
The concentration index $n$ is covariant with the effective radius parameter, $r_{\mathrm{eff}}$.\newline
The \textit{pdf} of the power law model with $q(R_{\mathrm{gc}})$ shows covariance, and the \textit{pdf} is strongly distorted from a Gaussian distribution.\newline
For the Einasto profile with $q(R_{\mathrm{gc}})$, the \textit{pdf} is more complex and skewed. The fitting parameters $r_0$, $q_0$, $q_{\infty}$ show a covariance, but their marginalizations have a Gaussian-like appearance.

Among models with constant flattening, the distribution of the sources is reasonably well fit by a power law model with $n=4.40^{+0.05}_{-0.04}$ and a halo flattening of $q=0.918^{+0.016}_{-0.014}$. Allowing for a break in the power-law profile, we find a break radius of $r_{\mathrm{break}}=38.7^{+0.69}_{-0.58}$, a halo flattening of $q=0.908^{+0.008}_{-0.006}$, and the inner and outer slopes $n_{\mathrm{inner}}=4.97^{+0.02}_{-0.05}$ and $n_{\mathrm{outer}}=3.93^{+0.05}_{-0.04}$, respectively. The distance distribution is fit comparably well by a model with an Einasto profile with $n=9.53^{+0.27}_{-0.28}$, an effective 
radius $r_{\mathrm{eff}}=1.07 \pm 0.10 \; \mathrm{kpc}$  and a halo flattening of $q_0=0.923 \pm 0.007$.
If we allow for a radius-dependent flattening $q(R_{\mathrm{gc}})$, we find the best-fit parameters for a power law model with $q(R_{\mathrm{gc}})$ as $r_0=25.0^{+1.7}_{-1.8}\; \mathrm{kpc}$, $n=4.61 \pm 0.03$, $q_0=0.773^{+0.017}_{-0.016}$, $q_{\infty}=0.998^{+0.002}_{-0.001}$.
The best-fit parameters for an Einasto profile with $q(R_{\mathrm{gc}})$ are $r_0=26.7^{+2.2}_{-2.0}\; \mathrm{kpc}$,  
$q_0=0.779 \pm 0.018$, $q_{\infty}=0.998^{+0.001}_{-0.002}$, $r_{\mathrm{eff}}=1.04^{+0.25}_{-0.13}\; \mathrm{kpc}$, $n=8.78^{+0.33}_{-0.30}$.

We find here $q_0<q_{\infty}$ for both models with variable flattening, indicating that the inner halo is more flattened than the outer halo. 
Assuming a constant flattening $q$ instead, its best-fit value is also consistent among the power law and Einasto profile models. 

For all five models, the best-fit values along with their $1\sigma$ uncertainties are summarized in Tab. \ref{tab:bestfit}.

Our results confirm that if a varying flattening is assumed, the halo profile has a $r_0$ close to 20 kpc and the inner halo is more flattened than the outer. This is also consistent with results by \cite{Carollo2007} and \cite{Carollo2010}, as well as \cite{Xue2015}, \cite{Das2016} and \cite{Iorio2017}.
For a BPL, we cannot confirm \cite{Deason2014} result of a steepening found beyond 65 kpc.
We discuss our results in comparison with previous attempts in more detail in Sec. \ref{sec:Comparison}.

\subsection{Comparing Models}
\label{sec:ComparingModels}

We have estimated the best-fitting parameters for each model. In addition to that, it is important to compare the results of different models to
determine which of them gives the best description of the data.

The most reliable way would be to compute the ratio of the Bayesian evidence, which is defined as the integral of the likelihood over all of the parameter
space, for each model in order to compare them. Especially in higher-dimensional parameter spaces, like the ones we deal with here, this turns out to be too
computationally expensive. However, under the assumption that the posterior distributions are almost Gaussian, an approximation can be used, called the Bayesian information criterion \citep[BIC]{Schwarz1978}. 

The BIC takes into account both the statistical goodness of fit, as well as the number of parameters that have to be estimated to achieve this particular degree of fit, by imposing a penalty for increasing the number of parameters in order to avoid overfitting. The BIC is defined as

\begin{equation}
\mathrm{BIC} = \mathrm{dim}(\boldsymbol{\theta}) \ln(N) -2 \ln (\mathcal{L}_{\mathrm{max}})
\end{equation}
where $\boldsymbol{\theta}$ are the model parameters, $N$ is the number of objects in the sample, and $\mathcal{L}_{\mathrm{max}}$ is the maximum likelihood, where we defined the likelihood function in Equ. \eqref{eq:posterior} as $\ln p (\mathcal{D} \vert \boldsymbol{\theta})$.

Using the BIC for selecting a best-fit model, the model with lowest BIC is preferred.

We have computed the BIC for all of our models, and show them in Tab. \ref{tab:bestfit} along with the best-fit parameters.

According to the BIC, we find the best-fit model to be the power law with $q(R_{\mathrm{gc}})$, followed by the Einasto profile with $q(R_{\mathrm{gc}})$, the constant-flattening power law, the constant-flattening Einasto profile and finally the BPL. As the values of BIC in Tab. \ref{tab:bestfit} indicate, allowing for flattening variations makes for distinctly better fits to the distribution of the RRab stars.

However, attention should be paid to the shape of the posterior distribution. When calculating the BIC, it is assumed that the posterior distributions are reasonable comparable to a Gaussian. As we see from Fig. \ref{fig:triangle_powerlaw} to \ref{fig:triangle_einasto_qr}, the power-law model as well as the Einasto profile have posterior distributions that compare well to a Gaussian distribution, whereas for the cases with $q(R_{\mathrm{gc}})$, the posterior distributions are somewhat distorted and show also a covariance between parameters.

Another issue is whether a difference in BIC is significant. A rating of the strength of the evidence against the model with the higher BIC value is given in \cite{Kass1995}: A $\Delta\mathrm{BIC}>10$ indicates a very strong evidence against the model with the higher BIC.

\subsection{Local Halo Properties}
\label{sec:LocalHaloProperties}

In Section \ref{sec:BestFitModelParameters}, we estimated best-fit parameters for the complete cleaned RRab sample which spans 3/4 of the sky. Here, we now estimate them on smaller parts of the sky. This will help us to resolve and identify possible local variations in the best-fit model, especially in the halo flattening $q$ and steepness $n$. We also look for previously unknown overdensities that we might find due to the spatial extent and depth of the RRab sample.

\subsubsection{Fitting Hemispheres and Pencil Beams}

We now fit the halo profile for both the north and south Galactic hemisphere independently, in order to explore what effects on our models -- of rather restrictive functional form -- are. The north hemisphere contains 6,880 sources, whereas the south hemisphere contains only 4,145 sources because of the PS1 3$\pi$ survey footprint.

The results of this fitting attempt are summarized in Tab. \ref{tab:hemispheres}. What we find is that the steepness parameters $n$ of all best-fit hemisphere models compare well for both the north and south Galactic hemisphere and also compare well with the fit for the complete halo. When taking a look at the flattening-related parameters, $q$, $q_0$, $q_{\infty}$, $r_{\mathrm{eff}}$, we find that for models with constant flattening (both the power law and Einasto profile models), $q_{\mathrm{south}} < q < q_{\mathrm{north}}$. In the case of models with $q(R_{\mathrm{gc}})$, we find the value of parameter $q_0$ being smaller for the south than for the north hemisphere,
$q_{0,\mathrm{south}} < q_0 < q_{0,\mathrm{north}}$, whereas the value of the parameter $q_{\infty}$ is similar for both hemispheres.
Furthermore, we find that $r_{0,\mathrm{north}}>r_{0,\mathrm{south}}>r_0$ for both the power law with $q(R_{\mathrm{gc}})$ and the Einasto profile with $q(R_{\mathrm{gc}})$.

The results of finding $q_{\mathrm{south}} < q < q_{\mathrm{north}}$ for models with constant flattening, and $q_{0,\mathrm{south}} < q_0 < q_{0,\mathrm{north}}$, $q_{\infty,\mathrm{south}}  \sim q_{\infty,\mathrm{north}} \sim q_{\infty,\mathrm{south}}$ $r_{0,\mathrm{north}}>r_{0,\mathrm{south}}>r_0$ in the case of a radius-dependent flattening, are consistent: by definition of $q(R_{\mathrm{gc}})$ (Eq. \eqref{eq:q_r}), $q_0$ is the flattening at center, $q_{\infty}$ is the flattening at large Galactocentric radii, and $r_0$ is the exponential scale radius over which the change of flattening occurs. A larger $r_0$ means that the flattening of the inner halo, where we find $q_0<q_{\infty}$, is in force out to a larger radius than for a smaller $r_{\mathrm{eff}}$, thus leading to a larger part of the halo being more flattened.

The generalized result is thus that the south Galactic hemisphere is somewhat more flattened than the north Galactic hemisphere. 

We also tried fitting models to the data in disjoint pencil beams $(\Delta l=30\arcdeg) \times (\Delta b=60\arcdeg)$, to further understand possible local variations in the best-fit model, especially in the halo flattening $q$ and steepness $n$.

The angular source number density for the cleaned RRab sample, given per $(\Delta l=30\arcdeg) \times (\Delta b=60\arcdeg)$ bin, is shown in Fig. \ref{fig:numberofsources}.

The resulting best-fit parameters for the power law model, power-law model with $q(R_{\mathrm{gc}})$, Einasto profile and Einasto profile with $q(R_{\mathrm{gc}})$ are shown in the Figures \ref{fig:powerlaw_60_heatmap} to \ref{fig:einasto_qr_60_heatmap}, and are given in the Tables Tab. \ref{tab:powerlaw_60} to \ref{tab:einasto_qr_60} along with their $1\sigma$ uncertainties.

The fitting procedure also works well with small pieces of the sky. As an example, we show the fitted models for two small patches on the sky,
$240\arcdeg < l < 270\arcdeg$, $-30 \arcdeg < b < 30 \arcdeg$ and $30\arcdeg < l < 60\arcdeg$, $-90\arcdeg < b < -30\arcdeg$ (see Fig. \ref{fig:halofit_patches}).
To illustrate the fitting performance further, in Fig. \ref{fig:triangle_powerlaw_qr_l270_b30_mockhalo} we give the posterior probability distribution in the case of fitting a power law with varying flattening $q(R_{\mathrm{gc}})$ (Equ. \eqref{eq:q_r}) to a $30 \arcdeg \times 60 \arcdeg$ patch of mock data.
The posterior distribution is comparable to those when fitting the same halo profile to the full cleaned sample (see Fig. \ref{fig:triangle_allsky_powerlaw_qr_mockhalo} for comparison).
However, the width of the posterior probability distribution increases compared to the full cleaned sample. This is also reflected in the $1\sigma$ intervals given along with the best-fit parameters in the top right part of the figure.

Starting with the results of fitting the power law model to the $(\Delta l=30\arcdeg) \times (\Delta b=60\arcdeg)$ patches, Fig. \ref{fig:powerlaw_60_heatmap}, we find 
that the flattening parameter $q$ is homogeneous over almost the complete sky. There are a few exceptions, i.e.  for $0\arcdeg < l < 30\arcdeg$, $-90\arcdeg<b<-30\arcdeg$, the resulting flattening parameter $q$ is suspiciously small. However, within that region, there are only 22 sources which makes a reliable fit difficult.

Again for $300<l<330$, $-30<b<30$, the resulting $q$, and here also the power-law index $n$, is small. Since there are only 2 sources within that region, we obviously have to exclude that fit.
Outside of these regions, the resulting $q$ and $n$ are relatively homogeneous, with a trend to smaller $n$ near the edges of the survey (see white empty region at $l>240 \arcdeg$ in the figures) and at very high latitudes.

For the Einasto profile, Fig. \ref{fig:einasto_60_heatmap}, we also find regions on the sky where the fitting parameters are considerably deviating.
For the parameter $n$, this is especially the case for $180\arcdeg<l<210\arcdeg$, $-30\arcdeg<b<90\arcdeg$, as well as at some regions at high latitudes. In those cases, the best-fitting $n$ is sometimes much higher and sometimes much smaller than for the power law model; however, this is a result of the different definition of $n$ in both models (see Equ. \eqref{eq:halomodel} vs. Equ. \eqref{eq:einasto}, and the steepness of the Einasto profile not being constant but changing continuously as given by Equ. \eqref{eq:einastosteepness}), and the fitted profiles look comparable.

In the case of a power law with $q(R_{\mathrm{gc}})$, as shown in Fig. \ref{fig:powerlaw_qr_60_heatmap}, the best-fit values for $q_0$, $q_{\infty}$ are more similar than for the fit of the complete halo.
In general, as is clearly visible in Fig. \ref{fig:powerlaw_qr_60_heatmap}, the distribution of the halo-flattening parameters $q_0$, $q_{\infty}$ and the power-law index $n$ shows more scatter than for a power law with constant halo flattening.
This might be caused by the model tending to overfit the data, a problem common to higher-dimensional models, overreacting to fluctuations in the underlying data set that should be fitted.

In the case of an Einasto profile with $q(R_{\mathrm{gc}})$, as shown in Fig. \ref{fig:einasto_qr_60_heatmap}, the best-fit values for $q_0$, $q_{\infty}$ are again more similar than for the fit of the complete halo. 
We find about the same deviations as reported for the other models, such as unreliable fits at $0\arcdeg < l < 30\arcdeg$, $-90\arcdeg<b<-30\arcdeg$, and $300<l<330$, $-30<b<30$, due to the small number of sources within that regions.

Within $180\arcdeg<l<210\arcdeg$, the best-fit value of $n$ is influenced by the presence of outskirts of the Sagittarius stream, which were not fully removed by our cuts. Within this region, a small number of stars from the stream appears to be present, and in general, the number of sources in this region of the sky is small after applying our cuts on overdensities.
This is also the case for $300\arcdeg<l<330\arcdeg$, $-30\arcdeg<b<90\arcdeg$. A higher-dimensional model is more affected by this than a lower-dimensional one; compare the extreme cases of the 2-dimensional power law model and the 5-dimensional Einasto profile with $q(R_{\mathrm{gc}})$.

Again, in those cases, the best-fitting $n$ is much higher than for the case of a power law model; however, this is a result of the different definition of $n$ in each model (see Equ. \eqref{eq:halomodel} vs. Equ. \eqref{eq:einasto}, and the steepness of the Einasto profile being not constant but changing continuously as given by Equ.  \eqref{eq:einastosteepness}), and the fitted profiles look comparable.

We give the mean and variance for the best-fit parameters on $\Delta l = 30 \arcdeg$, $\Delta b = 60 \arcdeg$ bins for all four models in Tab. \ref{tab:mean_variance_60deg}.

\subsubsection{Density Residuals and their Significance}
\label{sec:ResidualsAndSignificance}
Additionally, we compared the best-fit model to the PS1 3$\pi$ RR Lyrae sample by calculating the residuals of that model.
In Fig. \ref{fig:compare_predicted_observed}, we give density plots in the Cartesian reference frame $(X, Y, Z)$ (see Equ. \eqref{eq:cartesian}) for the best-fit model, as well as residuals for the observed cleaned sample of PS1 3$\pi$ RRab stars. Densities are each color-coded according to the legend.

\begin{figure*}
\begin{center}  
\subfigure[ ]{\includegraphics[trim=0.0inch 0.0inch 0.0inch 0.4inch, clip=true]{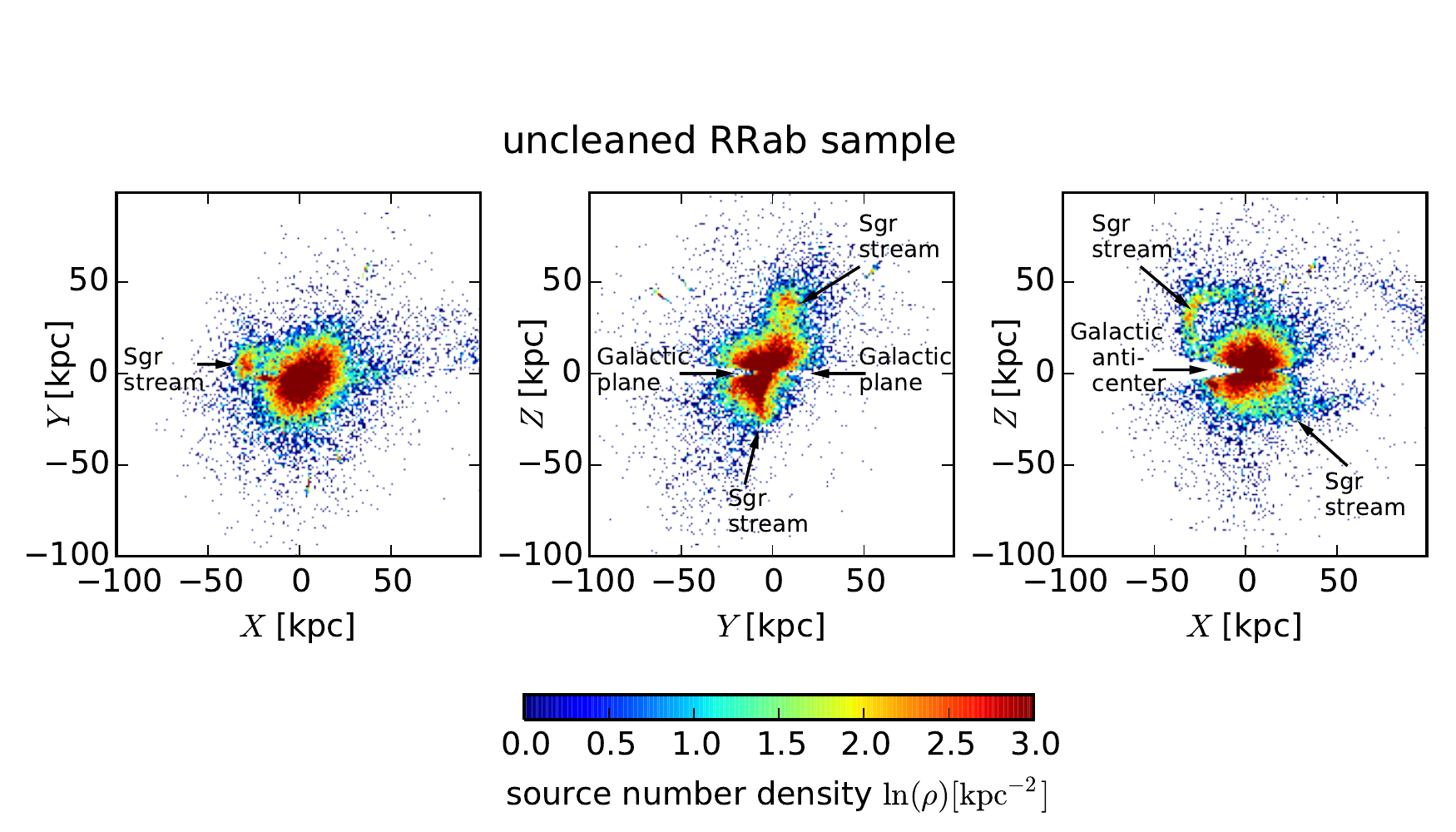}} 
\subfigure[ ]{\includegraphics[trim=0.0inch 0.0inch 0.0inch 0.4inch, clip=true]{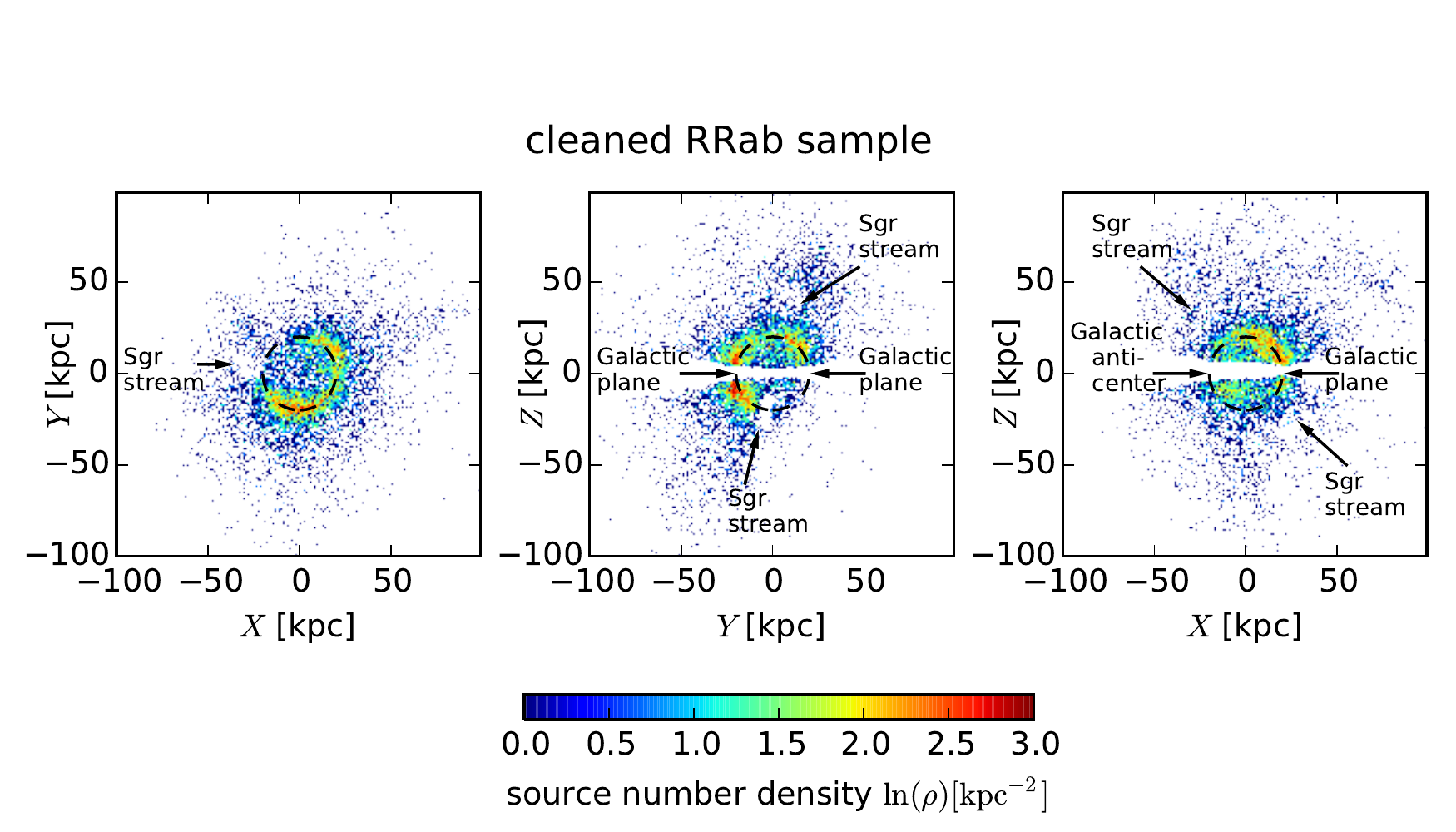}} 
\caption[The uncleaned and cleaned PS1 3$\pi$ sample of RRab stars, shown in the Cartesian reference frame $(X,Y,Z)$ as given in Equ. \eqref{eq:cartesian}]
{{The PS1 3$\pi$ sample of RRab stars, shown in the Cartesian reference frame $(X,Y,Z)$ as given in Equ. \eqref{eq:cartesian}. This reference frame is centered at the Galactic center, the Galactic disc is placed in the $(X,Y)$ plane, the $X$ axis is pointing to the Sun and the $Z$ axis to the North Galactic Pole.
The logarithmic source number density in each projection is given for 1 kpc$^2$ wide bins.\newline
Starting from a sample of 44,403 sources \citep{Sesar2017b}, containing overdensities like globular clusters and streams and affected by sample incompleteness near the Galactic plane and apocenter (here shown in the top panel as ``uncleaned sample"), we construct a sample of 11,025 sources outside of such known overdensities. To do so, we apply the selection cuts described in Sec. \ref{sec:SelectionFunction}, to geometrically excise such overdensities. The largest overdensities removed are the Sagittarius stream (we remove sources associated with the Sgr stream according to \cite{Hernitschek2017}), as well as the thick disc (we remove sources within $\vert b \vert <10 \arcdeg$) and close to the Galactic center and the bulge (we remove sources within $R_{\mathrm{gc}} \leq 20~{\mathrm{kpc}}$).
The effects of removing those sources is clearly visible in the lower panel, and are each labeled. The dashed circle here represents the 20 kpc cut. Sources within the circle but further away are seen due to projection effects; the distinctly higher density just after 20 kpc shows the stars that are no longer affected by this distance cut.
}
\label{fig:sample_map}}
\end{center}  
\end{figure*}

\begin{figure*}
\begin{center}  
\includegraphics[]{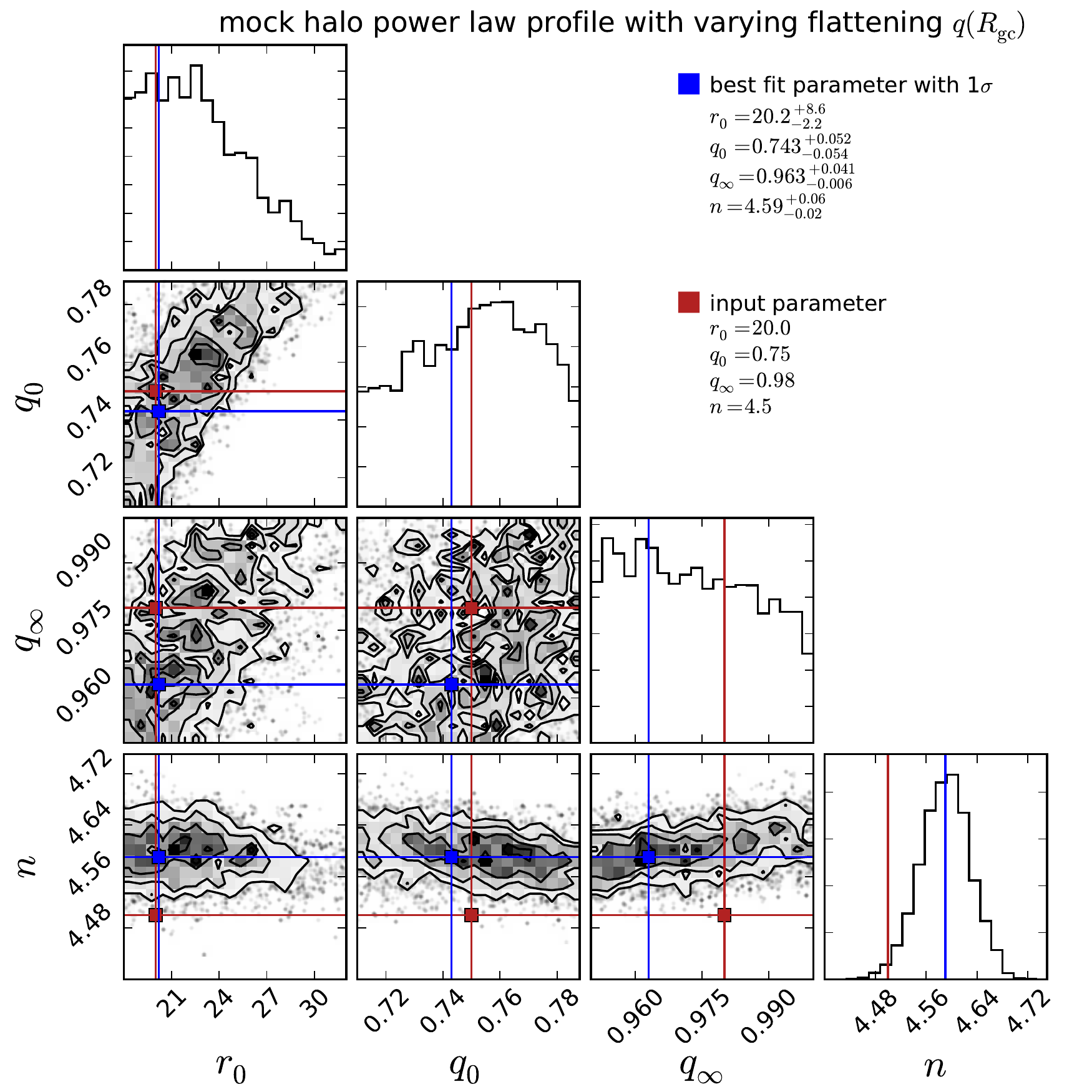}
\caption[One- and two-dimensional projections of the posterior probability distributions (\textit{pdf})
of parameters $(r_0,q_0,q_{\infty}, n)$ of the power law with varying flattening $q(R_{\mathrm{gc}})$ (Equ. \eqref{eq:q_r}) fitted to a mock sample.]
{{One- and two-dimensional projections of the posterior probability distributions (\textit{pdf})
of parameters $(r_0,q_0,q_{\infty}, n)$ of the power law with varying flattening $q(R_{\mathrm{gc}})$ (Equ. \eqref{eq:q_r}) fitted to a mock sample, used to test the methodology for fitting the halo density profile.
The blue lines and squares mark the maximum likely value of each parameter. The best-fit parameters are given along with their $1\sigma$ intervals in the top right part of the figure. The parameters used for generating the mock sample are indicated by dark red lines and squares and also given in the right part of the figure.}
\label{fig:triangle_allsky_powerlaw_qr_mockhalo}}
\end{center}  
\end{figure*}

\begin{figure*}
\begin{center}  
\subfigure[ ]{\includegraphics[trim=0.0inch 0.0inch 0.0inch 0.0inch, clip=true]{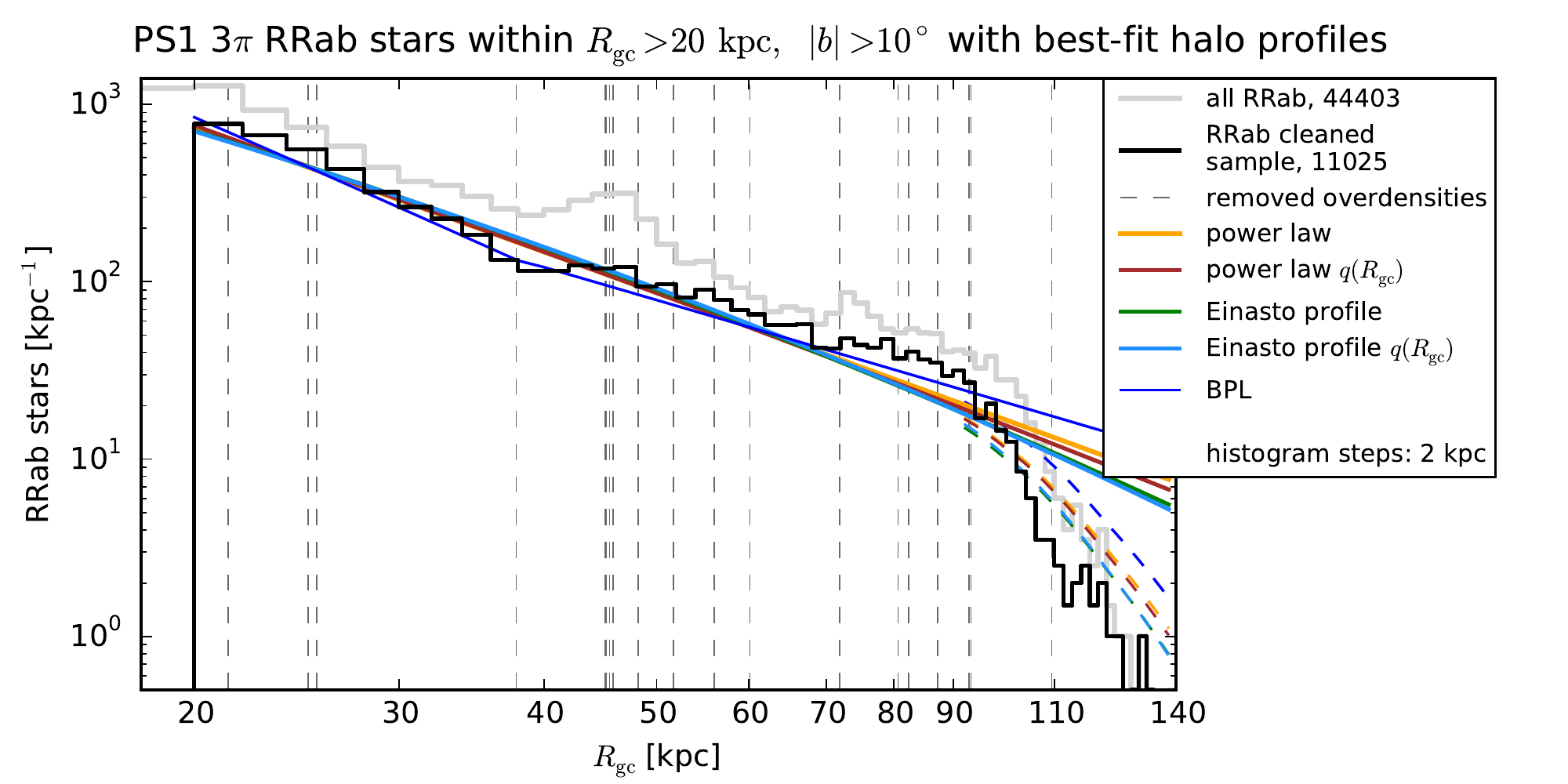}} 
\caption[Comparison between the observed distance distribution of the cleaned samples and the predicted distributions by the best-fit models, with the number density shown in a log plot.]
{{Comparison between the observed distance distribution of the cleaned samples and the predicted distributions by the best-fit models, with the number density shown in a log plot.\newline
The black histogram shows the Galactocentric distance distribution of our cleaned sample of 11,025 RRab stars, whereas the grey histogram gives the distance distribution of the full data set of 44,403 RRab stars from \cite{Sesar2017b}. Removed overdensities are highlighted with dashed lines, and are listed in Table \ref{tab:overdensities}.\newline
The overplotted solid lines represent the best-fit model for each of the five halo profiles. As a result of the selection function, these models don't follow a straight line in the log plot, but drop much more rapidly especially beyond a Galactocentric distance of 80 kpc.
For comparison, dashed lines, in the same color as the solid lines, represent each  $\rho_{\mathrm{halo}} \times \mathcal{S}(l,b,D)$, where $\mathcal{S}$ is the selection function as given in Equ. \eqref{eq:allselfunc}. We see that all each our five models can fit the distance distribution properly, and our assumption about the selection function $\mathcal{S}$ represents the true selection effects and overdensity cuts.\newline
The best-fit parameters for each of the models are given in Table \ref{tab:bestfit}.
}
\label{fig:halofit_allsky}}
\end{center}  
\end{figure*}

\begin{figure*}
\begin{center}  
\includegraphics[]{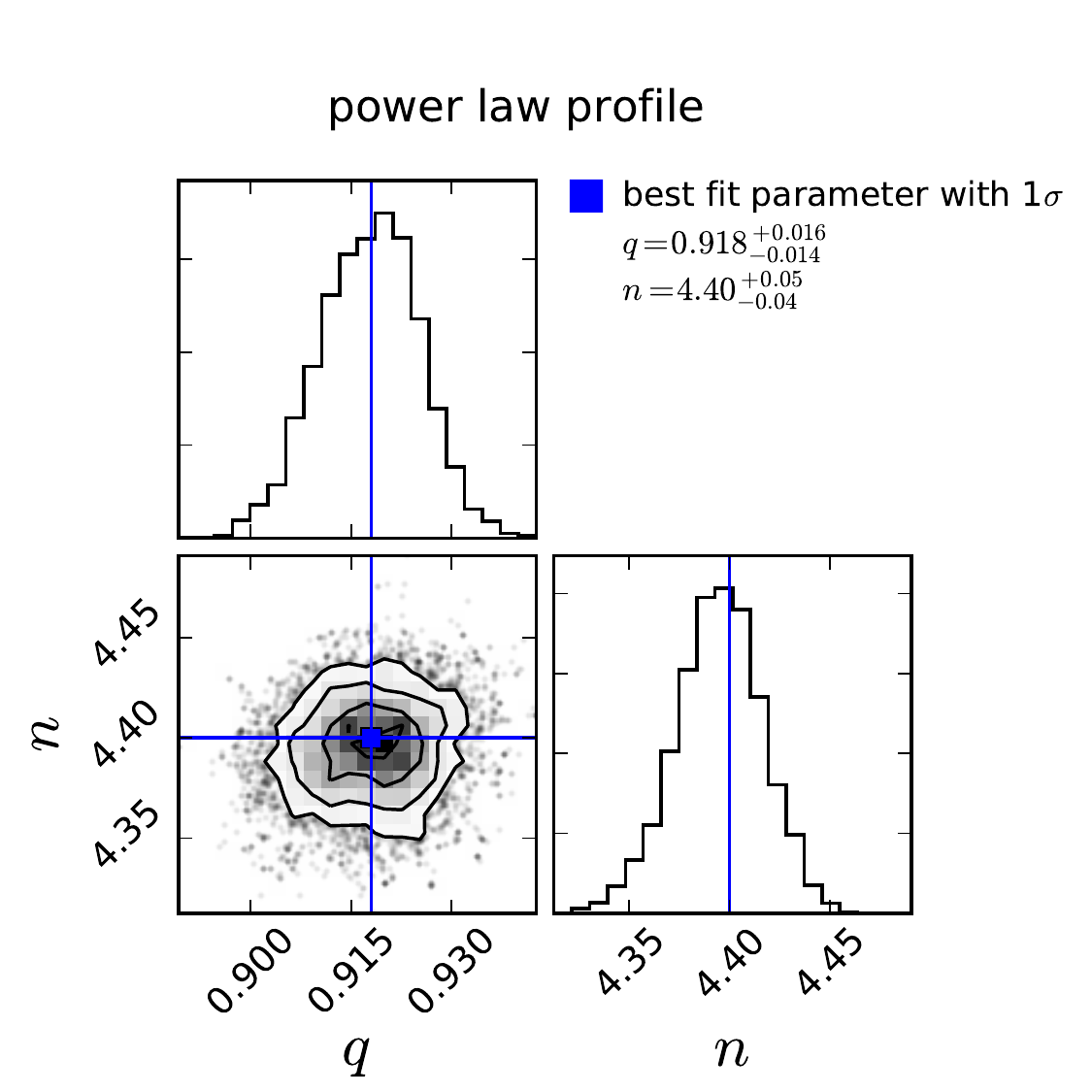}
\caption[One- and two-dimensional projections of the posterior probability distributions
of parameters $(q, n)$ of the power law (Equ. \eqref{eq:halomodel}) fitted to the cleaned sample.]
{{One- and two-dimensional projections of the posterior probability distributions
of parameters $(q, n)$ of the power law (Equ. \eqref{eq:halomodel}) fitted to the cleaned sample.
The blue lines and squares mark the median value of each parameter. The best-fit parameters are given along with their $1\sigma$ intervals in the top right part of the figure. Both the power-law index $n$ and the flattening parameter $q$ show an almost Gaussian distribution with no covariance.}
\label{fig:triangle_powerlaw}}
\end{center}  
\end{figure*}

\begin{figure*}
\begin{center}  
\includegraphics[]{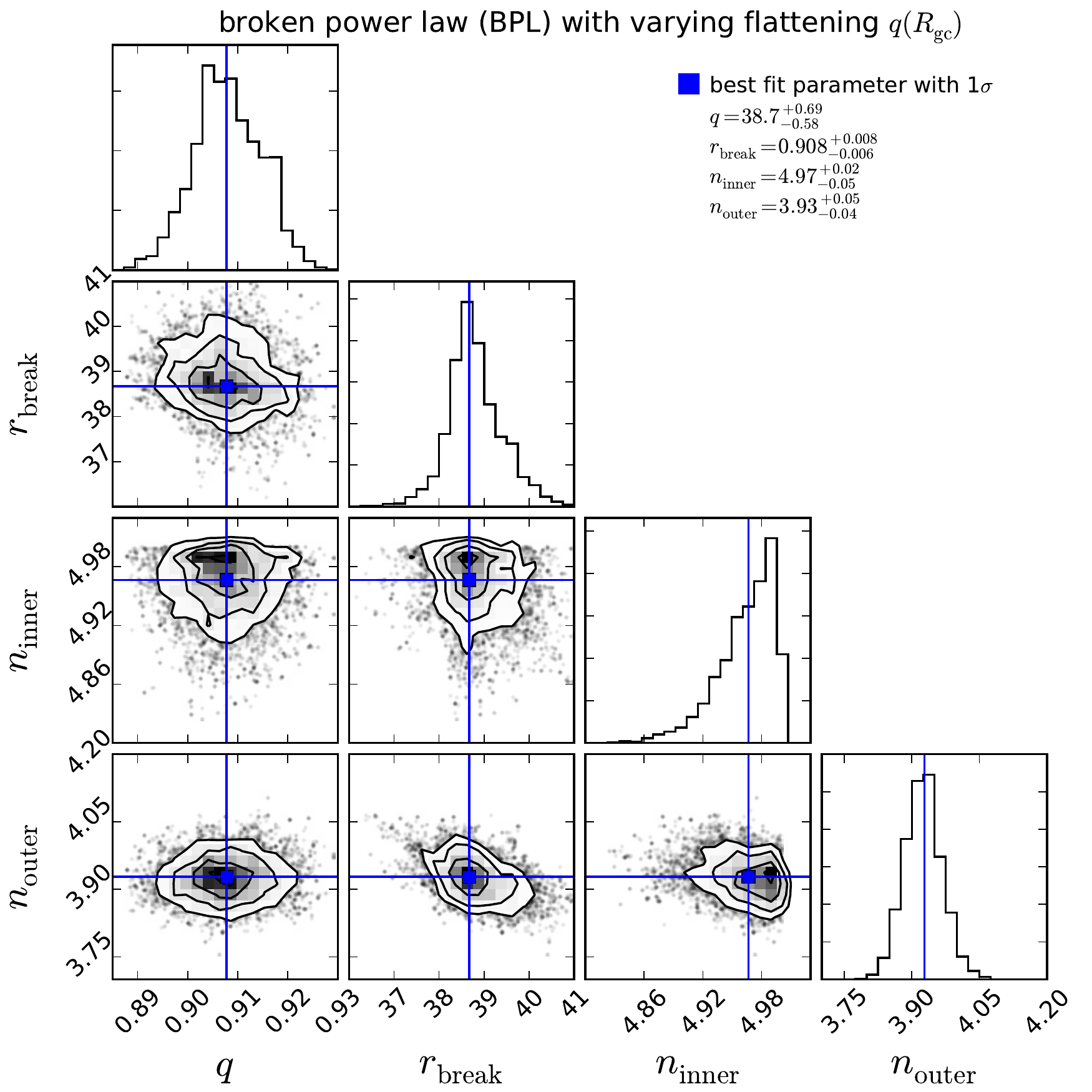}
\caption[One- and two-dimensional projections of the posterior probability distributions
of parameters $(q, r_{\mathrm{break}}, n_{\mathrm{inner}}, n_{\mathrm{outer}})$ of the BPL (Equ. \eqref{eq:BPL_halomodel}) fitted to the cleaned sample.]
{{One- and two-dimensional projections of the posterior probability distributions
of parameters $(q, r_{\mathrm{break}}, n_{\mathrm{inner}}, n_{\mathrm{outer}})$ of the power law (Equ. \eqref{eq:BPL_halomodel}) fitted to the cleaned sample.
The blue lines and squares mark the median value of each parameter. The best-fit parameters are given along with their $1\sigma$ intervals in the top right part of the figure. The parameters show an almost Gaussian distribution with no covariance.}
\label{fig:triangle_BPL}}
\end{center}  
\end{figure*}

\begin{figure*}
\begin{center}  
\includegraphics[]{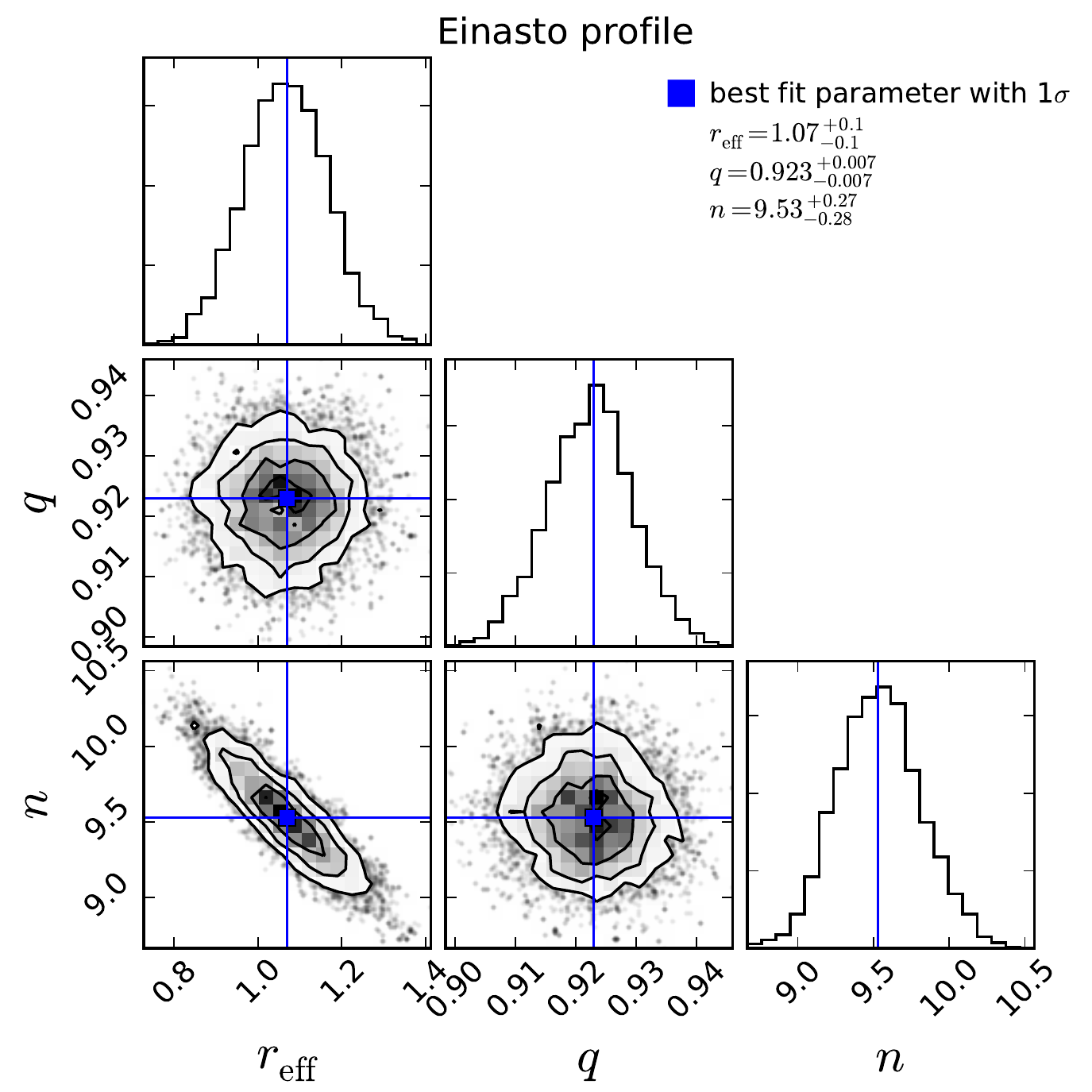}
\caption[One- and two-dimensional projections of the posterior probability distributions
of parameters $(r_{\mathrm{eff}},q, n)$ of the Einasto profile (Equ. \eqref{eq:einasto}) fitted to the cleaned sample.]
{{One- and two-dimensional projections of the posterior probability distributions
of parameters $(r_{\mathrm{eff}},q, n)$ of the Einasto profile (Equ. \eqref{eq:einasto}) fitted to the cleaned sample.
The blue lines and squares mark the median value of each parameter. The best-fit parameters are given along with their $1\sigma$ intervals in the top right part of the figure. As is the case for the power law, the concentration index $n$ and the flattening parameter $q$ show an almost Gaussian distribution with no covariance.
The concentration index $n$ is covariant with the effective radius parameter, $r_{\mathrm{eff}}$.
}
\label{fig:triangle_einasto}}
\end{center}  
\end{figure*}

\begin{figure*}
\begin{center}  
\includegraphics[]{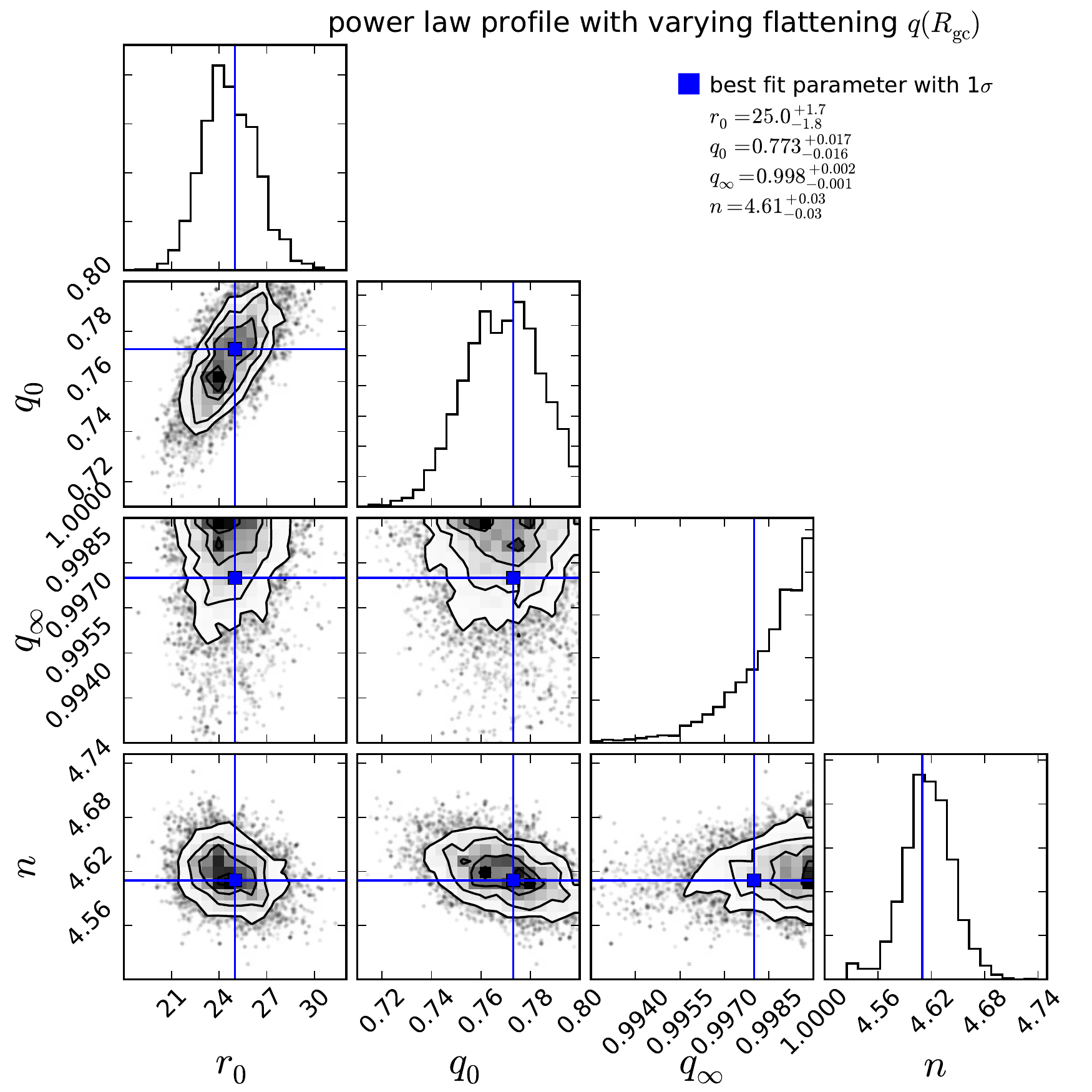}
\caption[One- and two-dimensional projections of the posterior probability distributions (\textit{pdf})
of parameters $(r_0,q_0,q_{\infty}, n)$ of the power law with varying flattening $q(R_{\mathrm{gc}})$ (Equ. \eqref{eq:q_r}) fitted to the cleaned sample.]
{{One- and two-dimensional projections of the posterior probability distributions (\textit{pdf})
of parameters $(r_0,q_0,q_{\infty}, n)$ of the power law with varying flattening $q(R_{\mathrm{gc}})$ (Equ. \eqref{eq:q_r}) fitted to the cleaned sample.
The blue lines and squares mark the maximum likely value of each parameter. The best-fit parameters are given along with their $1\sigma$ intervals in the top right part of the figure. The fitting parameters show a covariance and the \textit{pdf} is strongly distorted from a Gaussian distribution, including local maxima in the distribution of $r_0$.}
\label{fig:triangle_powerlaw_qr}}
\end{center}  
\end{figure*}

\begin{figure*}
\begin{center}  
\includegraphics[width=\textwidth]{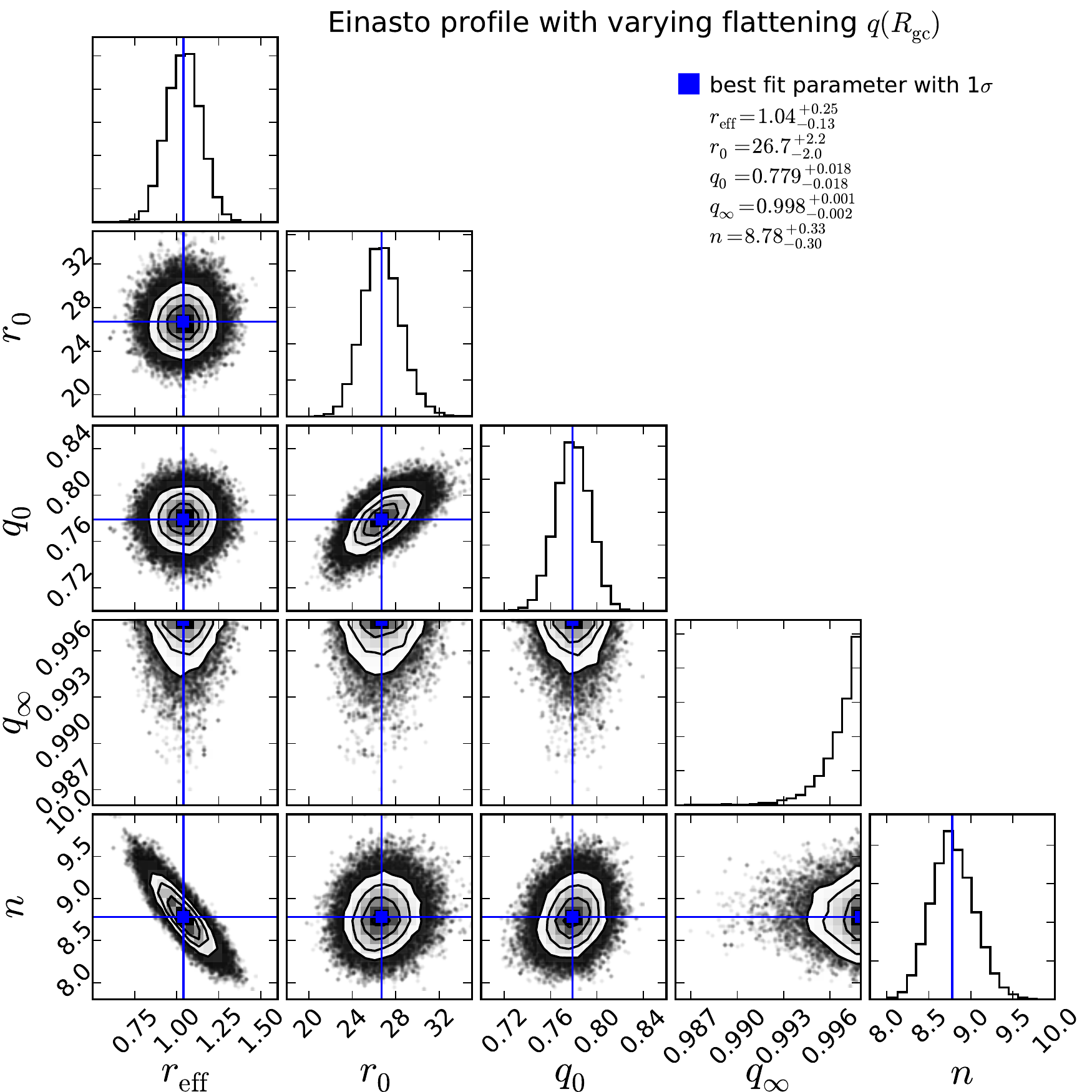}
\caption[One- and two-dimensional projections of the posterior probability distributions (\textit{pdf})
of parameters $(r_{\mathrm{eff}},r_0,q_0,q_{\infty}, n)$ of the Einasto profile with varying flattening $q(R_{\mathrm{gc}})$ (Equ. \eqref{eq:q_r}) fitted to the cleaned sample.]
{{One- and two-dimensional projections of the posterior probability distributions (\textit{pdf})
of parameters $(r_{\mathrm{eff}},r_0,q_0,q_{\infty}, n)$ of the Einasto profile with varying flattening $q(R_{\mathrm{gc}})$ (Equ. \eqref{eq:q_r}) fitted to the cleaned sample.
The blue lines and squares mark the maximum likely value of each parameter. The best-fit parameters are given along with their $1\sigma$ intervals in the top right part of the figure.\newline
The fitting parameters $r_0$, $q_0$, $q_{\infty}$ show a (partially strong) covariance and the \textit{pdf} is strongly distorted from a Gaussian distribution, including local maxima in the distribution of $r_0$, $q_0$, $q_{\infty}$. The best-fit model as well as the \textit{pdf} lead to $q_0 \sim q_{\infty}$, thus being quite similar to the Einasto profile with a constant flattening.
}
\label{fig:triangle_einasto_qr}}
\end{center}  
\end{figure*}

\begin{figure*}
\begin{center}  
\includegraphics[trim=1.0inch 0.0inch 0.0inch 2.31inch, clip=true]{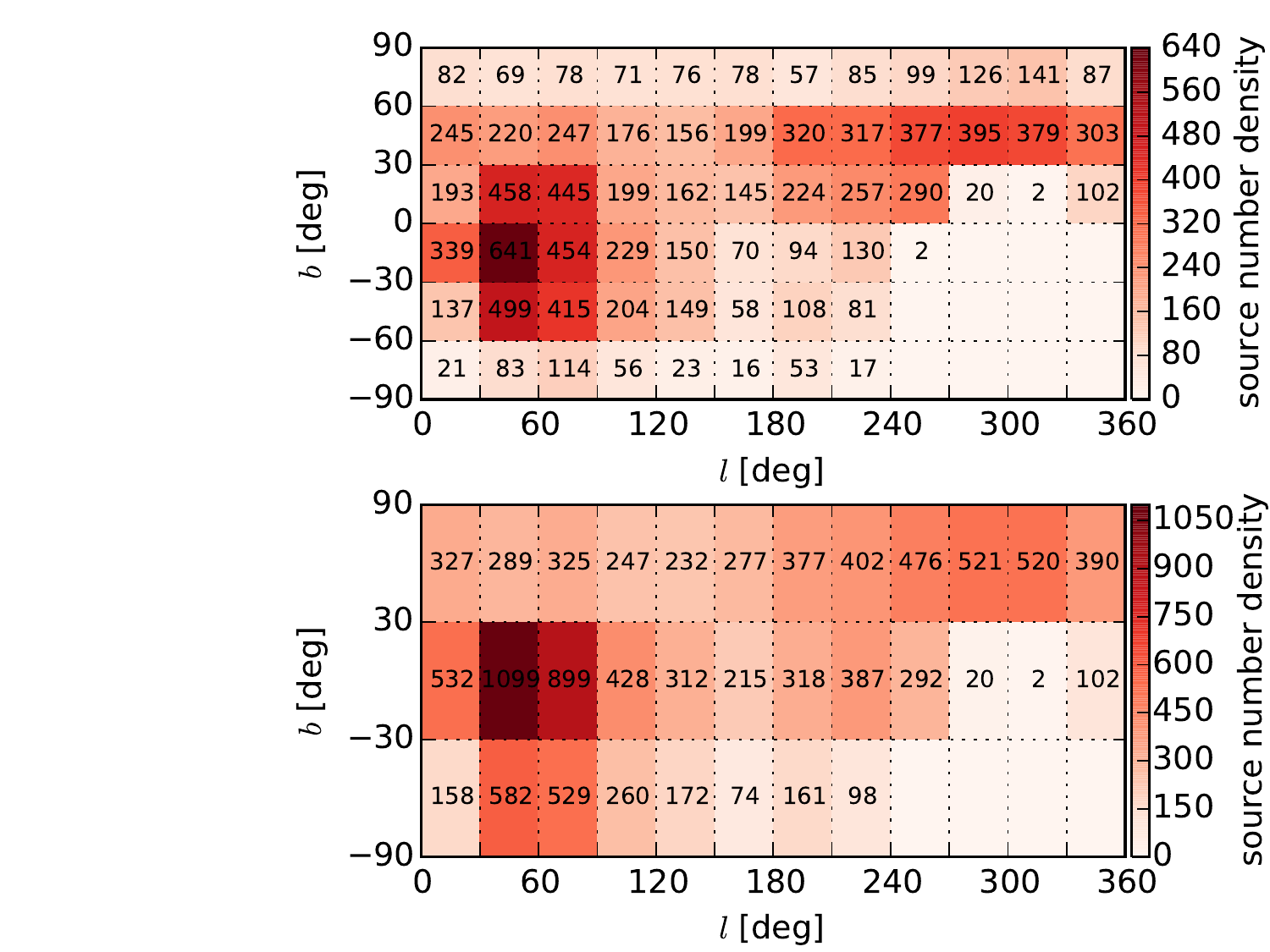}
\caption[The angular source number density for the cleaned RRab sample, given per $(\Delta l=30\arcdeg) \times (\Delta b=60\arcdeg)$ bin.]
{{The angular source number density for the cleaned RRab sample, given per $(\Delta l=30\arcdeg) \times (\Delta b=60\arcdeg)$ bin. This binning is used to fit for the local halo properties.
The number density is color-coded as well as given in numbers. Empty cells are outside the survey footprint.\newline
Away from the Galactic equator, the angular number density drops as the spanned area decreases.
We find that the number of sources is increased within $30\arcdeg<l<90\arcdeg$, $\vert b \vert<30\arcdeg$; stars not fully excised from the Galactic bulge as well as the crossing Sagittarius stream account for that. Also around $240\arcdeg<l<330\arcdeg$, $30\arcdeg<b<90\arcdeg$ we find an increase of sources, as we can remove most but not all stars from the Sgr stream by setting geometric cuts on their angle above the plane of the stream.
A significant increase near the Galactic center, where sources would fall into the bin $0\arcdeg<l<30\arcdeg$, $-30\arcdeg<b<30\arcdeg$, is not found, as we remove everything within $R_{\mathrm{gc}} \leq 20;\mathrm{kpc}$ and $\vert b\vert<10 \arcdeg$ well.
}
\label{fig:numberofsources}}
\end{center}
\end{figure*}

\begin{figure*}
\begin{center}  
\includegraphics[width=0.8\textwidth, trim=0.0inch 0.0inch 0.0inch 0.0inch, clip=true]{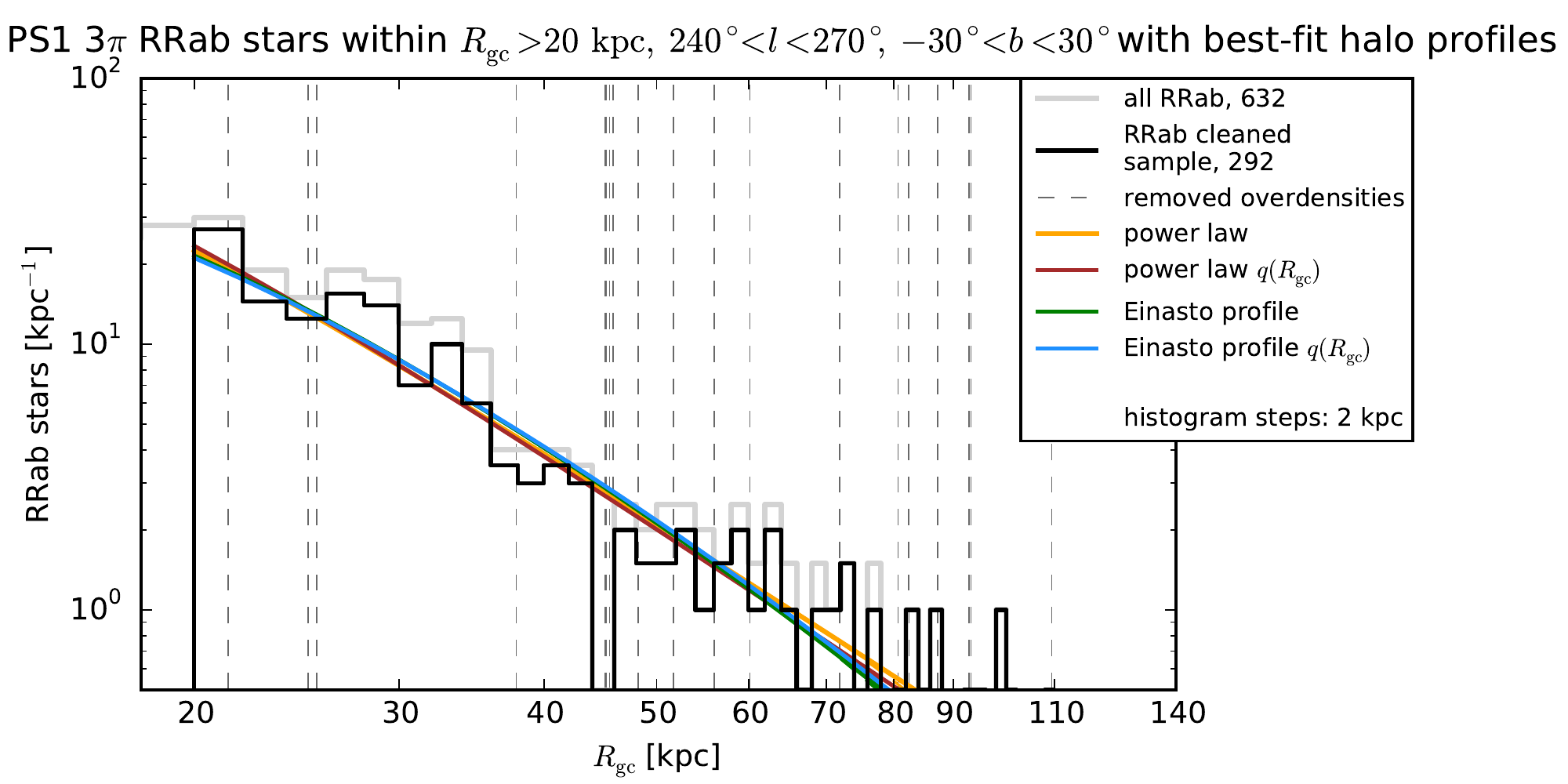}  
\includegraphics[width=0.8\textwidth, trim=0.0inch 0.0inch 0.0inch 0.0inch, clip=true]{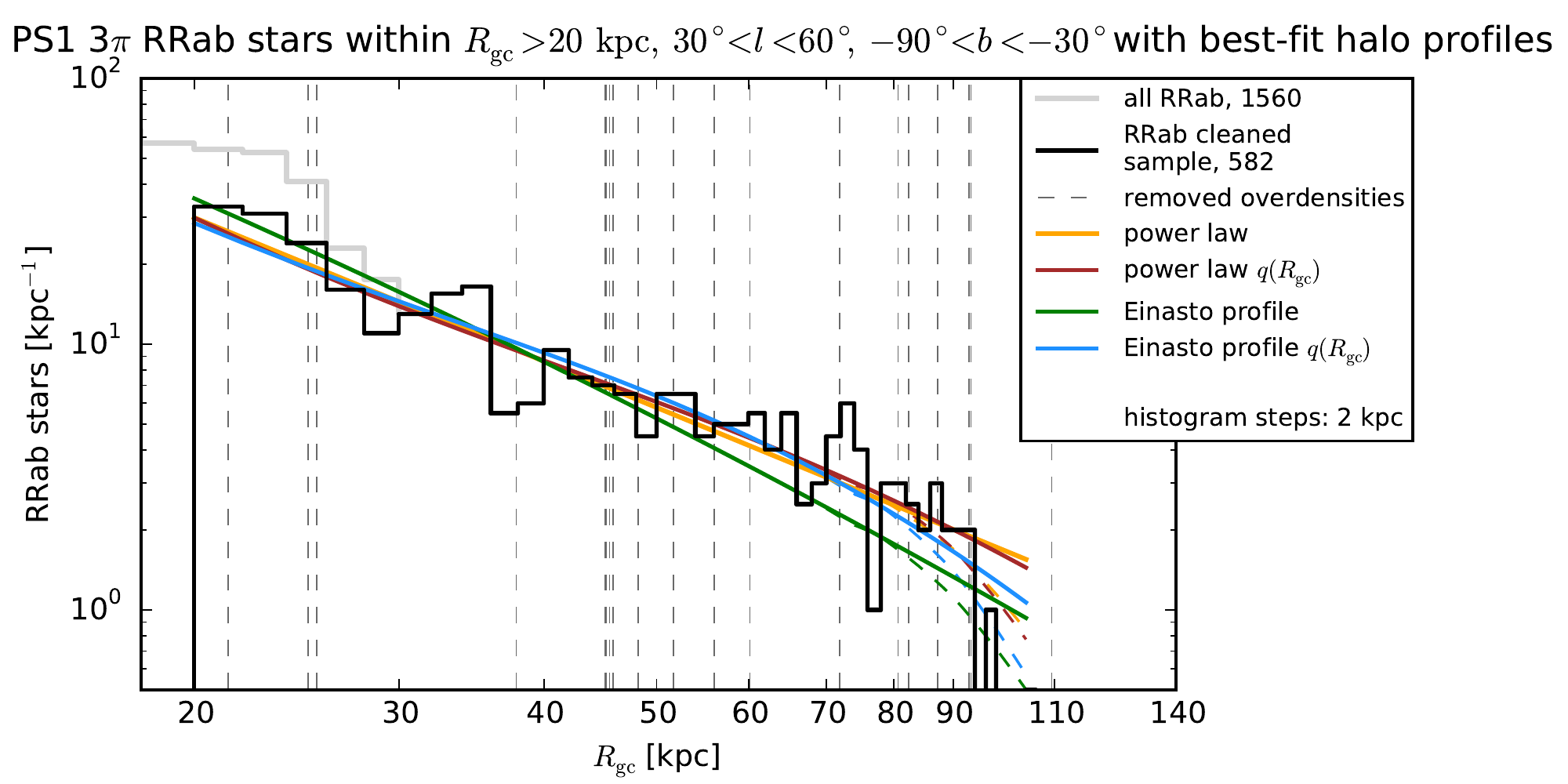} 
\caption{{
Comparison between the observed distance distribution of the cleaned samples and the predicted distributions by the best-fit models for $30 \arcdeg \times 60 \arcdeg$ patches on the sky, with the number density shown in a log plot. We find that the fitting prodecure also works reliable with small pieces on the sky.\newline
The black histogram shows the Galactocentric distance distribution of our cleaned sample of RRab stars within the given patch on the sky, whereas the grey histogram gives the distance distribution of all RRab stars within the given patch on the sky. Removed overdensities are highlighted with dashed lines, and are listed in Table \ref{tab:overdensities}.\newline
The overplotted solid lines represent the best-fit model for each of the four halo profiles. As a result of the selection function, these models don't follow a straight line in the log plot, but drop much more rapidly especially beyond a Galactocentric distance of 80 kpc.
For comparison, dashed lines, in the same color as the solid lines, represent each  $\rho_{\mathrm{halo}} \times \mathcal{S}(l,b,D)$, where $\mathcal{S}$ is the selection function as given in Equ. \eqref{eq:allselfunc}. 
All color-coding and lines are comparable to those in Fig. \ref{fig:halofit_allsky}.\newline
}
\label{fig:halofit_patches}}
\end{center}  
\end{figure*}

\begin{figure*}
\begin{center}  
\includegraphics[]{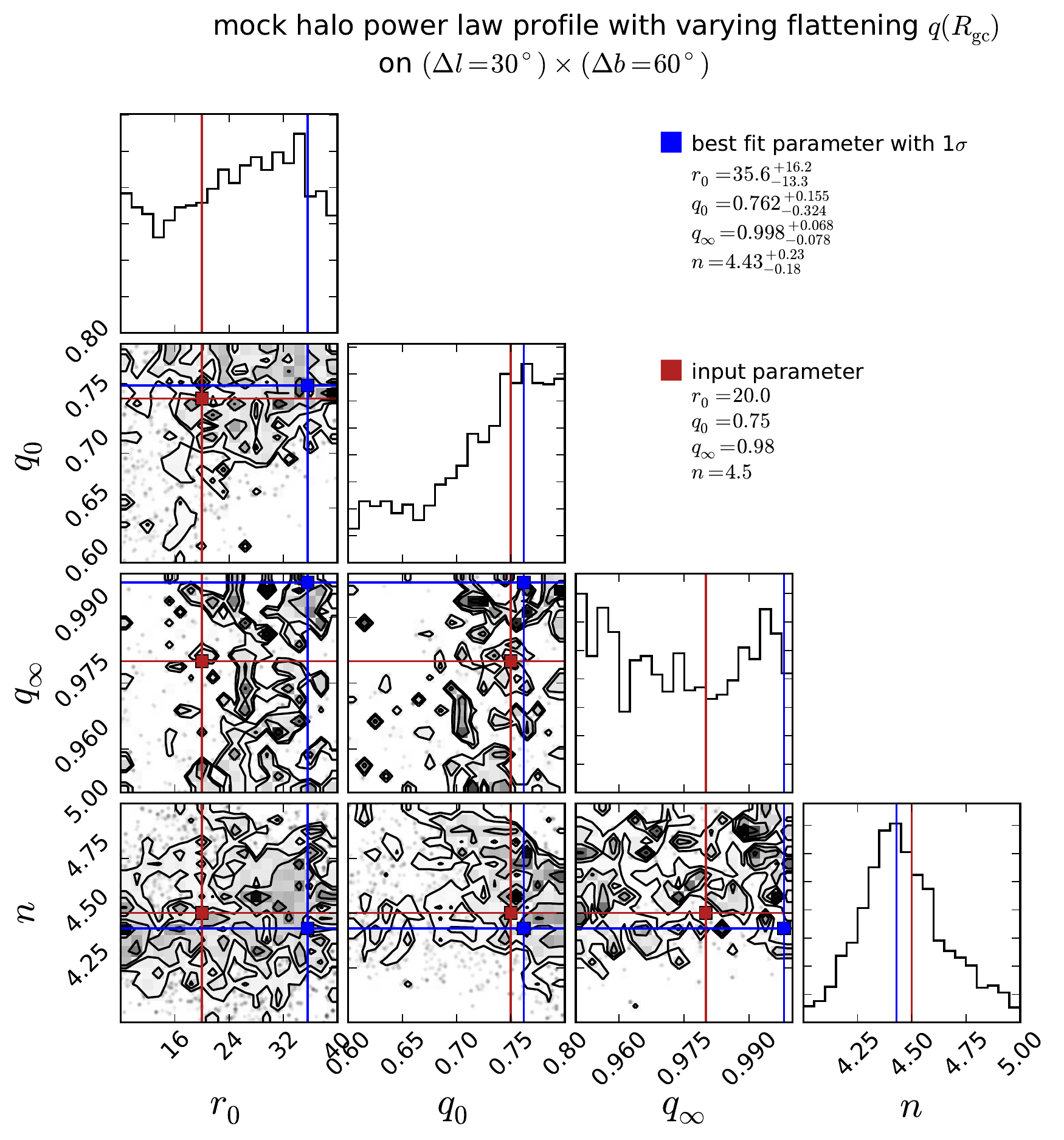}
\caption[One- and two-dimensional projections of the posterior probability distributions (\textit{pdf})
of parameters $(r_0,q_0,q_{\infty}, n)$ of the power law with varying flattening $q(R_{\mathrm{gc}})$ (Equ. \eqref{eq:q_r}) fitted to a $30 \arcdeg \times 60 \arcdeg$ patch of the cleaned sample.]
{{One- and two-dimensional projections of the posterior probability distributions (\textit{pdf})
of parameters $(r_0,q_0,q_{\infty}, n)$ of the power law with varying flattening $q(R_{\mathrm{gc}})$ (Equ. \eqref{eq:q_r}) fitted to a $30 \arcdeg \times 60 \arcdeg$  patch of the cleaned sample. The blue lines and squares mark the maximum likely value of each parameter. The same patch on the sky as in the lower panel of Fig. \ref{fig:halofit_patches} was chosen. The posterior distribution is comparable to those when fitting the same halo profile to the full cleaned sample (see Fig. \ref{fig:triangle_allsky_powerlaw_qr_mockhalo} for comparison). However, the width of the posterior probability distribution is increased in the case shown here, compared to the full cleaned sample. This is also reflected in the $1\sigma$ intervals given along with the best-fit parameters in the top right part of the figure.
The best-fit parameters are given along with their $1\sigma$ intervals in the top right part of the figure. The covariance of the parameters is comparable to those found when fitting the complete cleaned sample.}
\label{fig:triangle_powerlaw_qr_l270_b30_mockhalo}}
\end{center}  
\end{figure*}

\begin{figure*}
\begin{center}  
\includegraphics[trim=1.0inch 0.0inch 0.0inch 0.15inch, clip=true, scale=0.95]{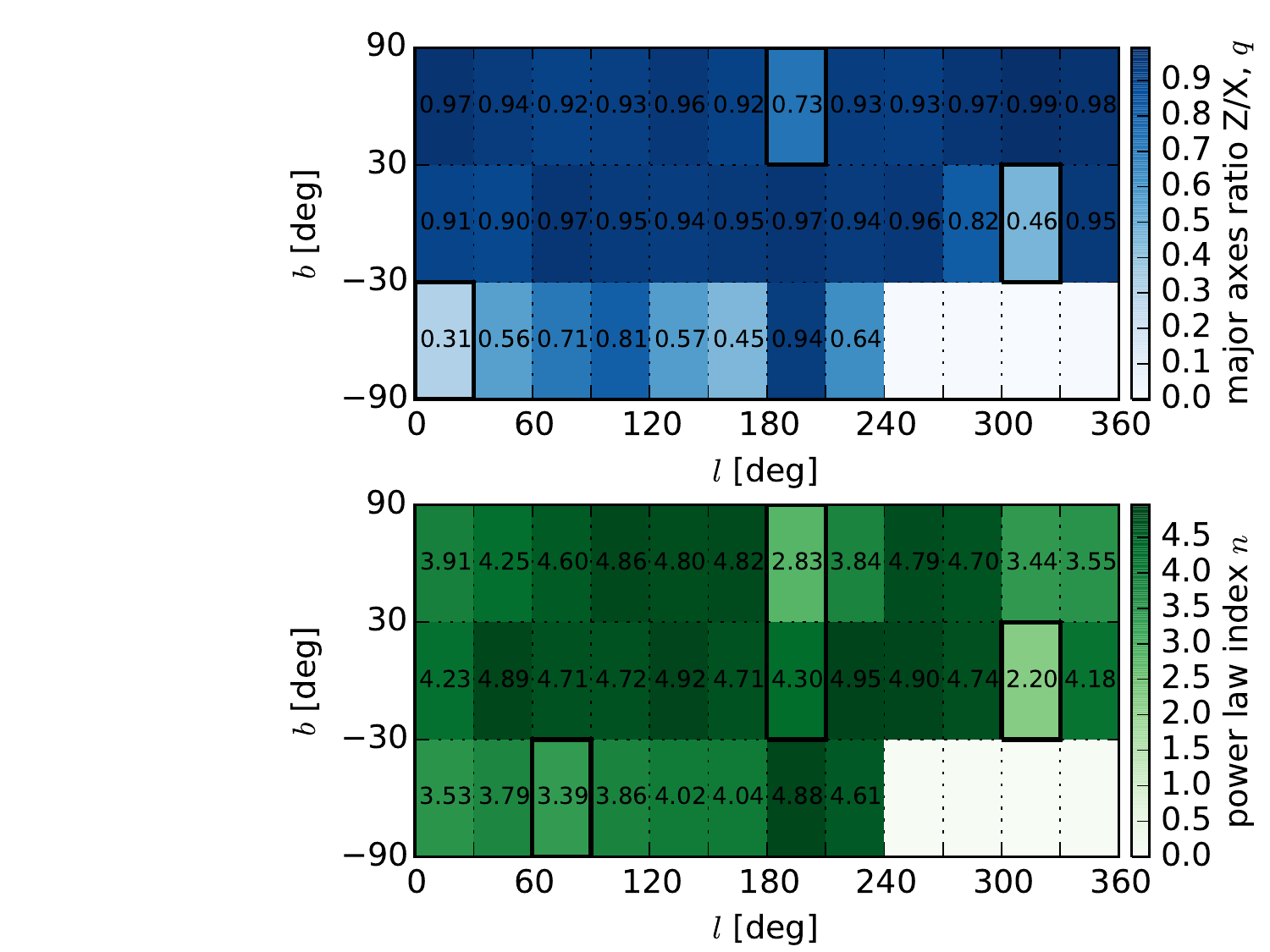} 
\caption[The angular distribution for the best-fit power law parameters $q$ and $n$, respectively, given per $(\Delta l=30\arcdeg) \times (\Delta b=60\arcdeg)$ bin.]
{{The angular distribution for the best-fit power law parameters $q$ and $n$, respectively, given per $(\Delta l=30\arcdeg) \times (\Delta b=60\arcdeg)$ bin. The number density is color-coded as well as given in numbers. Empty cells are outside of the survey area.\newline
Some cells show a large deviation of $q$ or $n$ from the mean or from the expected value. These cells are highlighted with a thick frame. Reasons for those deviations are discussed in Section \ref{sec:LocalHaloProperties}.\newline
The values for the cells in these plots are given in Table \ref{tab:powerlaw_60}.
}
\label{fig:powerlaw_60_heatmap}}
\end{center}  
\end{figure*}

\begin{figure*}
\begin{center}  
\includegraphics[trim=1.0inch 0.0inch 0.0inch 0.15inch, clip=true, scale=0.95]{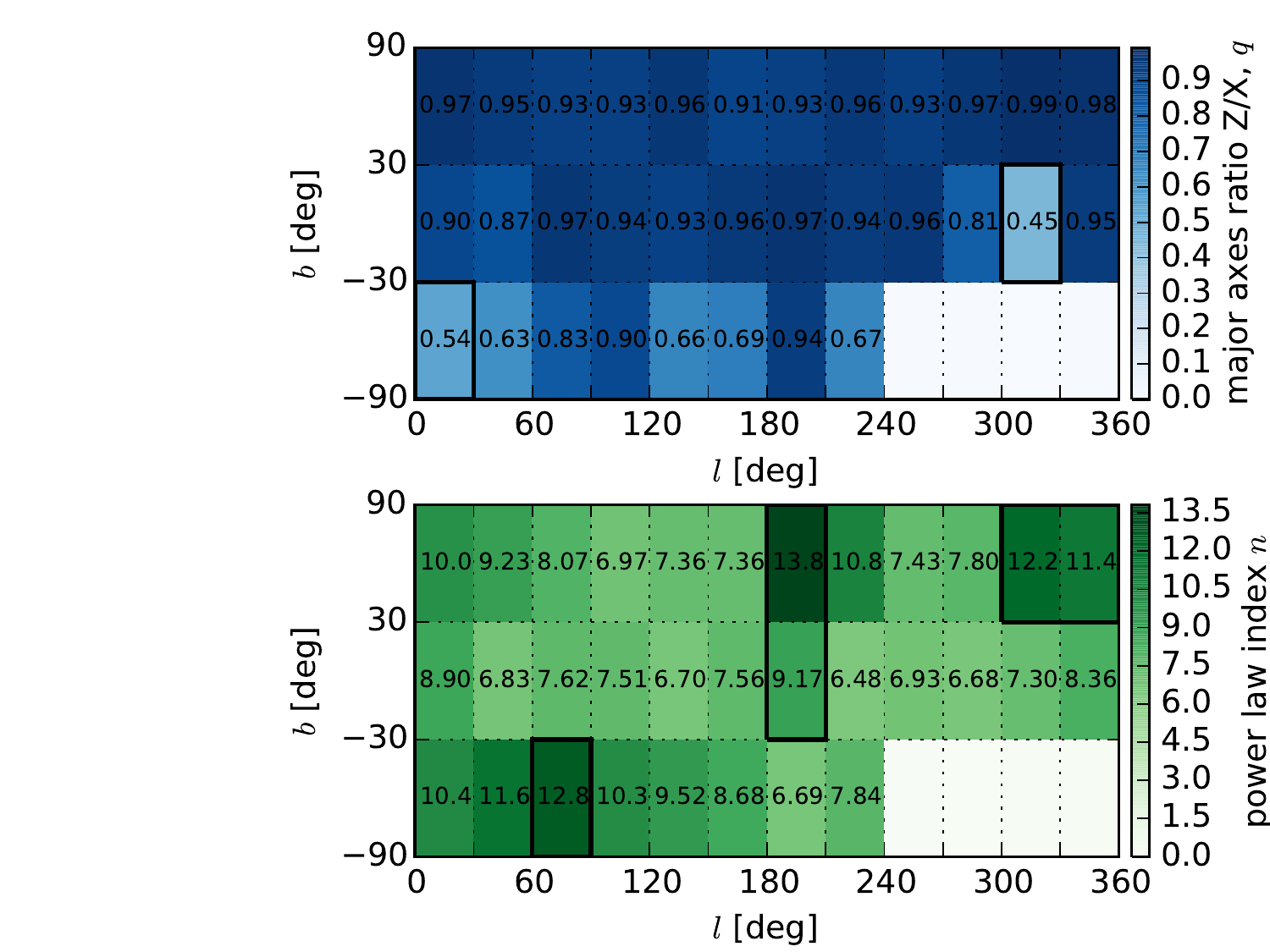}
\caption[The angular distribution for the best-fit Einasto profile parameters $q$ and $n$, respectively, given per $(\Delta l=30\arcdeg) \times (\Delta b=60\arcdeg)$ bin.]
{{The angular distribution for the best-fit Einasto profile parameters $q$ and $n$, respectively, given per $(\Delta l=30\arcdeg) \times (\Delta b=60\arcdeg)$ bin. The number density is color-coded as well as given in numbers. Empty cells are outside of the survey area. The Einasto profile parameter $r_{\mathrm{eff}}$ was neglected here for the sake of clarity, and as we compare mainly the results on oblateness and steepness of the halo profile.\newline
Some cells show a large deviation of $q$ or $n$ from the mean or from the expected value. These cells are highlighted with a thick frame. Reasons for those deviations are discussed in Section \ref{sec:LocalHaloProperties}. The values for the cells in these plots are given in Table \ref{tab:einasto_60}.
}
\label{fig:einasto_60_heatmap}}
\end{center}  
\end{figure*}

\begin{figure*}
\begin{center}  
\includegraphics[trim=1.0inch 0.0inch 0.0inch 0.0inch, clip=true]{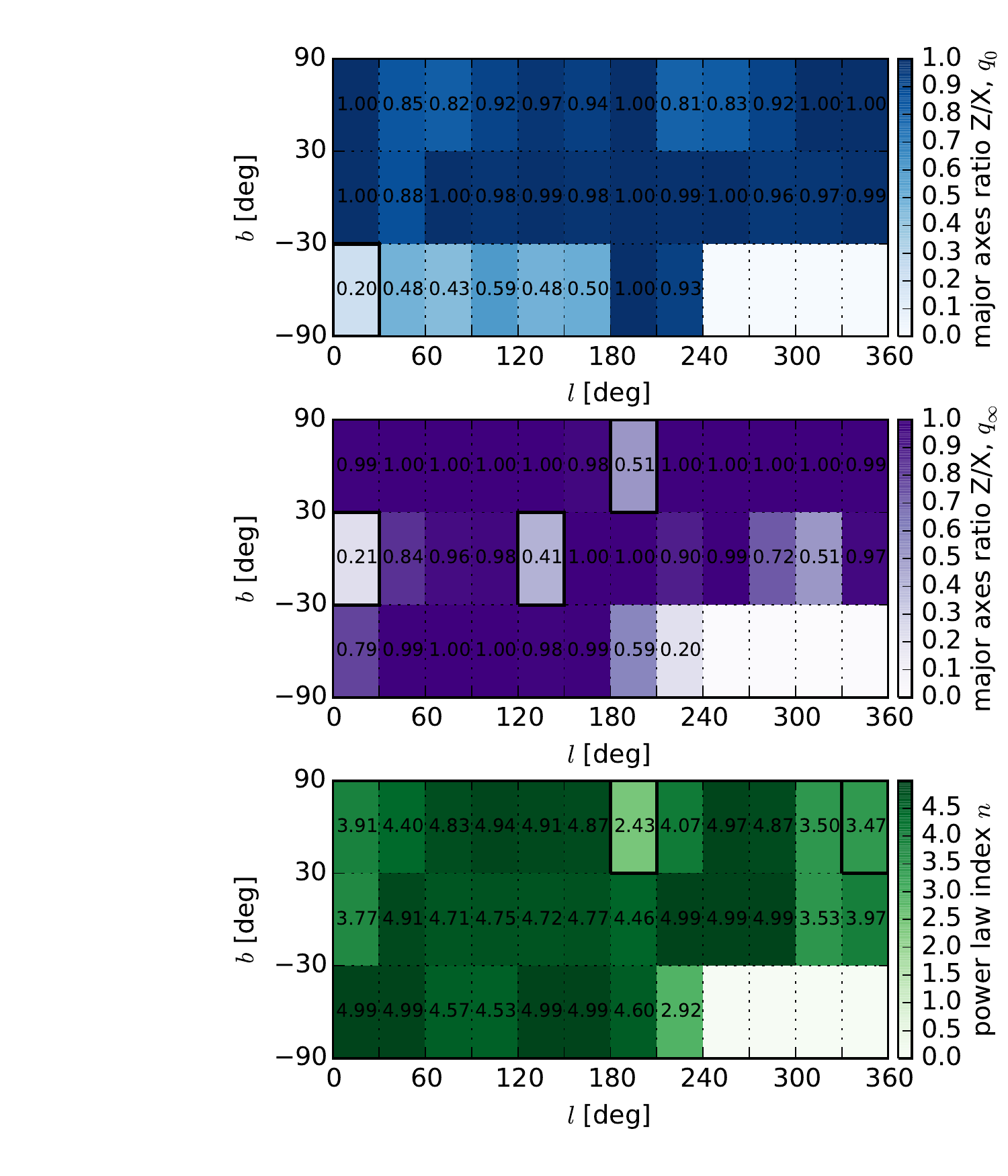}
\caption[The angular distribution for the best-fit parameters $q_0$, $q_{\infty}$ and $n$, respectively, of a power law model with $q(R_{\mathrm{gc}})$.]
{{The angular distribution for the best-fit parameters $q_0$, $q_{\infty}$ and $n$, respectively, of a power law model with $q(R_{\mathrm{gc}})$. The distribution is given on a $(\Delta l=30\arcdeg) \times (\Delta b=60\arcdeg)$ grid. The number density is color-coded as well as given in numbers. Empty cells are outside of the survey area.\newline
The power law parameter $r_0$ was neglected here for the sake of clarity, and as we compare mainly the results on oblateness and steepness of the halo profile.\newline
Some cells show a large deviation of $q_0$, $q_{\infty}$ or $n$ from the mean or from the expected value for the value in case. These cells are highlighted with a thick frame. Reasons for those deviations are discussed in Section \ref{sec:LocalHaloProperties}.\newline
The values for the cells in this plot are given in Table \ref{tab:powerlaw_qr_60}.
}
\label{fig:powerlaw_qr_60_heatmap}}
\end{center}  
\end{figure*}

\begin{figure*}
\begin{center}  
\includegraphics[trim=1.0inch 0.0inch 0.0inch 0.0inch, clip=true]{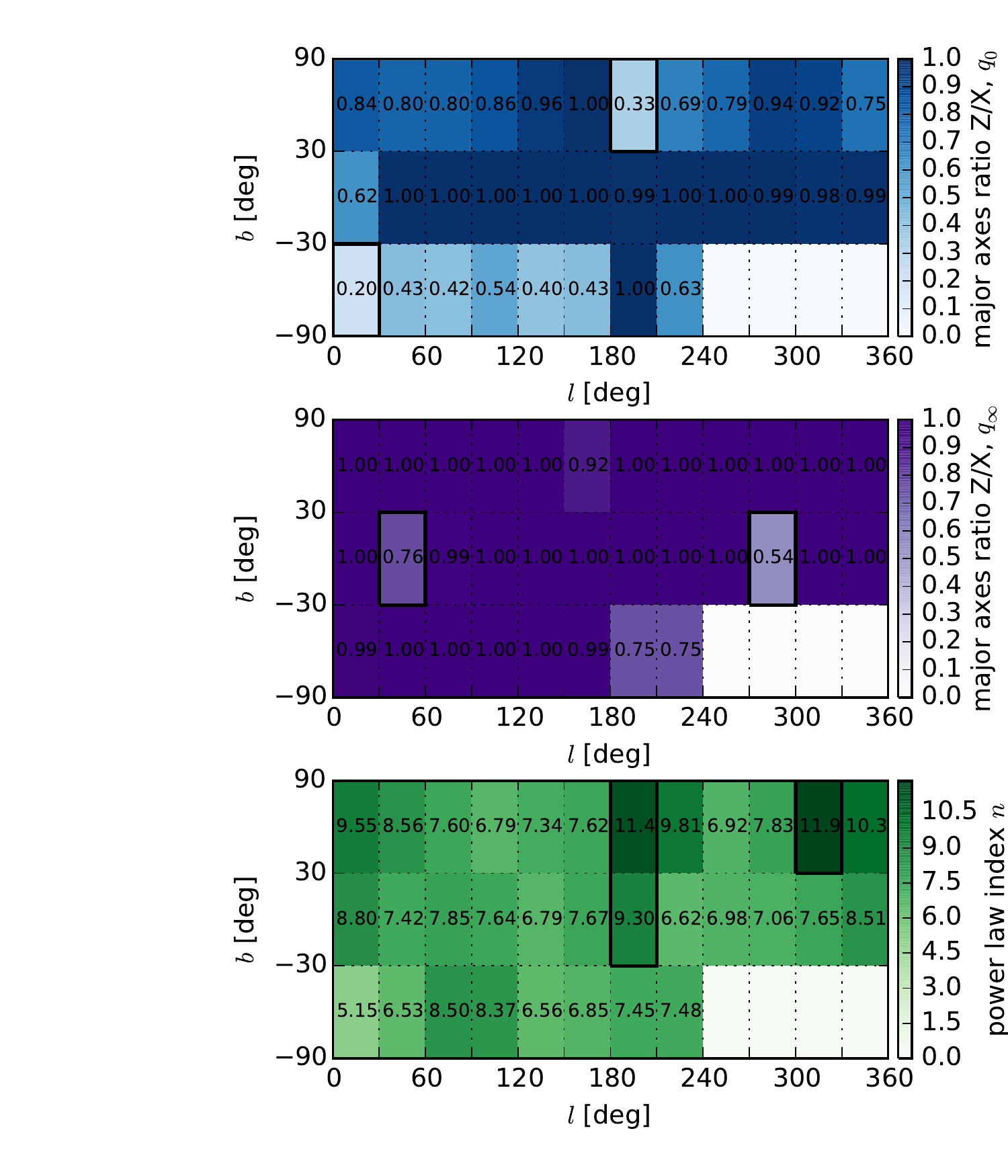}
\caption[The angular distribution for the best-fit parameters $q_0$, $q_{\infty}$ and $n$, respectively, of an Einasto profile with $q(R_{\mathrm{gc}})$.]
{{
The angular distribution for the best-fit parameters $q_0$, $q_{\infty}$ and $n$, respectively, of an Einasto profile with $q(R_{\mathrm{gc}})$. The distribution is given on a $(\Delta l=30\arcdeg) \times (\Delta b=60\arcdeg)$ grid. The number density is color-coded as well as given in numbers. Empty cells are outside of the survey area.\newline
The power law parameter $r_0$ was neglected here for the sake of clarity, and as we compare mainly the results on oblateness and steepness of the halo profile.\newline
Some cells show a large deviation of $q_0$, $q_{\infty}$ or $n$ from the mean or from the expected value for the value in case. These cells are highlighted with a thick frame. Reasons for those deviations are discussed in Section \ref{sec:LocalHaloProperties}.\newline
The values for the cells in this plot are given in Table \ref{tab:einasto_qr_60}.
}
\label{fig:einasto_qr_60_heatmap}}
\end{center}  
\end{figure*}

\begin{figure*}
\begin{center}  
\includegraphics[width=0.8\textwidth, trim=0.0inch 0.8inch 0.0inch 0.61inch, clip=true]{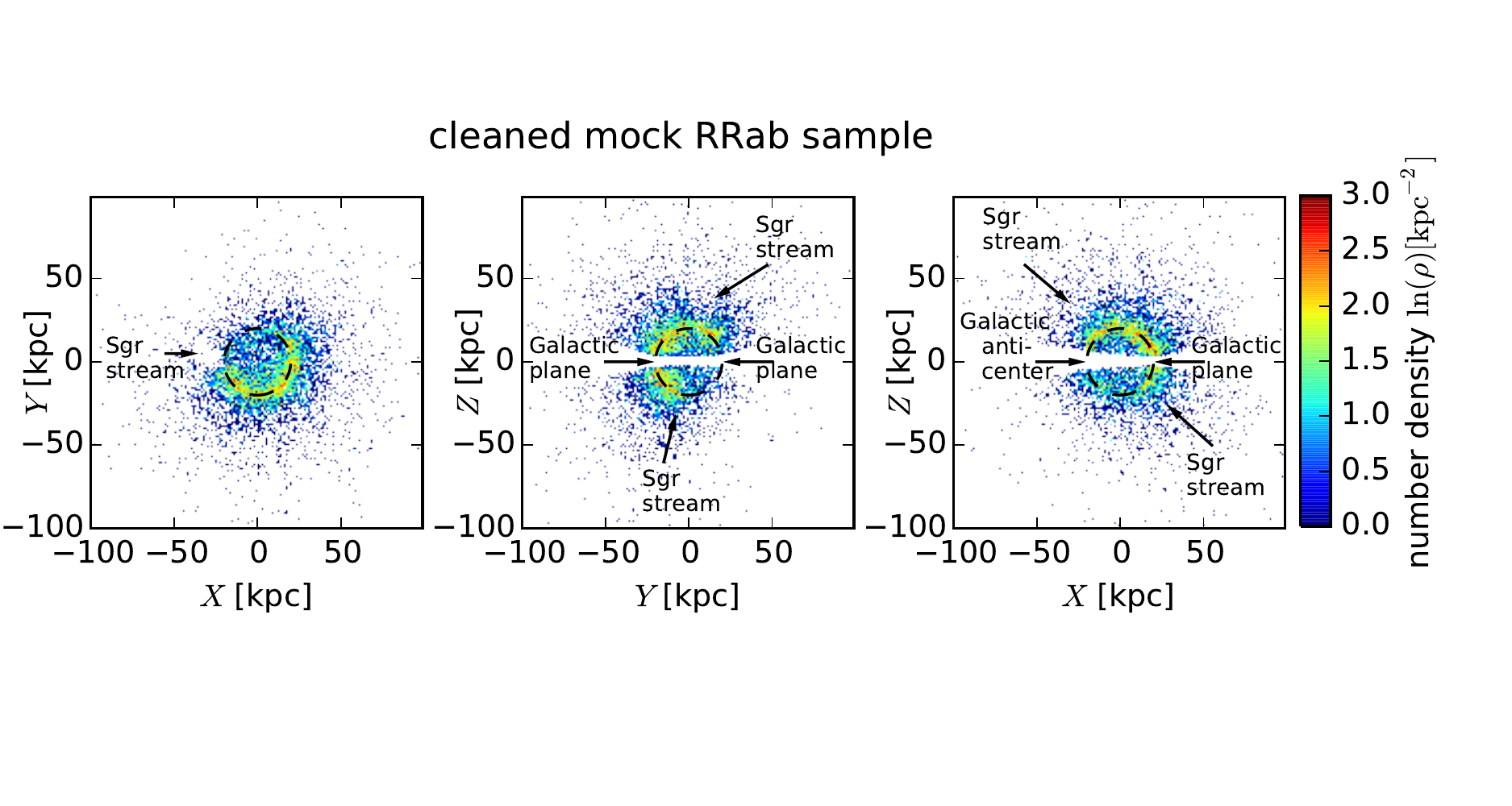}  
\includegraphics[width=0.8\textwidth, trim=0.0inch 0.8inch 0.0inch 0.61inch, clip=true]{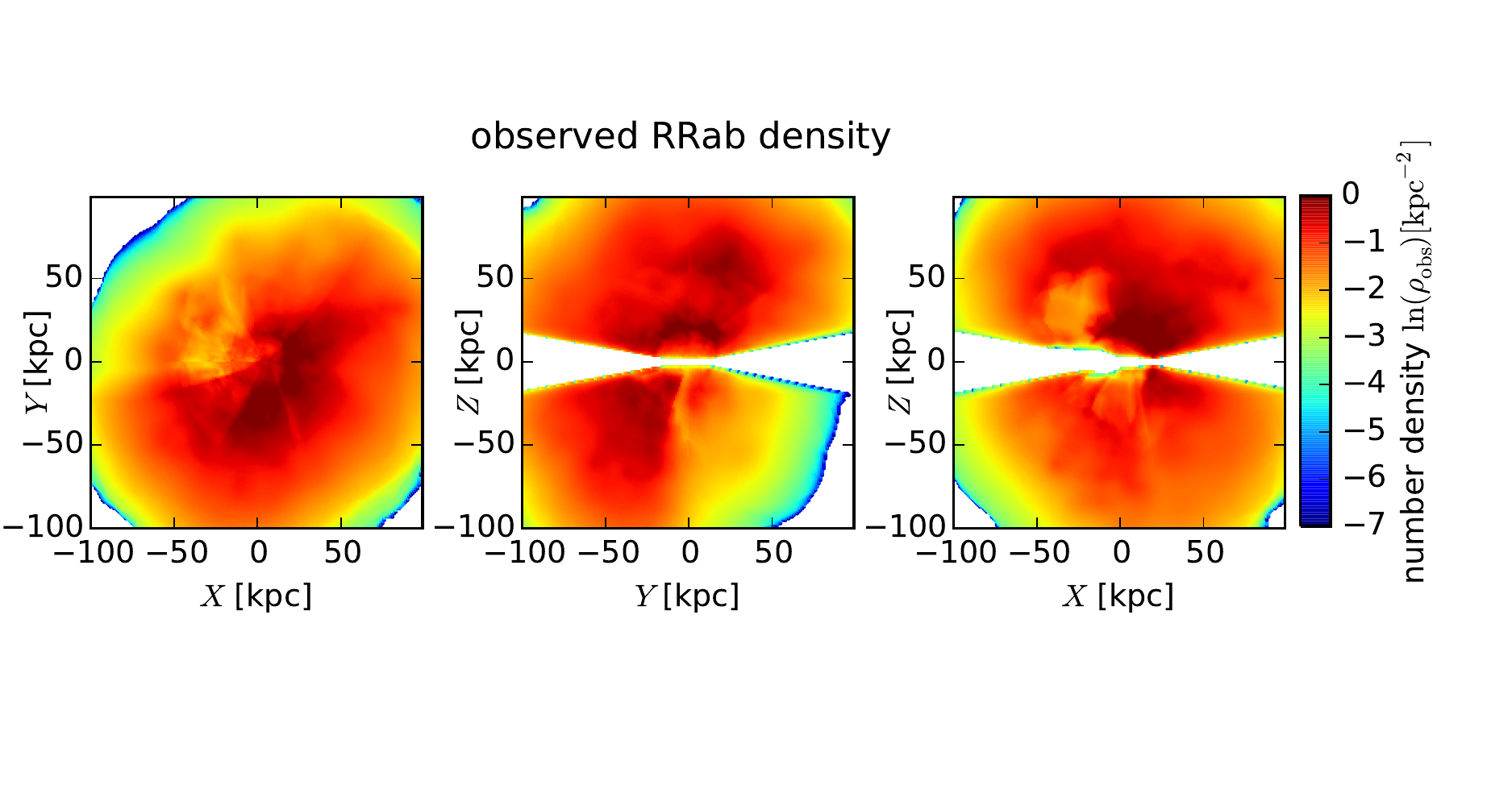} 
\includegraphics[width=0.8\textwidth, trim=0.0inch 0.8inch 0.0inch 0.61inch, clip=true]{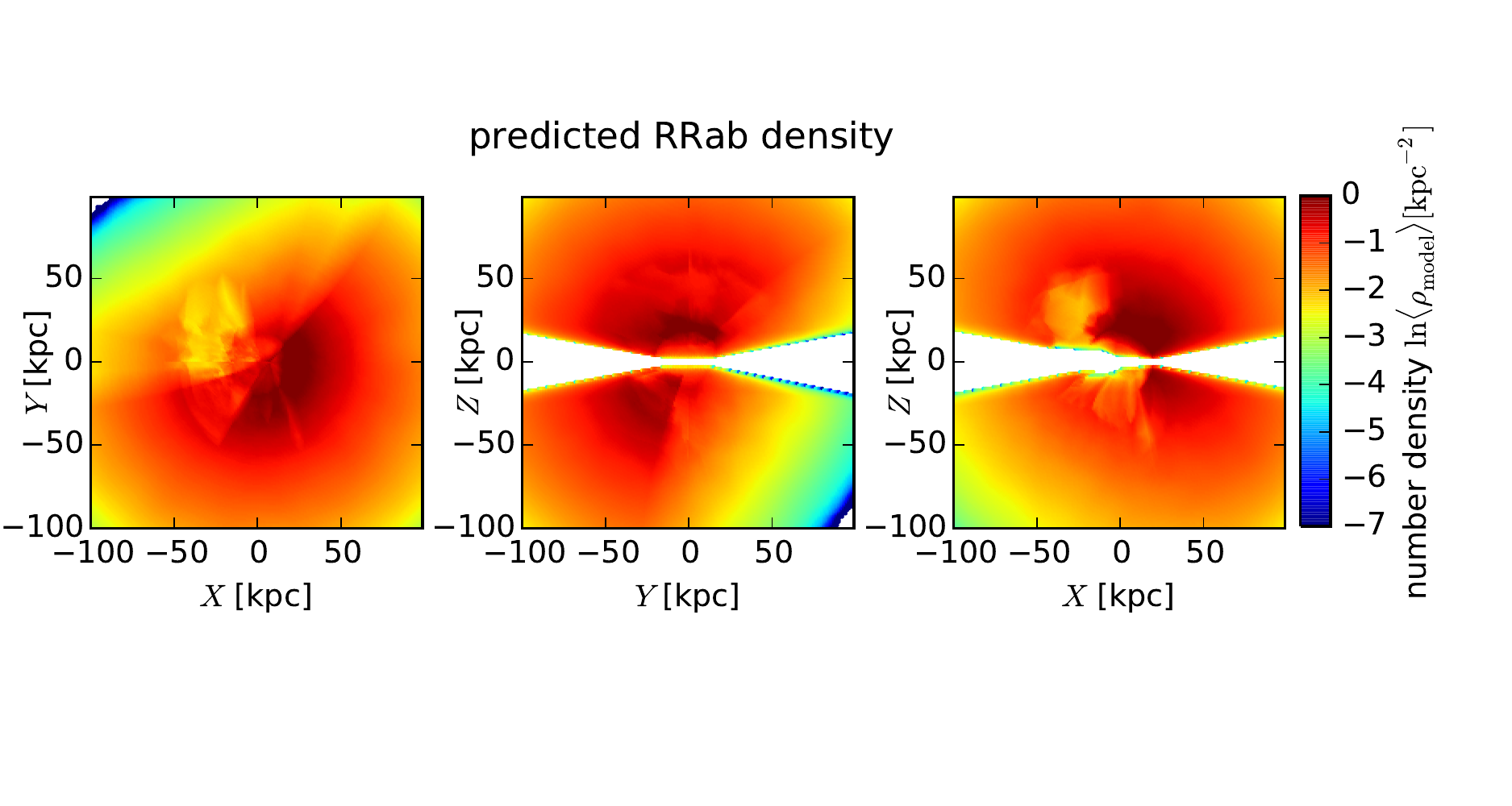}
\includegraphics[width=0.8\textwidth, trim=0.0inch 0.9inch 0.0inch 0.61inch, clip=true]{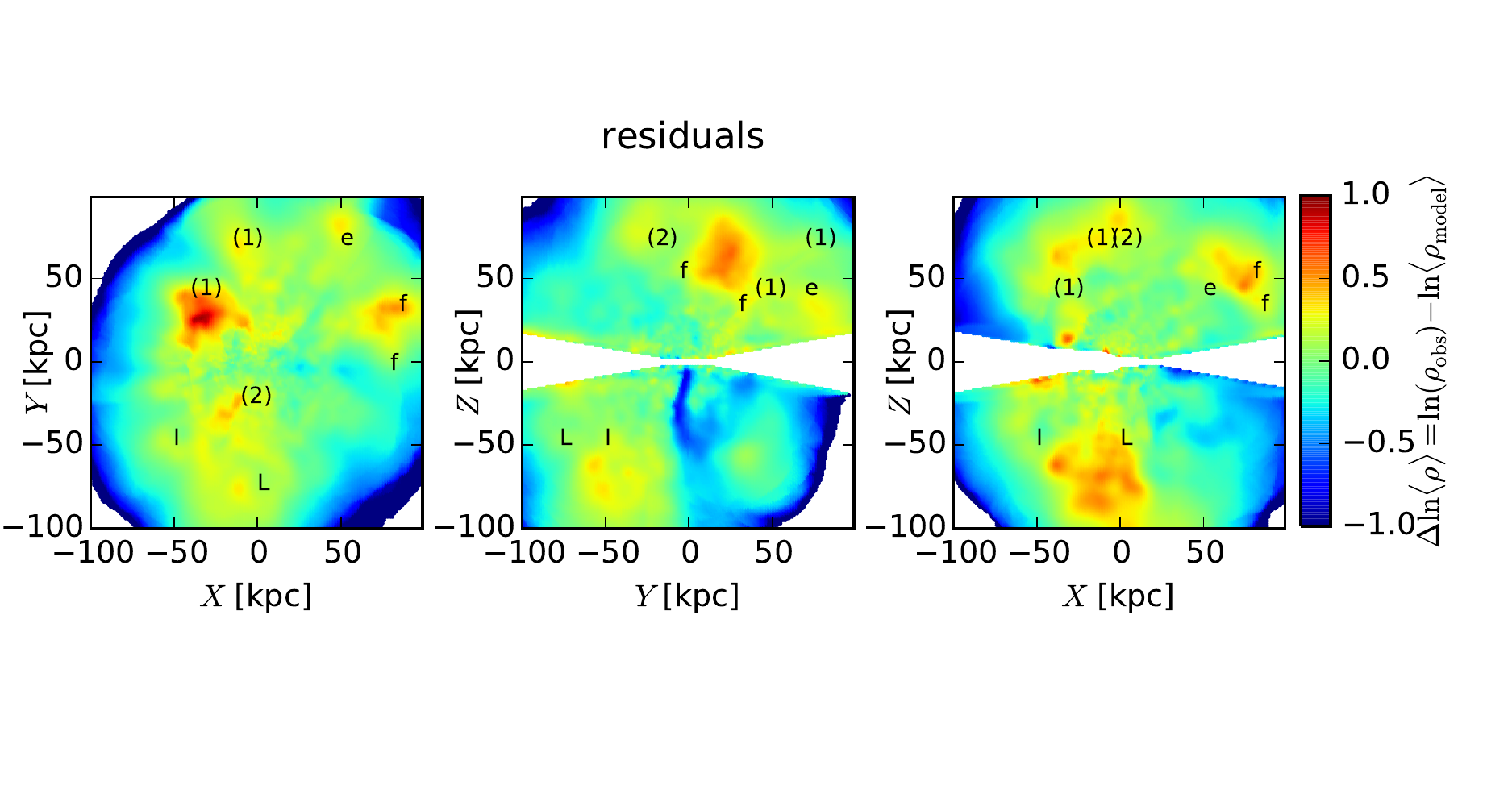}
\caption{{
Density plots in the Cartesian reference frame $(X, Y, Z)$ for the best-fit model, as well as its residuals. Densities are each color-coded according to the legend. Our results are described in detail in Section \ref{sec:LocalHaloProperties}.\newline
First row: A realization of a mock ``cleaned sample'' from the best-fit model with the selection function applied.
This sample consists of 11,025 sources, the same number of sources as in the observed cleaned sample.
Second row: Number density of the observed cleaned sample at each $(X,Y,Z)$, given in Fig. \ref{fig:sample_map}, using a nearest-neighbor approach.
Third row: Mean model density from applying the same estimation of the number density to 10 realizations of mock samples from the best-fit model, where a single mock sample looks like those given in the first row. 
Last row: The logarithmic residuals of the best-fit model. A $\Delta \ln \langle \rho \rangle<0$ indicates the best-fit model overestimates the number densities, whereas a $\Delta \ln \langle \rho \rangle>0$ means that it underestimates the number density. The green color ($\Delta \ln \langle \rho \rangle = 0$) corresponds to the density predicted by our best-fit halo model. 
We find $\Delta \ln \langle \rho \rangle \sim 0$ over wide ranges, but also regions where the model underestimates the number density (yellow to red) and thus are overdensities. The dark blue regions are edge effects when the samples become sparse at the survey's outskirts. The overdensities are of further interest; we label them accordingly to \cite{Sesar2007}.
}
\label{fig:compare_predicted_observed}}
\end{center}  
\end{figure*}

The first row of Fig. \ref{fig:compare_predicted_observed} shows a realization of a mock ``cleaned sample'' of 11,025 sources (the same number of sources as in the observed cleaned sample), sampled from the best-fit model, a power law with $q(R_{\mathrm{gc}})$ with $r_0=25.0\; \mathrm{kpc}$, $q_0=0.773$, $q_{\infty}=0.998$, $n=4.61$, with applied selection function.\newline
This mock sample looks very comparable to the observed cleaned sample, Fig. \ref{fig:sample_map}(b). The position of the Galactic plane and Sgr stream, as removed by the selection function, are indicated. The dashed circle represents the $R_{\mathrm{gc}} >20 \; \mathrm{kpc}$ cut. Sources within the circle but further away than 20 kpc are seen due to projection effects; the distinctly higher density just after 20 kpc shows the stars that are no longer affected by this distance cut.\newline
We calculated the number density of our observed cleaned sample (given in Fig. \ref{fig:sample_map}) at each $(X,Y,Z)$, using a nearest-neighbor based
adaptive Bayesian density estimator \citep{Sesar2013b, Ivezic2005} yielding $\ln (\rho_{\mathrm{obs}})$. The result is shown in the second row of Fig. \ref{fig:compare_predicted_observed}.\newline
We then applied the same estimation of the 3D number density to 10 realizations of mock samples from the best-fit model; the resulting mean density is given in the third row of the figure as $\ln \langle \rho_{\mathrm{model}} \rangle $.\newline
The logarithmic residuals of the best-fit model were calculated by subtracting the $\ln$ model mean number density (third row) from the observed number density (second row), yielding $\Delta \ln \langle \rho \rangle = \ln (   \rho_{\mathrm{obs}} ) - \ln \langle \rho_{\mathrm{model}}    \rangle $, as given in the last row of this figure. A $\Delta \ln \langle \rho \rangle<0$ indicates the best-fit model overestimates the number densities, whereas a $\Delta \ln \langle \rho \rangle>0$ means that it underestimates the number density.\newline
We find that the best-fit model leads, as expected, to a $\Delta \ln \langle \rho \rangle \sim 0$ over wide ranges, but also shows 
regions where the model underestimates the number density (yellow to red). This can be due to selection effects from the PS1 3$\pi$ observing strategy, but can also be an indicator for unknown structure  and overdensities, as well as a more distorted halo shape. Also our finding that the flattening is different for both hemispheres points toward a halo structure that is more complex than just an ellipsoid.\newline
There are also regions showing slightly negative, near-zero residuals. As we draw the same number of stars from the mock sample as were observed, the overall density is naturally slightly overpredicted if there are underpredicted regions (regions in the PS1 RRab sample containing previously unknown overdensities) in order to match the total number of sources.
This leads to slightly negative residuals when comparing the observed and the mock sample. A similar behavior is shown in \cite{Sesar2013b}, Fig. 10. They illustrate that this behavior is also found when fitting a mock data sample consisting of an underlying halo profile with added diffuse overdensities: slightly negative residuals are found
over a wide area due to a clumpy halo, i.e. a halo with diffuse overdensities. As indicated when discussing the selection function in Sec. \ref{sec:SelectionFunction}, our estimation of purity and completeness are rather uncertain beyond distances of 90 kpc.\newline
The dark blue regions in this plot occur due to edge effects when the samples become sparse at the survey's outskirts.
The diffuse overdensities thus revealed are of further interest; we will discuss them in more detail in Sec. \ref{sec:Overdensities} and thus label them in Fig. \ref{fig:compare_predicted_observed} accordingly to \cite{Sesar2007}.

\begin{figure*}
\begin{center}  
\includegraphics[width=0.8\textwidth, trim=0.0inch 0.8inch 0.0inch 0.5inch, clip=true]{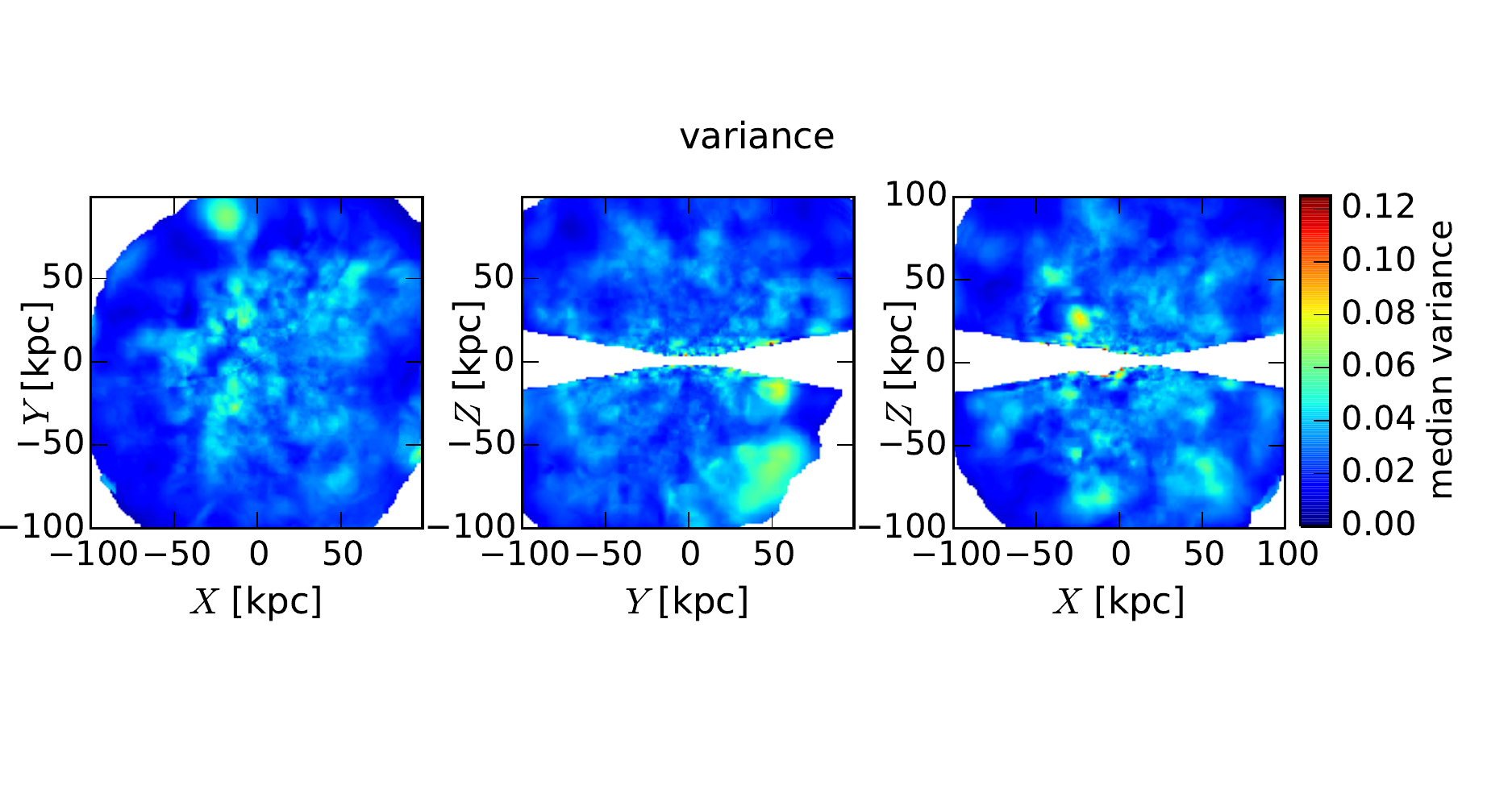}
\includegraphics[width=0.8\textwidth, trim=0.0inch 0.8inch 0.0inch 0.5inch, clip=true]{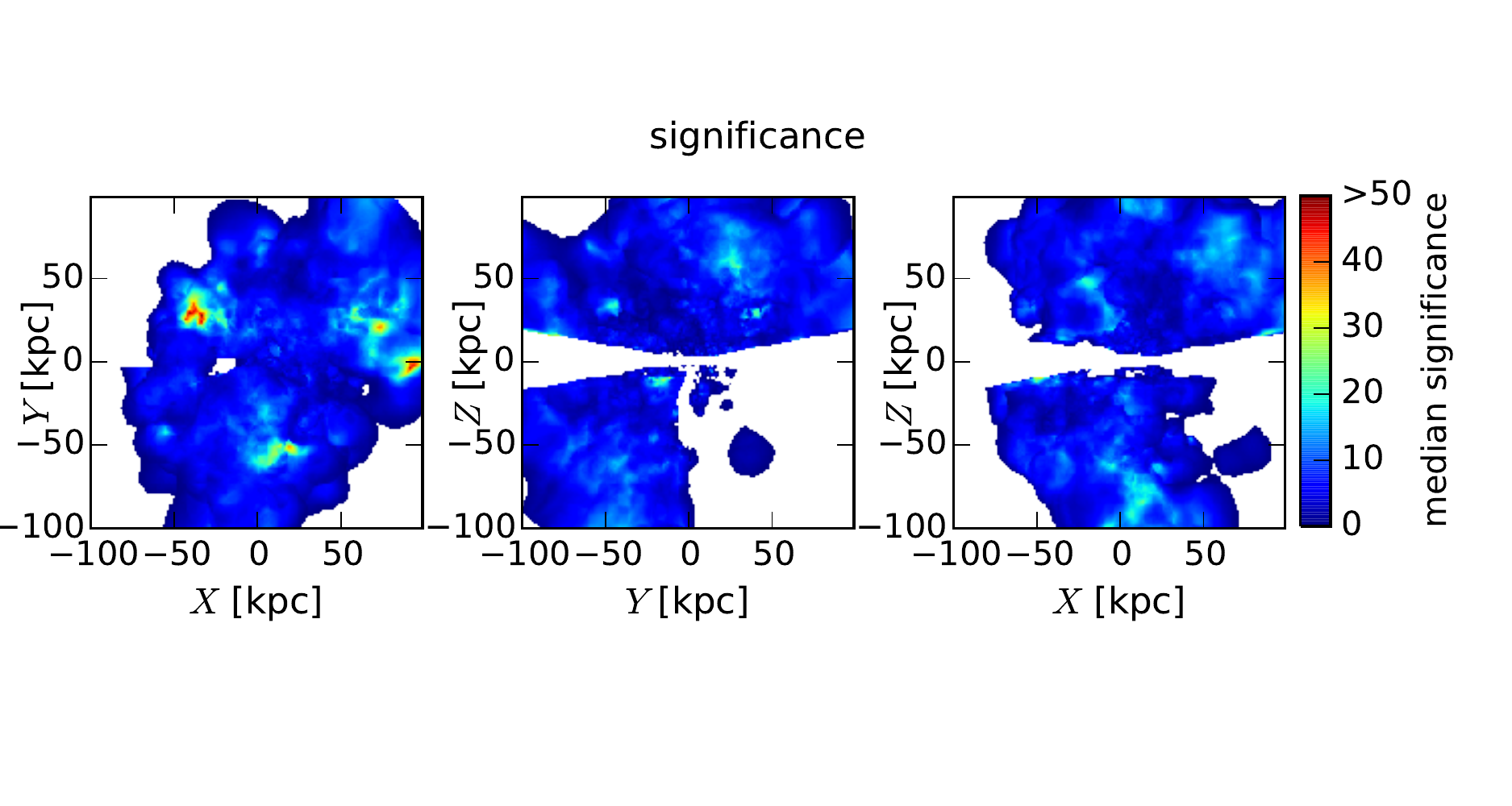}
\caption{{
In order to estimate the significance of the overdensities shown in Fig. \ref{fig:compare_predicted_observed}, we calculate their variance and significance 
as described in Sec. \ref{sec:ResidualsAndSignificance}.}
\label{fig:variance_significance}}
\end{center}  
\end{figure*}

Also, as the relatively sparseness of our cleaned RRab sample (11,025 sources within almost 3/4 of the sky and an extent of $20 \; \mathrm{kpc} < R_{\mathrm{gc}} < 131 \; \mathrm{kpc}$) introduces local number density fluctuations even for a smooth underlying density distribution, we have to estimate the significance of these overdensities. To do so, we carry out the following approach:

\begin{itemize}
\item We bootstrap the observed RRab sample $N=50$ times. 
We estimate the density of each of these bootstrapped samples, using the density estimator by \cite{Ivezic2005}, resulting in $\rho_{\mathrm{obs},i}$ for $i=1...N$. 
We fit each of these bootstrapped samples, sample each of them 10 times and get the mean model density using the density estimator. This yields $\langle \rho_{\mathrm{model}}\rangle_i$ for $i=1...N$, and further $\Delta \left(\ln \langle \rho \rangle \right)_i \equiv \ln(\rho_{\mathrm{obs},i}) - \ln \langle \rho_{\mathrm{model},i}\rangle$ for $i=1...N$.
\item From the above, we can construct the variance
   $\sigma\left(\Delta \left(\ln \langle \rho \rangle \right) \right) \equiv \mathrm{Var} \langle  \Delta \left(\ln \langle \rho \rangle \right)_i \rangle$.
\item The 3D significance is then $\Delta \left(\ln \langle \rho \rangle \right)  / \sigma \left(\Delta \left(\ln \langle \rho \rangle \right) \right)$.
\end{itemize}
The resulting variance and significance are shown in Fig. \ref{fig:variance_significance}, each projected using the mean. Per definition, the significance is 
0 where $\Delta \left(\ln \langle \rho \rangle  \right)=0$.

We find a significance of ${\sim}20$ to $>50$ at regions that coincide with the lower row of panels in Fig. \ref{fig:compare_predicted_observed}, and the variance 
is small and does not exceed 0.04 to 0.08 within these overdense regions. We count this as a strong indicator of these overdensities being real and not caused by Poisson number density fluctuations.

\subsubsection{Overdensities}
\label{sec:Overdensities}

We compare the overdensities found by us with those discovered previously by \cite{Sesar2007} and \cite{Sesar2010}. 

In their studies they analyzed the spatial distribution of candidate RR Lyrae stars discovered by SDSS Stripe 82 along the Celestial Equator. They had used 634 RR Lyrae candidates from SDSS Stripe 82 and 296 RR Lyrae candidates from \cite{Ivezic2000} in their 2007 analysis \citep{Sesar2007}, and later on cleaned the SDSS Stripe 82 sample of RR Lyrae \citep{Sesar2010}, using then 366 highly probable RRab stars.

\begin{figure*}
\begin{center}  
\includegraphics[trim=0.5inch 0.0inch 0.0inch 0.0inch, clip=true]{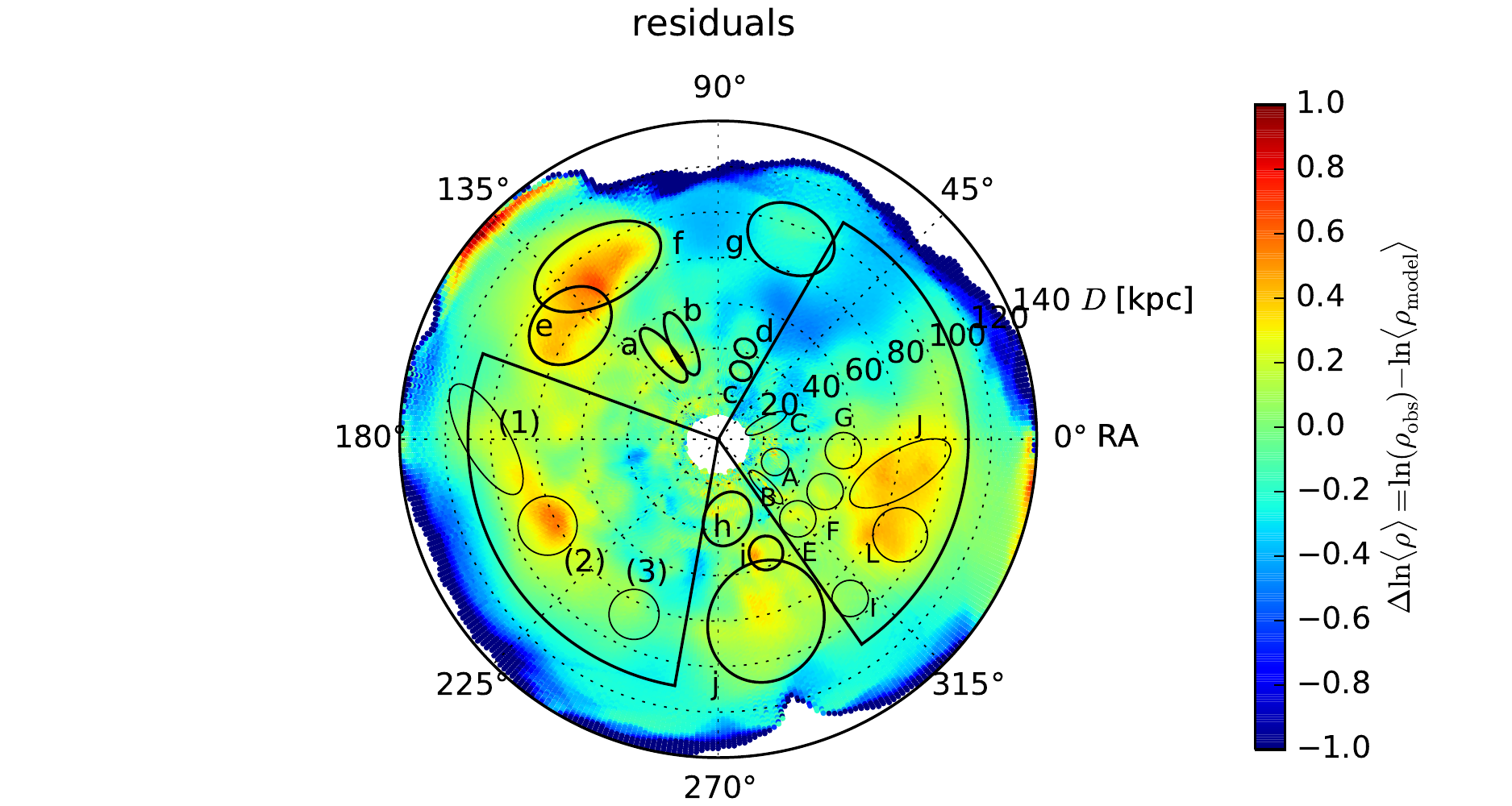}
\caption{{
Plot of the overdensities in a $(\mathrm{RA}, D)$ projection similar to \cite{Sesar2007}. The green color ($\Delta \ln \langle \rho \rangle = 0$) corresponds to the model density, yellow and red regions are overdensities, and blue regions are underdensities, analogous to Fig. \ref{fig:compare_predicted_observed}.
Upper-case letters denote overdensities found in the SDSS sample of \cite{Sesar2007} and \cite{Sesar2010}, numbers denote overdensities found in their analysis of the \cite{Ivezic2005} sample (not numbered in \cite{Sesar2007}), and lower-case letters denote overdensities we found in regions not covered by the analysis of \cite{Sesar2010}.
A detailed discussion is given in Sec. \ref{sec:Overdensities}.}
\label{fig:overdensities}}
\end{center}  
\end{figure*}

In Fig. \ref{fig:overdensities}, we plot the overdensities in a $(\mathrm{RA}, D)$ projection similar to \cite{Sesar2007} (see their Fig. 13, see also Fig. 11 in \cite{Sesar2010}), using our full range in declination and highlighting the region covered by their analysis. Upper-case letters denote overdensities found in the SDSS sample of \cite{Sesar2007} and \cite{Sesar2010}, numbers denote overdensities found in their analysis of the \cite{Ivezic2000} sample (not numbered in \cite{Sesar2007}), and lower-case letters denote overdensities we found in regions not covered by the analysis of \cite{Sesar2010}.

We can recover most of the overdensities found by \cite{Sesar2007} and \cite{Sesar2010}, i.e. we recover their overdensities A, B, C, E, F, G, I, J, L.
Among them, \cite{Sesar2010} claim that they do not find overdensities I and L they had found in their previous analysis and attribute this to their then better, cleaned sample of RR Lyrae stars. However, we find the overdensities I and L, where especially L stands out.
We could verify that some overdensities found in \cite{Sesar2007} will disappear in a more cleaned sample, as shown in \cite{Sesar2010}: consistent with \cite{Sesar2010}, we don't find the overdensities D, H, K and M. However, the overdensitiy D appears very small in \cite{Sesar2007}, so we cannot say for sure if we are able to identify anything at this position. Our sample does not cover exactly the same extent in RA and Dec, so the overdensities D, H, K, M as given in \cite{Sesar2007} lie in regions we don't cover. We have checked adjacent slices in declination to see if we might detect those overdensities, but cannot find them. So our conclusion regarding those four overdensities is that either they are not real, as assumed by \cite{Sesar2007}, or they have a small extent in declination.

The left wedge of Fig. \ref{fig:overdensities} compares to the left wedge of Fig. 13 in \cite{Sesar2007}, where they used RR Lyrae from \cite{Ivezic2005}. We label these overdensities (1), (2), (3).

In regions not covered by \cite{Sesar2007}, we detect many new overdensities out to $D \gtrsim 100 \; {\mathrm{kpc}}$ continuing the overall distribution of overdensities found before. We label them by lower-case letters.

The strongest clump in the left wedge, (2), stems from stars belonging to the Sgr stream not being fully excised by our cuts. The same holds for the small and sparse overdensity C being part of the stream' trailing arm \citep{Sesar2007,Ivezic2003a}.

\cite{Sesar2010} claim that the overdensity J is most probably a stellar stream, the Pisces oversensity \citep[see also][]{Watkins2009}.
In contrast to \cite{Sesar2007} and \cite{Sesar2010}, we find that the overdensities J and L might be connected.

\section{Discussion}
\label{sec:Discussion}

Before discussing possible implications of the results, it is worth to discuss some potential sources of bias.

Carrying out the described modeling of the halo profile for the complete cleaned sample, we find that among models
with constant flattening, the distribution of the sources is reasonably well fit
by a power law model with $n=4.40^{+0.05}_{-0.04}$ and a halo flattening of
$q=0.918^{+0.016}_{-0.014}$. The distance distribution is fit comparably well by
a model with an Einasto profile with $n=9.53^{+0.27}_{-0.28}$, an effective
radius $r_{\mathrm{eff}}=1.07 \pm 0.10 \; \mathrm{kpc}$ and a halo
flattening of $q=0.923 \pm 0.007$.
If we allow for a radius-dependent flattening $q(R_{\mathrm{gc}})$, we find the
best-fit parameters for a power law model with $q(R_{\mathrm{gc}})$ as
$r_0=25.0^{+1.8}_{-1.7}\; \mathrm{kpc}$, $n=4.61 \pm 0.03$,
$q_0=0.773^{+0.017}_{-0.016}$, $q_{\infty}=0.998^{+0.002}_{-0.001}$.
The best-fit parameters for an Einasto profile with $q(R_{\mathrm{gc}})$ are
$r_0=26.7^{+2.2}_{-2.0}\; \mathrm{kpc}$, $q_0=0.779 \pm 0.018$, $q_{\infty}=0.998^{+0.001}_{-0.002}$,
$r_{\mathrm{eff}}=1.04^{+0.25}_{-0.13}\; \mathrm{kpc}$, $n=8.78^{+0.33}_{-0.30}$.
Allowing for a break in the power-law profile, we find a break radius of $r_{\mathrm{break}}=38.7^{+0.69}_{-0.58}$, a halo flattening of $q=0.908^{+0.008}_{-0.006}$, and the inner and outer slopes $n_{\mathrm{inner}}=4.97^{+0.02}_{-0.05}$ and $n_{\mathrm{outer}}=3.93^{+0.05}_{-0.04}$, respectively.

From these fits, we find two robust effects to emerge: There is evidence for the
stellar halo being distinctly more flattened at small radii ($q \sim 0.8$), and
more spherical at large radii ($q \sim 1$). The flattening is consistent among
all halo profiles we explored. There is no evidence for a steepening of the halo profile beyond 65 kpc as found by \cite{Deason2014}, neither from our fits nor our data.

We also fitted the halo profile for both the north and south Galactic hemisphere
independently, in order to look for possible local variations in the best-fit
model, especially in the halo flattening $q$ and steepness $n$. The north
hemisphere contains 6,880 sources, while the south hemisphere contains only
4,145 sources because of the PS1 3$\pi$ survey footprint.
We find that the steepness parameters $n$ of all best-fit hemisphere models
compare well for each the north and south Galactic hemisphere and also compare
well with the fit for the complete halo. We further find that for models with
constant flattening (thus the power law and Einasto profile models),
$q_{\mathrm{south}} < q < q_{\mathrm{north}}$, and the same applying for $q_0$ in the case of models with
$q(R_{\mathrm{gc}})$. For those models, we find the value of the parameter $q_{\infty}$ being similar for the
north and south hemisphere and the complete halo. However, we find that
$r_{0,\mathrm{north}}>r_{0,\mathrm{south}}>r_0$ for both the power law with
$q(R_{\mathrm{gc}})$ and the Einasto profile with $q(R_{\mathrm{gc}})$.

The results $q_{\mathrm{south}} < q < q_{\mathrm{north}}$ for models with
constant flattening, and 
$q_{0,\mathrm{south}} < q_0 < q_{0,\mathrm{north}}$, $q_{\infty,\mathrm{south}}  \sim
q_{\infty,\mathrm{north}} \sim q_{\infty,\mathrm{south}}$
$r_{0,\mathrm{north}}>r_{0,\mathrm{south}}>r_0$ in the case of a
radius-dependent flattening, are consistent: by definition of
$q(R_{\mathrm{gc}})$ (see Eq. \eqref{eq:q_r}), $q_0$ is the flattening at the
center, whereas $q_{\infty}$ is the flattening at large Galactocentric radii,
and $r_0$ is the exponential scale radius for the flattening. A larger $r_0$
means that the flattening of the inner halo, where we find $q_0<q_{\infty}$, 
in effect extends out to a larger Galactocentric radius than for a smaller
$r_{\mathrm{eff}}$, thus leading to a larger fraction of the halo being more
flattened.

Thus, the generalized result suggests that the south Galactic hemisphere is
somewhat more flattened than the north Galactic hemisphere.

Our finding that a best-fit model requires a flattened, and especially a
varying-flattened, halo with a smaller $q$ (minor-to-major-axis ratio) in the
inner parts and a larger value, $q{\sim}1$, in the outskirts is supported by many other
studies.

In our subsequent analysis, we found that the halo might be more irregular than
only being influenced by a flattened halo or flattened inner and outer halo
(``dichotomy'' of the halo).
From calculating the residuals of our best-fit model, we find that there is some
deviation of the real halo structure from our best-fit model.

Using our best-fit halo model, we then continued by computing the residuals in
order to find local deviations from the smooth halo described by the best-fit
model. We found striking overdensities and compare them to the ones discovered
by \cite{Sesar2007} and \cite{Sesar2010}. Additionally, we find new
overdensities  in regions of the sky not accessible to \cite{Sesar2007} and
\cite{Sesar2010}.

After describing the outcome of our study and its scientific relevance, we now
briefly discuss possible sources of bias in the maximum likelihood analysis.
Removing known overdensities such as streams, globular clusters and dwarf
galaxies, as well as the Milky Way disc and bulge, from our sample, thus producing the
``cleaned sample'', was crucial. Our cuts on the disc and bulge are fairly
broad. For the Sgr stream, we tried to find a good tradeoff between removing
most of the stream and not removing too many background stars, and thus decided
to remove sources based on their probability for being associated with the Sgr stream as shown in our previous work \citep[][see Equ. (11) therein]{Hernitschek2017}.

Another crucial point is our assumption on the cleanness of the RRab sample, as
well on its distance precision. For both we refer to \cite{Sesar2017b}, who
claim, based on extensive testing,
a high-latitude sample purity of 90\%, a sample completeness of $90\%$ within 60~kpc and
$\geq80\%$ at 80~kpc, and a distance precision of 3\%. To account for the
distance-dependent completeness, we introduced a term in our selection function.

Regions with high dust extinction can add severe uncertainties in the study of the
distribution of stars in the Galaxy. We account for that with our cut on
the region around the Galactic disc, $\vert b \vert \geq 10 \arcdeg$.

The halo fit can also be influenced by up to now unknown overdensities. On the
other hand, we can use the halo fit to identify such overdensities, as well as
to infer the distortion of the halo from an ellipsoid that is flattened in the $Z$ direction.

\subsection{Comparison With Previous Results}
\label{sec:Comparison}

We now compare our results -- especially halo steepness and
flattening -- to earlier findings from the many other groups that have attacked this important and interesting problem.
In previous efforts to determine the halo shape, RR Lyrae and BHB stars have
often been used as tracers because they are found in old populations, have
precise distances and are bright enough to be observed at radii out to
${\sim}$100 kpc \citep[see e.g.][]{Xue2008, Xue2011, Sesar2010, Deason2011, Deason2013, Deason2014}.

The major issue with BHB stars is potential confusion with blue stragglers and with QSOs. Samples of BHB stars must be carefully vetted 
to ensure that contamination is minimalized. This is not easy, especially as one moves further out to fainter objects, see e.g. \cite{Deason2012}, 
where in an effort to build a sample of BHB stars beyond 80 kpc, of 48 candidates selected photometrically, after the acquisition of low spectral
resolution VLT-FORS2 spectra, only 7 turned out to be bona fide BHB stars. RR Lyrae, on the other hand, can be identified and verified with just photometric
light curves, available from the application of modern machine learning techniques to the
 databases of the large multi-epoch photometric surveys that have been carried out over the past decade.

It is important to remember that our sample selection procedure which we apply to the Pan-STARRS1 3$\pi$ database takes advantage of
the experience gained by attempting to use the SDSS \citep{Sesar2007,Sesar2010} and then to analyze the more difficult Palomar Transient Facility \citep{Cohen2017} with few observations 
taken with a random cadence. Our PS1 sample of RRab stars \citep{Sesar2017b} is unique in that it contains a large number (44,403) of RR Lyrae in total, of which 17,452 lie within the radial
range of $R_{\mathrm{gc}}$ from 20 up to 130~kpc, all with highly precise photometry.
This yields a sample of high purity and completeness exceeding 80\% out to $R_{\mathrm{gc}} = 80 \; \mathrm{kpc}$. 
Out of these sources, 11,025 are found outside of dense regions such as stellar streams, the Galactic disc and bulge or globular clusters and stellar streams.

Furthermore, \cite{Sesar2017b} have quantified this completeness with extensive testing throughout the entire radial range.
Confusion with QSOs and with blue stragglers is eliminated by requiring a light curve characteristic of a RR Lyrae star.

First glimpses of the variation of the halo shape were already caught by
\cite{Kinman1966}, based on RR Lyrae stars as halo tracer. As a first attempt,
\cite{Preston1991} argued that the flattening changes from $q=0.5$ at 1 kpc to
$q \sim 1$ at 20 kpc. However, later work by \cite{SluisArnold1998} shows a
constant flattening of $q \sim 0.5$ without any evidence for a radius-depending
flattening. A recent work by \cite{DePropris2010} utilizing 666 BHB stars from
the 2dF QSO Redshift Survey states that the halo is approximately spherical with
a power-law index of ${\sim}$2.5 out to 100 kpc. \cite{Sesar2010} who studied
main-sequence turn-off (MSTO) stars from the Canada-France-Hawaii Telescope
Legacy Survey find that the flattening is approximately constant at $q \sim 0.7$
out to 35 kpc.

\cite{Carollo2007} and \cite{Carollo2010} found that the inner halo is highly
flattened with axis ratios of $q\sim 0.6$, whereas the outer halo is more
spherical with axis ratios of $q\sim 0.9$. In contrast to us, they include stars
as close as 2 kpc into their fitting, and thus get a more pronounced flattening
within about $5-10$ kpc.

Others also find evidence that at least the innermost part of the halo is quite
flattened: \cite{Sesar2010} fit the Galactic halo profile based on ${\sim}$5000
RR Lyrae stars from the recalibrated LINEAR data set, spanning $5
\;{\mathrm{kpc}} < D < 30 \;{\mathrm{kpc}}$ over $\sim$8000 deg$^2$ of the sky.
They find for their best-fit model an oblate ellipsoid with an axis ratio of
$q=0.63$, and a double power-law model with $q=0.65$, $n_{\mathrm{inner}} =1$,
$n_{\mathrm{outer}} =2.7$, $r_{\mathrm{break}} =16 \; \mathrm{kpc}$.

So, based on the different distances span by the aforementioned work, there is much evidence that the innermost part of the halo is quite flattened, the
outer part of the halo is more spherical, and our results confirm that.

As a reason for the smaller minor-to-major axis ratio $q$ found for the inner
halo, \cite{Deason2011} state that
inner-halo stars possess generally high orbital eccentricities, and exhibit a
modest prograde rotation around the Galactic center. In contrast, stars in the
outer halo exhibit a much more spherical spatial distribution as they cover a
wide range of orbital eccentricities, and show a retrograde rotation about the
Galactic center.

For the density slope of the halo profile, many studies
\citep[e.g.][]{Sesar2007, Watkins2009, Sesar2011, Deason2011} find that it shows
a shift from a relatively shallow one, as described by $n {\sim}2.5$, to a much
steeper one outside of about 20 to 30 kpc that is consistent with $n {\sim} 4$.
The earliest evidence for that is from \cite{Saha1985} who found that RR Lyrae
are well described by a broken power law (BPL) with $n \sim 3$ out to 25 kpc,
and $n \sim 5$ beyond.

Subsequently, \cite{Sesar2011} used 27,544 near-turnoff MS stars out to ~35 kpc
selected from the Canada-France-Hawaii Telescope Legacy Survey to find a
flattening of the stellar halo of 0.7 and the density distribution to be
consistent with a BPL with an inner slope of 2.62 and an outer slope of 3.8 at
the break radius of 28 kpc, or an equally good Einasto profile with a
concentration index of 2.2 and an effective radius of 22.2 kpc.

\cite{Xue2015} probe the Galactic halo at $10 \; \mathrm{kpc} < R_{\mathrm{gc}}
< 80 \; \mathrm{kpc}$ using 1,757 stars from the SEGUE K-giant Survey. The
majority of their sources are found at $R_{\mathrm{gc}}<30\; \mathrm{kpc}$,
whereas in our sample, 1,093 RRab stars exist beyond a Galactocentric distance of
80 kpc, and 238 beyond 100 kpc.
They find that they can fit their sample by an Einasto profile with $n=3.1$,
$r_{\mathrm{eff}}$ = 15 kpc, $q = 0.7$, by an equally flattened broken power law
with $n_{\mathrm{in}}$ = 2.1, $n_{\mathrm{out}}$ = 3.8, $r_{\mathrm{break}}$ =
18 kpc (this is something we had not applied), and when fitting by an Einasto
profile with $q(R_{\mathrm{gc}})$, they find the halo being considerably more
flattened as $q$ changes from $0.55 \pm 0.02$ at 10 kpc to $0.8 \pm 0.03$ at
large radii.

\cite{Bell2008} used ${\sim}$4 million color-selected MS turnoff stars from DR5
of the SDSS out to 40 kpc, and find a best-fit flattening of the stellar halo of
$0.5 - 0.8$, and the density profile of the stellar halo is approximately
described by a power law with index of $2 - 4$.

Other estimates of the power law index, or slope, of the halo give break radii
or effective radii of ${\sim}20-30$ kpc, power-law slopes of $n\sim3$
\citep[e.g.][]{Sesar2011, Deason2011, Deason2014, Xue2015}. For example,
\cite{Sesar2011} fit the Galactic halo within heliocentric distances of $< 35$
kpc, steeper at $R_{\mathrm{gc}} > 28 \; \mathrm{kpc}$, 
$n_{\mathrm{inner}} = 2.62$,
$n_{\mathrm{outer}} = 3.8$, or a best-fit Einasto profile with $n=2.2$,
$r_{\mathrm{e}} = 22.2 \; \mathrm{kpc}$, $q = 0.7$, where they found no evidence
that it changes across the range of probed distances.
Subsequently, \cite{Deason2014} found a very steep outer halo profile with a
power law of 6 beyond 50 kpc, and yet steeper slopes of 6 - 10 at larger radii.

Our findings of $n=4.40 - 4.61$ for a power-law model, or $n=8.78 - 9.53$
for an Einasto profile(keep in mind the different definitions of $n$) for our
sample starting at $R_{\mathrm{gc}} = 20 \; \mathrm{kpc}$ are thus in good
agreement with most previous results, assuming no break radius and thus a
constant density slope or the slope variation as introduced by the Einasto
profile (see Equ. \eqref{eq:einastosteepness}). We claim that the estimate of
the power-law parameter from \cite{DePropris2010}, $n{\sim}2.5$, is too shallow,
as well as that the estimate of the power-law parameter beyond 65 kpc from
\cite{Deason2014}, $n=6-10$, is too steep, as we don't see such a drop from our fit nor from our data.

We cannot verify results showing a break radius near 20 kpc, as our sample starts at 20 kpc and thus only a small number of sources would be found, if at
all, within the break radius. However, the extent of our sample enables us to check the finding by \cite{Deason2014}. They find a BPL with three ranges of subsequently steepening slope:
2.5 for $10 \; \mathrm{kpc} < R_{\mathrm{gc}}<25 \; \mathrm{kpc}$, 4.5 for $25
\; \mathrm{kpc} < R_{\mathrm{gc}}<65 \; \mathrm{kpc}$, 10 for $65 \;
\mathrm{kpc} < R_{\mathrm{gc}}<100 \; \mathrm{kpc}$. Whereas in the distance
range $25 \; \mathrm{kpc} < R_{\mathrm{gc}}<65 \; \mathrm{kpc}$, their profile
agrees well with our results, we cannot probe the inner part as our sample
starts at 20 kpc, and our data argue against a significant steeper profile at
$R_{\mathrm{gc}}>65 \; \mathrm{kpc}$ as long as the PS1 selection function is
applied. We thus fit a BPL to our sample, allowing for a break radius beyond 20 kpc, and {find a slope of 3.93 beyond 39 kpc all the way out to the limit of our sample}. There is no evidence for a steepening of the halo profile beyond 65 kpc as found by \cite{Deason2014}, or any other indication that there is a truncation or break in the halo profile within the range we probe by the RRab sample. We roughly confirm their power-law slope of 4.8 for the region $25 \; \mathrm{kpc}< R_{\mathrm{gc}} < 45\; \mathrm{kpc}$, as we find a power-law slope of 4.97 for $20 \; \mathrm{kpc}< R_{\mathrm{gc}} < 39\; \mathrm{kpc}$. The small change in slope near 39 kpc may be related to how, in detail, the Sgr stream is removed. Beyond 40 kpc we find a much less steep power-law slope than do \cite{Deason2014}.

\cite{Deason2013} interpret the presence or absence of a break as linked to the
details of the stellar accretion history. They state that a prominent break can
arise if the stellar halo is dominated by the debris from an accretion event
that is massive, single and early. \newline

\cite{Xue2015} and \cite{Slater2016} find a halo profile that is shallower than
our best-fit models; it is difficult to know if this difference results from
methological differences or some intrinsic difference in the
distribution between RR Lyrae and giants \citep{Slater2016} or K giants
\citep{Xue2015}.

\cite{Iorio2017} very recently carried out an attempt to map the
Galactic inner halo with $R_{\mathrm{gc}} < 28 \; \mathrm{kpc}$ based on a
sample of 21,600 RR Lyrae from the Gaia and 2MASS surveys.
They found that the best-fit model to describe the halo distribution is a power
law with $n=2.96 \pm 0.05$, and flattening is present resulting in an triaxial
ellipsoid.

\begin{figure*}
\begin{center}  
\includegraphics[]{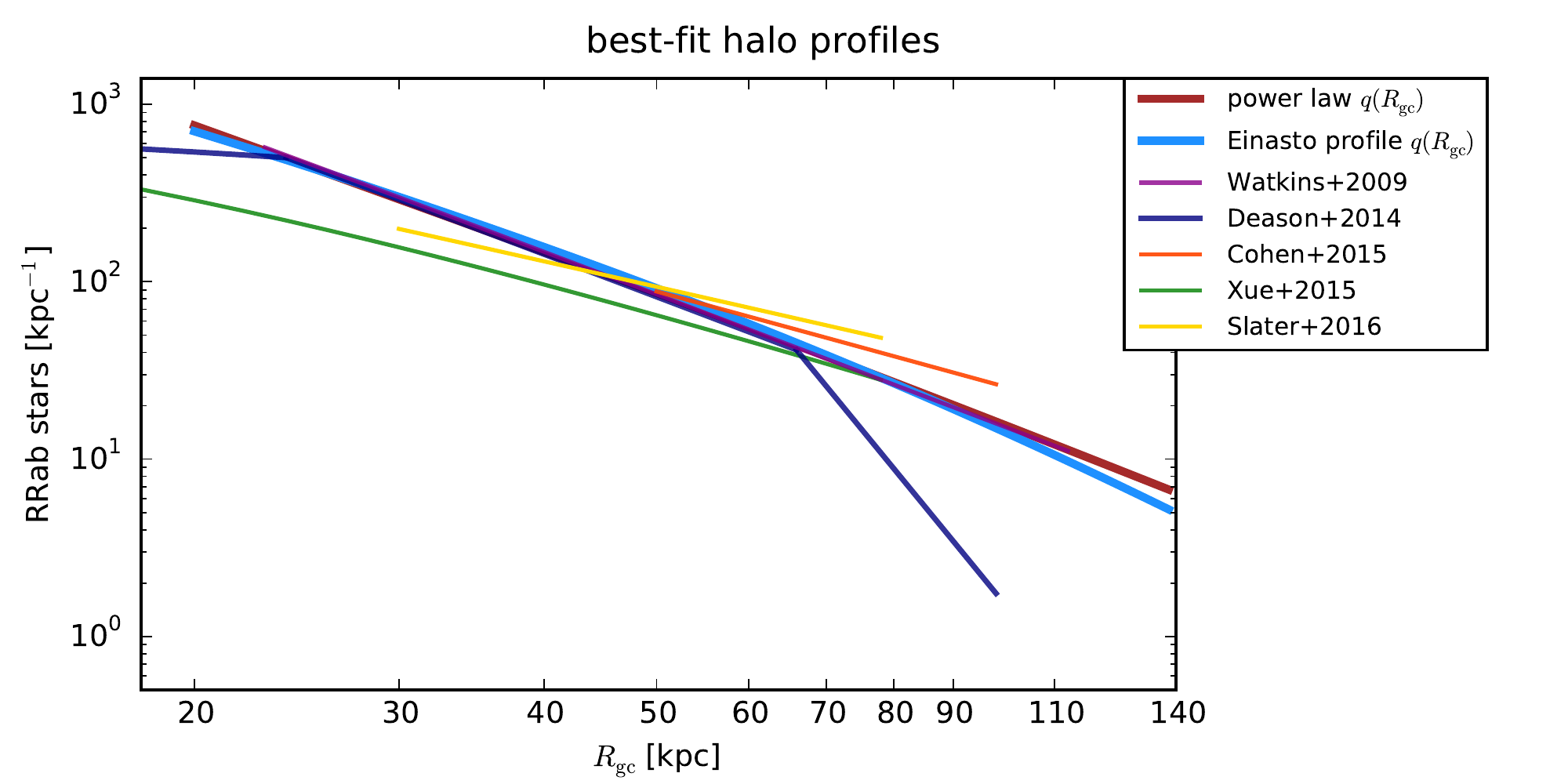}
\caption{{
Comparison of two of our best-fit models, the power law and the Einasto profile each with radius-dependent $q(R_{\mathrm{gc}})$, to best-fit models of other work.
Their model parameters are given in  Tab. \ref{tab:compare_models_selected}, whereas the parameters of our best-fit models are given in Tab. \ref{tab:bestfit}.\newline
Most best-fit models compare well to ours within their distance range. However, the models by \cite{Xue2015} and \cite{Slater2016} are slightly shallower, and the model by \cite{Deason2014} shows a broken power-law (BPL) shape that we find neither from our fits nor our data.}
\label{fig:compare_models_selected}}
\end{center}  
\end{figure*}

In Table \ref{tab:compare_models_selected}, we give the best-fit parameters for the
Galactic halo as found in other work, along with the distance range over which
they were estimated. These models are visualized, together with our best-fit
models, in Fig. \ref{fig:compare_models_selected}.

Fig. \ref{fig:compare_models_selected} shows a remarkably different slope for the models based on giants 
(the yellow, orange, and green lines in this figure) in contrast to those
from using horizontal branch and RRab stars (all other lines in this figure, including our halo fits). We interpret this as a sign that older stellar populations (RR Lyrae and BHB stars) are distributed in a more concentrated
way than giants that should span a wider age spread, and thus gives information on the assembly
process of the halo: by definition, very early accretion contains only old stellar populations,
and these could be more concentrated because only more bound orbits accreted onto
the young and lighter proto Milky Way, whereas more giants originate from later accretions of dwarf galaxies 
that had prolonged star formation and therefore formed more stars but not more RR Lyrae.

We find that models of stellar accretions supporting these ideas, especially \cite{Font2008} stating that 
the larger spread in ages found in the outer halo results from the late assembly of those stars compared to those in the inner halo
\citep{Bullock2005,Font2006}. The outer haloes in the models studied by \cite{Font2008} tend to show a larger spread in the ages and
metallicities of their stellar populations than the inner haloes, and this suggests that the outer halo should have a significantly larger fraction of 
intermediate-age versus old stars than the inner halo.

In our subsequent analysis, we found that even after removing all known prominent substructures, the halo might be more irregular than
only being influenced by a flattened halo or flattened inner and outer halo
(``dichotomy'' of the halo). We find that the south Galactic hemisphere is somewhat more flattened than the north Galactic hemisphere.
From calculating the residuals of our best-fit model, we find that there is some
deviation of the real halo structure from our best-fit model.

\section{Summary}
\label{sec:Summary}

We used a sample of of 44,403 PS1 RRab stars from \cite{Sesar2017b} in order to determine the spatial structure of the Galactic halo using Pan-STARRS 1 3$\pi$ RR Lyrae.
We excluded known overdensities, among them the Sagittarius stream, dwarf galaxies
such as the Draco dSph, and globular clusters. Also, we cut out sources too close to the Galactic plane ($\vert b \vert < 10 \arcdeg$), or too close to the
Galactic center ($R_{\mathrm{gc}} \leq 20 \; \mathrm{kpc}$). We end up with a sample of 11,025 RRab in the Galactic stellar halo, called the ``cleaned
sample".
Each RRab star has a highly precise distance (3\% uncertainty). The sample is very pure and with high completeness. Each star has a photometric light curve which
resembles that of a RRab, guaranteeing a very low level of interlopers.

A forward modeling approach using different density models
for the Galactic halo profile, as well as a selection function of the sample describing the aforementioned cuts to exclude overdensities, as well as the distance-depending completeness,was then applied to this sample.

Our basic result is that the stellar Galactic halo, when described purely by RRab stars outside of known overdensities, can be characterized by a power law with an exponent of
$n=4.61 \pm 0.03$, and a varying flattening ($q(R_{\mathrm{gc}})$) with a more oblate inner halo and an almost spherical outer halo, as described by the parameters
 $r_0=25.0^{+1.8}_{-1.7}\; \mathrm{kpc}$, $q_0=0.773^{+0.017}_{-0.016}$, $q_{\infty}=0.998^{+0.002}_{-0.001}$.
From our halo fits, we find three robust effects to emerge: There is evidence for the stellar halo being distinctly more flattened at small radii ($q \sim 0.8$), and
more spherical at large radii ($q \sim 1$). The flattening is consistent among all halo profiles we explored. We have no indication that there is a truncation or break in the halo profile within the range we probe by the RRab sample.
As discussed in Sec. \ref{sec:Comparison}, broadly speaking the results of our work are largely consistent with most earlier work.
However, we do not find a broken power-law halo claimed by \cite{Deason2014}, in particular we cannot reproduce their extreme power law of $n=6-10$
beyond 65 kpc, but do confirm their power-law slope of 4.8 for the region $25 \; \mathrm{kpc}< R_{\mathrm{gc}} < 45\; \mathrm{kpc}$.

Further, we claim that the estimate of the power-law parameter given by \cite{DePropris2010}, $n{\sim}2.5$, is too small to agree with our results.

To explore further, we fitted the halo profile for both the north and south Galactic hemisphere independently, in order to look for possible local variations in the best-fit
model, especially in the halo flattening $q$ and steepness $n$.
Our generalized result suggests that the south Galactic hemisphere is somewhat more flattened than the north Galactic hemisphere.

The final step in our analysis of the structure
of the outer halo of the Milky Way was to compute the residuals from our results as compared to
the smooth halo described by the best fit model.  This difference map was used
to find local deviations from the smooth halo described by the best-fit
model. We found striking overdensities and compare them to the ones discovered
earlier by \cite{Sesar2007} and \cite{Sesar2010}. Additionally, we find new
overdensities which are in regions of the sky not accessable to
\cite{Sesar2007} and \cite{Sesar2010}.

The overaching goals of studies of the Milky Way's outer halo are to
determine the total mass of the Galaxy and, to the extent possible,
the origin and importance of substructure from which one can infer
clues regarding the accretion history of the formation of the Galaxy.
Having established the spatial profile of the Galactic halo and evaluated
at least partially the local deviations from smooth structure as present
but not overwhelming over the regime from $R_{\mathrm{gc}}= 20$ to 130~kpc,
the next step towards these goals is a study of the kinematics of the outer halo.
We need 6D information, i.e. positions, distances, proper motions and radial velocities.
We already have the first two of this list.  However, 
unfortunately the accuracy of Gaia proper motions of RRab in the outer halo
at the large $R_{\mathrm{gc}}$ we have probed is poor in Gaia DR1, and even
after the end of the Gaia mission, is still not as accurate as might be desired.
Determination of the radial velocity distribution as a function of $R_{GC}$
out to these large distances will require
a massive dedicated spectroscopic program at a large telescope. 
Such an effort has been initiated.

\acknowledgments

H.-W.R. acknowledges funding from the
European Research Council under the European Unions
Seventh Framework Programme (FP 7) ERC Grant Agreement n. [321035].

The Pan-STARRS1 Surveys (PS1) have been made possible through contributions of the Institute for Astronomy, the University of Hawaii, the Pan-STARRS Project Office, the Max-Planck Society and its participating institutes, the Max Planck Institute for Astronomy, Heidelberg and the Max Planck Institute for Extraterrestrial Physics, Garching, The Johns Hopkins University, Durham University, the University of Edinburgh, Queen's University Belfast, the Harvard-Smithsonian Center for Astrophysics, the Las Cumbres Observatory Global Telescope Network Incorporated, the National Central University of Taiwan, the Space Telescope Science Institute, the National Aeronautics and Space Administration under Grant No. NNX08AR22G issued through the Planetary Science Division of the NASA Science Mission Directorate, the National Science Foundation under Grant No. AST-1238877, the University of Maryland, and Eotvos Lorand University (ELTE) and the Los Alamos National Laboratory. 

\clearpage

\appendix

\section{Figures}
\label{sec:Figures}

\clearpage

\clearpage

\clearpage

\clearpage

\clearpage

%

\clearpage

\section{Tables}
\label{sec:Tables}
\newpage

\capstartfalse
\begin{deluxetable*}{lrrrrrrrr}
\tablecolumns{9}
\tablecaption{Removed Overdensities within $R_{\mathrm{gc}}>20$ kpc\label{tab:overdensities}}
\tablewidth{0pt}
\tablehead{
\colhead{name} &  \colhead{$R_{\mathrm{gc}}^\mathrm{a}$} & \colhead{remove} & \colhead{remove} & \colhead{remove} & \colhead{remove} & \colhead{remove} & \colhead{remove} & \colhead{removed} \\
\colhead{} & \colhead{(center)} & \colhead{$l$ min [$\arcdeg$]} & \colhead{$l$ max [$\arcdeg$]} & \colhead{$b$ min [$\arcdeg$]} & \colhead{$b$ max [$\arcdeg$]} & \colhead{$D$ min [kpc]} & \colhead{$D$ max [kpc]} & \colhead{sources}
}
\startdata
Bootes III dSph & 37.87 & 32 & 34.1 & 74.5 & 75.4 & 45.9 & 46.5 & 3\\
Sextans dSph & 45.14 & 242 & 245 & 41 & 44 & 60 & 120 & 99 \\
NGC 292 Bootes I dSph & 45.88 & 357 & 359 & 69 & 70 & 55 & 70 & 4\\
UMa 1 dSph &  60.17 & 150 & 160 & 54 & 54.6 & 90 & 120 & 4 \\
Draco dSph & 80.70 & 84 & 87 & 33.5 & 35.5 & 65 & 100 & 191 \\
UMi dSph & 48.23 & 100 & 110 & 40 & 50 & 60 & 80 & 53 \\
NGC 7089  M 2 & 25.08 & 53 & 53.5 & -36 & -35.5 & 11 & 12.5 & 4\\
NGC 6626  M 28 & 71.90 & 7.8 & 8 & -6 & -5.3 & 5.3 & 5.8 & 3*\\
Pal 3  & 45.25 & 240 & 240.2 & 41.8 & 42 & 80 & 100 & 3 \\
Laevens 3 & 45.56 & 63.58 & 63.602 & -21.2 & -21.13 & 55 & 62 & 2\\
NGC 2419  & 56.06 & 178 & 183 & 24 & 26 & 76 & 84 & 8* \\
NGC 6293 & 82.37 & 357 & 359 & 7 & 9 & 9 & 10 & 16*\\
NGC 6402  M 14 &93.26 & 21 & 21.8 & 14.5 & 15.2 & 8 & 10 & 6\\
NGC 6171 M 107 & 109.44 & 2.8 & 3.8 & 22.1 & 23.7 & 5.5 & 8 & 7*\\
Pisces Overdensity  & 51.67 & 87.3 & 87.4 & -58.2 & -57.9 & 79 & 82 & 1\\
RR10$^{\dag}$ & 25.49 & 186.37 & 186.38 & 51.5 & 51.6 & 41.2 & 41.3 & 1*\\
\enddata
\tablecomments{Overdensities are grouped by dwarf galaxies, globular clusters, and others (Pisces Overdensity, and the single RRab star RR10). In each group, they are ordered by $R_{\mathrm{gc}}$.}
\tablenotetext{a}{the center of the removed overdensity}
\tablenotetext{*}{these sources are also removed by other cuts}
\tablenotetext{$\dag$}{this RRab is a member of the Orphan stream, \cite{Sesar2013a}}
\end{deluxetable*}   
\capstarttrue

\capstartfalse
\begin{deluxetable*}{lllll}
\tablecolumns{5}
\tablecaption{Best-Fitting Halo Models\label{tab:bestfit}}
\tablehead{
\colhead{density model} & \colhead{best-fit parameters} & \colhead{$\ln (\mathcal{L}_{\mathrm{max}})$} & \colhead{BIC} & \colhead{$\Delta$BIC}
}
\startdata
power law model & $q=0.918 ^{+0.016}_{-0.014}$, $n=4.40 ^{+0.05}_{-0.04}$ & -157625 & 315269 & 203 \\
\tableline
BPL model & $r_{\mathrm{break}}=38.7^{+0.69}_{-0.58}$, $q=0.908^{+0.008}_{-0.006}$, & & \\
&  $n_{\mathrm{inner}}=4.97^{+0.02}_{-0.05}$, $n_{\mathrm{outer}}=3.93^{+0.05}_{-0.04}$ & -214222  & 428464  & 113398\\
\tableline
Einasto profile & $r_{\mathrm{eff}}=1.07 \pm 0.10 \; \mathrm{kpc}$, & & \\
& $q=0.923 \pm 0.007$, $n=9.53^{+0.27}_{-0.28}$ &  -157685 &  315388 & 322
 \\
\tableline
power law model with $q(R_{\mathrm{gc}})$ & $r_0=25.0^{+1.8}_{-1.7}\; \mathrm{kpc}$, $q_0=0.773^{+0.017}_{-0.016}$, & & \\
&  $q_{\infty}=0.998^{+0.002}_{-0.001}$, $n=4.61 \pm 0.03$ & -157524  &  315066 & 0\\
\tableline
Einasto profile with $q(R_{\mathrm{gc}})$ & $r_0=26.7^{+2.2}_{-2.0}\; \mathrm{kpc}$, $q_0=0.779 \pm 0.018$,  & & \\
& $q_{\infty}=0.998^{+0.001}_{-0.002}$, $r_{\mathrm{eff}}=1.04^{+0.25}_{-0.13}\; \mathrm{kpc}$, & & \\
&  $n=8.78^{+0.33}_{-0.30}$ & -157582  & 315182 &  116 \\	
\enddata
\tablecomments{Summary of our best-fitting halo density models. For each model, we give the type of the density model, its best-fitting parameters along with their 1$\sigma$
uncertainties estimated as the 15.87th and 84.13th percentiles, the maximum log likelihood $\ln (\mathcal{L}_{\mathrm{max}})$ and the BIC. $\Delta$BIC gives the difference between the BIC of the best-fit model, the power law model with $q(R_{\mathrm{gc}})$, and the model used.}
\end{deluxetable*}   
\capstarttrue

\capstartfalse
\begin{deluxetable*}{lll}
\tablecolumns{3}
\tablecaption{Best-Fitting Halo Models for each Hemisphere\label{tab:hemispheres}}
\tablehead{
\colhead{density model} & \colhead{best-fit parameters north Galactic hemisphere} & \colhead{best-fit parameters south Galactic hemisphere} 
}
\startdata
power law model & $q=0.925_{-0.009}^{+0.010}$, $n=4.36 \pm 0.03$ & $q=0.852_{-0.011}^{+0.010}$, $n=4.40 \pm 0.04$\\
\tableline
Einasto profile & $r_{\mathrm{eff}}=1.11^{+0.09}_{-0.10} \; \mathrm{kpc}$, & $r_{\mathrm{eff}}=1.18^{+0.09}_{-0.11} \; \mathrm{kpc}$  \\
& $q=0.934_{-0.010}^{+0.009}$, $n=9.59^{+0.30}_{-0.26}$ &  $q=0.851^{+0.013}_{-0.011}$, $n=9.10 ^{+0.31}_{-0.28}$
 \\
\tableline
power law model with $q(R_{\mathrm{gc}})$ & $r_0= 29.2 \pm 4.4\; \mathrm{kpc}$, $q_0=0.831^{+0.031}_{-0.017}$, & $r_0=18.8^{+1.4}_{-1.6}\; \mathrm{kpc}$, $q_0=0.515^{+0.027}_{-0.058}$ \\
&  $q_{\infty}=0.997^{+0.004}_{-0.001}$, $n=4.53^{+0.04}_{-0.06}$ &  $q_{\infty}=0.998^{+0.004}_{-0.002}$, $n=4.88_{-0.03}^{+0.06}$\\
\tableline
Einasto profile with $q(R_{\mathrm{gc}})$ & $r_0=31.9^{+3.9}_{-3.1}\; \mathrm{kpc}$, $q_0=0.837^{+0.036}_{-0.017}$,  & $r_0=20.9^{+2.3}_{-2.0}\; \mathrm{kpc}$, $q_0=0.545^{+0.040}_{-0.066}$\\
& $q_{\infty}=0.998^{+0.001}_{-0.003}$, $r_{\mathrm{eff}}=1.00^{+0.v}_{-0.11}\; \mathrm{kpc}$, & $q_{\infty}=0.998^{+0.001}_{-0.005}$, $r_{\mathrm{eff}}=1.14^{+0.16}_{-0.12}\; \mathrm{kpc}$ \\
&  $n=9.07^{+0.35}_{-0.28}$ & $n=7.57^{+0.40}_{-0.23}$ \\	
\enddata
\tablecomments{Summary of our best-fitting halo density models. For each model, we give the type of the density model, its best-fitting parameters along with their 1$\sigma$
uncertainties estimated as the 15.87th and 84.13th percentiles.}
\end{deluxetable*}   
\capstarttrue

\clearpage

\startlongtable
\capstartfalse
\begin{deluxetable*}{rrrll}
\tablecolumns{5}
\tablecaption{Best-Fit Parameters for the Power Law Model on $\Delta l = 30 \arcdeg$, $\Delta b = 60 \arcdeg$ Bins\label{tab:powerlaw_60}}
\tablehead{
\colhead{$l$} & \colhead{$b$} & \colhead{sources}  & \colhead{$q$} & \colhead{$n$}
}
\startdata
$0$ & $-90$ & $158$ & $(0.307_{-0.077}^{+0.126}) $ & $3.53_{-0.19}^{+0.18}$ \\
$0$ & $-30$ & $532$ & $0.912_{-0.068}^{+0.057} $ & $4.23_{-0.11}^{+0.12}$ \\
$0$ & $30$ & $327$ & $0.973_{-0.039}^{+0.020}$ & $3.914_{-0.127}^{+0.131}$ \\
$30$ & $-90$ & $582$ & $0.557_{-0.055}^{+0.052} $ & $3.79_{-0.09}^{+0.10}$ \\
$30$ & $-30$ & $1099$ & $0.896_{-0.051}^{+0.050} $ & $4.89_{-0.08}^{+0.06}$ \\
$30$ & $30$ & $289$ & $0.943_{-0.053}^{+0.04}$ & $4.25_{-0.14}^{+0.16}$ \\
$60$ & $-90$ & $529$ & $0.714_{-0.063}^{+0.061}$ & $(3.39_{-0.09}^{+0.08})$ \\
$60$ & $-30$ & $899$ & $0.967_{-0.041}^{+0.024} $ & $4.71_{-0.08}^{+0.08}$ \\
$60$ & $30$ & $325$ & $0.919_{-0.054}^{+0.048} $ & $4.60_{-0.14}^{+0.13}$ \\
$90$ & $-90$ & $260$ & $0.813_{-0.076}^{+0.076} $ & $3.86_{-0.14}^{+0.14}$ \\
$90$ & $-30$ & $428$ & $0.949_{-0.058}^{+0.037} $ & $4.72_{-0.12}^{+0.12}$ \\
$90$ & $30$ & $247$ & $0.927_{-0.056}^{+0.0476} $ & $4.862_{-0.129}^{+0.094}$ \\
$120$ & $-90$ & $172$ & $0.568_{-0.085}^{+0.081} $ & $4.02_{-0.17}^{+0.19}$ \\
$120$ & $-30$ & $312$ & $0.935_{-0.084}^{+0.046} $ & $4.92_{-0.09}^{+0.06}$ \\
$120$ & $30$ & $232$ & $0.960_{-0.042}^{+0.027} $ & $4.80_{-0.17}^{+0.12}$ \\
$150$ & $-90$ & $74$ & $0.447_{-0.159}^{+0.168} $ & $4.04_{-0.28}^{+0.32}$ \\
$150$ & $-30$ & $215$ & $0.952_{-0.065}^{+0.036} $ & $4.71_{-0.17}^{+0.14}$ \\
$150$ & $30$ & $277$ & $0.919_{-0.054}^{+0.048} $ & $4.82_{-0.13}^{+0.11}$ \\
$180$ & $-90$ & $161$ & $0.940_{-0.061}^{+0.041} $ & $4.88_{-0.13}^{+0.09}$ \\
$180$ & $-30$ & $318$ & $0.967_{-0.051}^{+0.026} $ & $(4.30_{-0.12}^{+0.12})$ \\
$180$ & $30$ & $377$ & $(0.730_{-0.067}^{+0.068}) $ & $(2.83_{-0.11}^{+0.10})$ \\
$210$ & $-90$ & $98$ & $0.640_{-0.099}^{+0.100} $ & $4.61_{-0.25}^{+0.22}$ \\
$210$ & $-30$ & $387$ & $0.942_{-0.077}^{+0.042} $ & $4.95_{-0.06}^{+0.04}$ \\
$210$ & $30$ & $402$ & $0.934_{-0.052}^{+0.042} $ & $3.84_{-0.11}^{+0.11}$ \\
$240$ & $-30$ & $292$ & $0.959_{-0.054}^{+0.030} $ & $4.90_{-0.11}^{+0.07}$ \\
$240$ & $30$ & $476$ & $0.933_{-0.043}^{+0.039} $ & $4.79_{-0.11}^{+0.11}$ \\
$270$ & $-30$ & $20$ & $0.818_{-0.193}^{+0.128} $ & $4.74_{-0.38}^{+0.20}$ \\
$270$ & $30$ & $521$ & $0.967_{-0.033}^{+0.023} $ & $4.70_{-0.11}^{+0.11}$ \\
$300$ & $-30$ & $2$ & $(0.458_{-0.191}^{+0.318}) $ & $(2.20_{-0.93}^{+1.66})$ \\
$300$ & $30$ & $520$ & $0.992_{-0.012}^{+0.006} $ & $3.44_{-0.10}^{+0.10}$ \\
$330$ & $-30$ & $102$ & $0.953_{-0.066}^{+0.036} $ & $4.18_{-0.28}^{+0.29}$ \\
$330$ & $30$ & $390$ & $0.976_{-0.031}^{+0.018} $ & $3.55_{-0.11}^{+0.11}$ \\
\enddata
\tablecomments{Best-fit values for the power law model, when carried out on $\Delta l = 30 \arcdeg$, $\Delta b = 60 \arcdeg$ bins. The table give the bin limits from $(l,b)$ to $(l+\Delta l, b+\Delta b)$, the number of sources contained, as well as the best-fit model parameters. Only bins containing sources are listed. Values in brackets are unreliable for reasons mentioned in Sec. \ref{sec:LocalHaloProperties}.\newline
The $(l,b)$ distribution of these table values is depicted in Fig. \ref{fig:powerlaw_60_heatmap}.}
\end{deluxetable*}   
\capstarttrue

\clearpage

\newpage
\startlongtable
\capstartfalse
\begin{deluxetable*}{rrrlll}
\tablecolumns{6}
\tablecaption{Best-Fit Parameters for the Einasto Profile on $\Delta l = 30 \arcdeg$, $\Delta b = 60 \arcdeg$ Bins\label{tab:einasto_60}}
\tablehead{
\colhead{$l$} & \colhead{$b$} & \colhead{sources} & \colhead{$r_{\mathrm{eff}}$} & \colhead{$q$} & \colhead{$n$}
}
\startdata
$0$ & $-90$ & $158$ & $1.26_{-0.10}^{+0.09} $ & $(0.539_{-0.095}^{+0.094})$ & $10.4_{-0.7}^{+0.7}$ \\
$0$ & $-30$ & $532$ & $1.22_{-0.09}^{+0.09} $ & $0.901_{-0.071}^{+0.063} $ & $8.90_{-0.52}^{+0.60}$ \\
$0$ & $30$ & $327$ & $1.25_{-0.10}^{+0.10} $ & $0.972_{-0.034}^{+0.020} $ & $10.0_{-0.6}^{+0.7}$ \\
$30$ & $-90$ & $582$ & $1.27_{-0.10}^{+0.09} $ & $0.629_{-0.048}^{+0.046} $ & $11.6_{-0.6}^{+0.6}$ \\
$30$ & $-30$ & $1099$ & $1.18_{-0.11}^{+0.10} $ & $0.865_{-0.051}^{+0.052} $ & $6.83_{-0.26}^{+0.30}$ \\
$30$ & $30$ & $289$ & $1.23_{-0.10}^{+0.10} $ & $0.949_{-0.050}^{+0.034} $ & $9.23_{-0.58}^{+0.63}$ \\
$60$ & $-90$ & $529$ & $1.34_{-0.10}^{+0.10} $ & $0.830_{-0.052}^{+0.050} $ & $(12.8_{-0.6}^{+0.6})$ \\
$60$ & $-30$ & $899$ & $1.18_{-0.10}^{+0.10} $ & $0.966_{-0.047}^{+0.026} $ & $7.62_{-0.33}^{+0.37}$ \\
$60$ & $30$ & $325$ & $1.20_{-0.10}^{+0.09} $ & $0.929_{-0.049}^{+0.043} $ & $8.07_{-0.48}^{+0.54}$ \\
$90$ & $-90$ & $260$ & $1.25_{-0.09}^{+0.09} $ & $0.896_{-0.063}^{+0.060} $ & $10.3_{-0.6}^{+0.7}$ \\
$90$ & $-30$ & $428$ & $1.19_{-0.09}^{+0.10} $ & $0.943_{-0.071}^{+0.041} $ & $7.51_{-0.39}^{+0.46}$ \\
$90$ & $30$ & $247$ & $1.17_{-0.10}^{+0.10} $ & $0.923_{-0.057}^{+0.045} $ & $6.97_{-0.45}^{+0.51}$ \\
$120$ & $-90$ & $172$ & $1.24_{-0.10}^{+0.09} $ & $0.665_{-0.070}^{+0.071} $ & $9.52_{-0.65}^{+0.69}$ \\
$120$ & $-30$ & $312$ & $1.184_{-0.101}^{+0.099} $ & $0.926_{-0.076}^{+0.053} $ & $6.70_{-0.35}^{+0.41}$ \\
$120$ & $30$ & $232$ & $1.18_{-0.10}^{+0.10} $ & $0.964_{-0.040}^{+0.025} $ & $7.356_{-0.457}^{+0.488}$ \\
$150$ & $-90$ & $74$ & $1.21_{-0.10}^{+0.10} $ & $0.694_{-0.111}^{+0.117} $ & $8.68_{-0.73}^{+0.76}$ \\
$150$ & $-30$ & $215$ & $1.20_{-0.10}^{+0.10} $ & $0.956_{-0.064}^{+0.032} $ & $7.56_{-0.50}^{+0.56}$ \\
$150$ & $30$ & $277$ & $1.19_{-0.11}^{+0.09} $ & $0.914_{-0.049}^{+0.049} $ & $7.36_{-0.45}^{+0.53}$ \\
$180$ & $-90$ & $161$ & $1.17_{-0.09}^{+0.11} $ & $0.938_{-0.059}^{+0.043} $ & $6.69_{-0.51}^{+0.58}$ \\
$180$ & $-30$ & $318$ & $1.23_{-0.10}^{+0.09} $ & $0.972_{-0.044}^{+0.021} $ & $(9.17_{-0.52}^{+0.54})$ \\
$180$ & $30$ & $377$ & $1.37_{-0.09}^{+0.09} $ & $0.931_{-0.048}^{+0.043} $ & $(13.8_{-0.6}^{+0.7})$ \\
$210$ & $-90$ & $98$ & $1.19_{-0.10}^{+0.10} $ & $0.666_{-0.088}^{+0.093} $ & $7.84_{-0.71}^{+0.71}$ \\
$210$ & $-30$ & $387$ & $1.17_{-0.10}^{+0.11} $ & $0.944_{-0.073}^{+0.041} $ & $6.48_{-0.36}^{+0.37}$ \\
$210$ & $30$ & $402$ & $1.26_{-0.10}^{+0.09} $ & $0.961_{-0.040}^{+0.026} $ & $10.8_{-0.56}^{+0.62}$ \\
$240$ & $-30$ & $292$ & $1.16_{-0.10}^{+0.09} $ & $0.961_{-0.048}^{+0.028} $ & $6.93_{-0.39}^{+0.47}$ \\
$240$ & $30$ & $476$ & $1.19_{-0.10}^{+0.10} $ & $0.931_{-0.040}^{+0.041} $ & $7.43_{-0.36}^{+0.41}$ \\
$270$ & $-30$ & $20$ & $1.19_{-0.10}^{+0.10} $ & $0.814_{-0.194}^{+0.129} $ & $6.68_{-1.01}^{+0.99}$ \\
$270$ & $30$ & $521$ & $1.18_{-0.10}^{+0.10} $ & $0.966_{-0.034}^{+0.023} $ & $7.80_{-0.42}^{+0.45}$ \\
$300$ & $-30$ & $2$ & $1.20_{-0.10}^{+0.10} $ & $(0.453_{-0.186}^{+0.330})$ & $7.30_{-0.97}^{+1.03}$ \\
$300$ & $30$ & $520$ & $1.30_{-0.09}^{+0.09} $ & $0.993_{-0.010}^{+0.005} $ & $(12.2_{-0.7}^{+0.6}$ \\
$330$ & $-30$ & $102$ & $1.21_{-0.10}^{+0.10} $ & $0.947_{-0.073}^{+0.037} $ & $8.36_{-0.76}^{+0.77})$ \\
$330$ & $30$ & $390$ & $1.28_{-0.09}^{+0.10} $ & $0.981_{-0.027}^{+0.014} $ & $(11.4_{-0.6}^{+0.6})$ \\
\enddata
\tablecomments{Best-fit values for the Einasto profile, when carried out on $\Delta l = 30 \arcdeg$, $\Delta b = 60 \arcdeg$ bins. The table give the bin limits from $(l,b)$ to $(l+\Delta l, b+\Delta b)$, the number of sources contained, as well as the best-fit model parameters. Only bins containing sources are listed. Values in brackets are unreliable for reasons mentioned in Sec. \ref{sec:LocalHaloProperties}.\newline
The $(l,b)$ distribution of these table values is depicted in Fig. \ref{fig:einasto_60_heatmap}.}
\end{deluxetable*}   
\capstarttrue

\clearpage

\newpage
\startlongtable
\capstartfalse
\begin{deluxetable*}{rrrllll}
\tablecolumns{7}
\tablecaption{Best-Fit Parameters for the Power Law Model with $q(R_{\mathrm{gc}})$ on $\Delta l = 30 \arcdeg$, $\Delta b = 60 \arcdeg$ Bins\label{tab:powerlaw_qr_60}}
\tablehead{
\colhead{$l$} & \colhead{$b$} & \colhead{sources} & \colhead{$r_0$} & \colhead{$q_0$} & \colhead{$q_{\infty}$}  & \colhead{$n$}
}
\startdata
$0$ & $-90$ & $159$ & $18.2_{-2.4}^{+2.7}$ & $(0.204_{-0.020}^{+0.051})$ & $0.789_{-0.111}^{+0.122}$ & $4.99_{-0.32}^{+0.14}$ \\
$0$ & $-30$ & $517$ & $39.5_{-8.3}^{+21.9}$ & $0.996_{-0.418}^{+0.210}$ & $(0.211_{-0.261}^{+0.099})$ & $3.77_{-0.19}^{+0.188}$ \\
$0$ & $30$ & $325$ & $32.2_{-3.5}^{+20.7}$ & $0.999_{-0.410}^{+0.200}$ & $0.991_{-0.042}^{+0.022}$ & $3.91_{-0.15}^{+0.15}$ \\
$30$ & $-90$ & $577$ & $34.6_{-3.8}^{+4.5}$ & $0.472_{-0.024}^{+0.027}$ & $0.998_{-0.001}^{+0.002}$ & $4.99_{-0.03}^{+0.14}$ \\
$30$ & $-30$ & $1086$ & $39.4_{-9.1}^{+19.7}$ & $0.878_{-0.487}^{+0.110}$ & $0.841_{-0.089}^{+0.068}$ & $4.91_{-0.096}^{+0.063}$ \\
$30$ & $30$ & $287$ & $39.1_{-16.0}^{+16.2}$ & $0.855_{-0.418}^{+0.110}$ & $0.999_{-0.058}^{+0.042}$ & $4.40_{-0.17}^{+0.18}$ \\
$60$ & $-90$ & $529$ & $31.1_{-3.5}^{+4.0}$ & $0.432_{-0.047}^{+0.050}$ & $0.999_{-0.025}^{+0.012}$ & $4.57_{-0.209}^{+0.209}$ \\
$60$ & $-30$ & $903$ & $34.8_{-3.3}^{+19.8}$ & $0.996_{-0.393}^{+0.255}$ & $0.960_{-0.054}^{+0.028}$ & $4.71_{-0.09}^{+0.09}$ \\
$60$ & $30$ & $326$ & $34.7_{-10.2}^{+19.8}$ & $0.821_{-0.382}^{+0.110}$ & $0.999_{-0.060}^{+0.045}$ & $4.83_{-0.180}^{+0.169}$ \\
$90$ & $-90$ & $264$ & $37.8_{-5.9}^{+4.8}$ & $0.590_{-0.065}^{+0.074}$ & $0.999_{-0.045}^{+0.019}$ & $4.53_{-0.23}^{+0.22}$ \\
$90$ & $-30$ & $431$ & $39.8_{-5.8}^{+17.9}$ & $0.976_{-0.423}^{+0.194}$ & $0.977_{-0.089}^{+0.041}$ & $4.75_{-0.13}^{+0.12}$ \\
$90$ & $30$ & $249$ & $39.4_{-4.4}^{+18.0}$ & $0.919_{-0.364}^{+0.223}$ & $0.997_{-0.059}^{+0.047}$ & $4.94_{-0.14}^{+0.09}$ \\
$120$ & $-90$ & $179$ & $37.0_{-7.9}^{+5.9}$ & $0.485_{-0.069}^{+0.062}$ & $0.984_{-0.108}^{+0.084}$ & $4.99_{-0.24}^{+0.10}$ \\
$120$ & $-30$ & $314$ & $39.7_{-12.2}^{+20.1}$ & $0.992_{-0.478}^{+0.135}$ & $(0.407_{-0.271}^{+0.085})$ & $4.72_{-0.13}^{+0.10}$ \\
$120$ & $30$ & $234$ & $34.8_{-4.3}^{+16.8}$ & $0.974_{-0.391}^{+0.210}$ & $0.999_{-0.046}^{+0.032}$ & $4.91_{-0.17}^{+0.13}$ \\
$150$ & $-90$ & $77$ & $39.7_{-6.0}^{+22.3}$ & $0.497_{-0.151}^{+0.270}$ & $0.992_{-0.171}^{+0.416}$ & $4.99_{-0.65}^{+0.82}$ \\
$150$ & $-30$ & $218$ & $38.0_{-7.0}^{+15.6}$ & $0.978_{-0.318}^{+0.302}$ & $0.998_{-0.068}^{+0.036}$ & $4.77_{-0.15}^{+0.16}$ \\
$150$ & $30$ & $280$ & $39.4_{-7.4}^{+21.4}$ & $0.940_{-0.497}^{+0.120}$ & $0.977_{-0.075}^{+0.049}$ & $4.87_{-0.17}^{+0.14}$ \\
$180$ & $-90$ & $166$ & $39.7_{-25.2}^{+7.3}$ & $0.997_{-0.451}^{+0.044}$ & $0.585_{-0.309}^{+0.211}$ & $4.60_{-0.55}^{+0.26}$ \\
$180$ & $-30$ & $327$ & $39.6_{-5.5}^{+17.0}$ & $0.998_{-0.357}^{+0.236}$ & $0.996_{-0.042}^{+0.022}$ & $4.46_{-0.12}^{+0.12}$ \\
$180$ & $30$ & $379$ & $39.7_{-15.5}^{+8.7}$ & $0.999_{-0.157}^{+0.099}$ & $(0.511_{-0.139}^{+0.116})$ & $(2.43_{-0.23}^{+0.23})$ \\
$210$ & $-90$ & $102$ & $14.1_{-10.2}^{+10.2}$ & $0.934_{-0.219}^{+0.192}$ & $0.203_{-0.182}^{+0.228}$ & $2.92_{-0.71}^{+0.65}$ \\
$210$ & $-30$ & $401$ & $37.2_{-6.4}^{+21.1}$ & $0.992_{-0.397}^{+0.191}$ & $0.899_{-0.110}^{+0.062}$ & $4.99_{-0.06}^{+0.03}$ \\
$210$ & $30$ & $412$ & $39.2_{-23.0}^{+9.3}$ & $0.808_{-0.148}^{+0.107}$ & $0.995_{-0.053}^{+0.028}$ & $4.07_{-0.16}^{+0.15}$ \\
$240$ & $-30$ & $299$ & $26.3_{-4.26}^{+18.2}$ & $0.998_{-0.450}^{+0.219}$ & $0.993_{-0.085}^{+0.034}$ & $4.99_{-0.099}^{+0.060}$ \\
$240$ & $30$ & $485$ & $32.2_{-11.3}^{+14.4}$ & $0.830_{-0.324}^{+0.093}$ & $0.999_{-0.043}^{+0.032}$ & $4.97_{-0.14}^{+0.10}$ \\
$270$ & $-30$ & $21$ & $39.1_{-14.2}^{+11.8}$ & $0.961_{-0.230}^{+0.135}$ & $0.723_{-0.211}^{+0.373}$ & $4.99_{-0.544}^{+0.270}$ \\
$270$ & $30$ & $524$ & $39.2_{-2.9}^{+19.3}$ & $0.920_{-0.456}^{+0.167}$ & $0.998_{-0.045}^{+0.027}$ & $4.87_{-0.12}^{+0.13}$ \\
$300$ & $-30$ & $2$ & $39.6_{-13.2}^{+14.0}$ & $0.969_{-0.187}^{+0.342}$ & $0.506_{-0.231}^{+0.322}$ & $3.53_{-1.13}^{+1.53}$ \\
$300$ & $30$ & $520$ & $39.6_{-1.8}^{+15.8}$ & $0.997_{-0.376}^{+0.314}$ & $0.999_{-0.017}^{+0.006}$ & $3.50_{-0.10}^{+0.10}$ \\
$330$ & $-30$ & $97$ & $38.6_{-8.1}^{+22.7}$ & $0.990_{-0.425}^{+0.139}$ & $0.971_{-0.143}^{+0.061}$ & $3.97_{-0.270}^{+0.277}$ \\
$330$ & $30$ & $388$ & $34.92072_{-2.63}^{+16.9}$ & $0.998_{-0.444}^{+0.243}$ & $0.993_{-0.036}^{+0.019}$ & $(3.47_{-0.12}^{+0.13})$ \\
\enddata
\tablecomments{Best-fit values for the power law model with $q(R_{\mathrm{gc}})$, when carried out on $\Delta l = 30 \arcdeg$, $\Delta b = 60 \arcdeg$ bins. The table give the bin limits from $(l,b)$ to $(l+\Delta l, b+\Delta b)$, the number of sources contained, as well as the best-fit model parameters. Only bins containing sources are listed. Values in brackets are unreliable for reasons mentioned in Sec. \ref{sec:LocalHaloProperties}.\newline
The $(l,b)$ distribution of these table values is depicted in Fig. \ref{fig:powerlaw_qr_60_heatmap}.}
\end{deluxetable*}   
\capstarttrue

\clearpage

\newpage
\startlongtable

\capstartfalse
\begin{deluxetable*}{rrrlllll}
\tablecolumns{8}
\tablecaption{Best-Fit Parameters for the Einasto Profile with $q(R_{\mathrm{gc}})$ on $\Delta l = 30 \arcdeg$, $\Delta b = 60 \arcdeg$ Bins\label{tab:einasto_qr_60}}
\tablehead{
\colhead{$l$} & \colhead{$b$} & \colhead{sources}& \colhead{$r_0$} & \colhead{$q_0$} & \colhead{$q_{\infty}$} & \colhead{$r_{\mathrm{eff}}$}  & \colhead{$n$}
}
\startdata
$0$ & $-90$ & $158$ & $23.4_{-4.8}^{+1.0}$ & $(0.200_{-0.01}^{+0.06})$ & $0.987_{-0.230}^{+0.031}$ & $1.18_{-0.10}^{+0.10}$ & $5.15_{-0.38}^{+2.2}$\\
$0$ & $-30$ & $532$ & $13.2_{-0.04}^{+6.5}$ & $0.621_{-0.032}^{+0.303}$ & $0.995_{-0.163}^{+0.017}$ & $1.21_{-0.08}^{+0.11}$ & $8.80_{-0.49}^{+0.73}$\\
$0$ & $30$ & $327$ & $25.3_{-10.6}^{+4.4}$ & $0.840_{-0.107}^{+0.070}$ & $0.999_{-0.045}^{+0.004}$ & $1.25_{-0.11}^{+0.08}$ & $9.55_{-0.61}^{+0.82}$\\
$30$ & $-90$ & $582$ & $35.6_{-3.9}^{+2.4}$ & $0.431_{-0.033}^{+0.034}$ & $1_{-0.049}^{+0.005}$ & $1.19_{-0.12}^{+0.08}$ & $6.53_{-0.22}^{+0.82}$\\
$30$ & $-30$ & $1099$ & $15.9_{-1.6}^{+6.7}$ & $0.997_{-0.189}^{+0.031}$ & $(0.762_{-0.048}^{+0.156})$ & $1.19_{-0.11}^{+0.09}$ & $7.42_{-0.60}^{+0.15}$\\
$30$ & $30$ & $289$ & $38.3_{-19.7}^{+1.7}$ & $0.800_{-0.04}^{+0.12}$ & $0.998_{-0.075}^{+0.006}$ & $1.21_{-0.10}^{+0.10}$ & $8.56_{-0.41}^{+0.97}$\\
$60$ & $-90$ & $529$ & $31.7_{-4.3}^{+2.6}$ & $0.418_{-0.049}^{+0.026}$ & $0.999_{-0.034}^{+0.002}$ & $1.198_{-0.095}^{+0.102}$ & $8.50_{-0.46}^{+0.80}$\\
$60$ & $-30$ & $899$ & $27.3_{-13.0}^{+6.0}$ & $0.999_{-0.113}^{+0.011}$ & $0.989_{-0.131}^{+0.006}$ & $1.21_{-0.12}^{+0.08}$ & $7.85_{-0.32}^{+0.42}$\\
$60$ & $30$ & $325$ & $37.3_{-20.3}^{-1.5}$ & $0.803_{-0.060}^{+0.099}$ & $0.999_{-0.088}^{+0.008}$ & $1.18_{-0.091}^{+0.105}$ & $7.60_{-0.403}^{+0.766}$\\
$90$ & $-90$ & $260$ & $36.1_{-8.59}^{+0.246}$ & $0.538_{-0.066}^{+0.054}$ & $0.999_{-0.047}^{+0.004}$ & $1.20_{-0.10}^{+0.09}$ & $8.37_{-0.55}^{+0.77}$\\
$90$ & $-30$ & $428$ & $37.7_{-23.7}^{+4.9}$ & $0.998_{-0.229}^{+0.023}$ & $0.999_{-0.189}^{+0.021}$ & $1.23_{-0.14}^{+0.06}$ & $7.64_{-0.29}^{+0.64}$\\
$90$ & $30$ & $247$ & $36.6_{-22.5}^{+4.6}$ & $0.860_{-0.122}^{+0.077}$ & $0.999_{-0.112}^{+0.014}$ & $1.193_{-0.113}^{+0.086}$ & $6.794_{-0.354}^{+0.790}$\\
$120$ & $-90$ & $172$ & $35.5_{-10.4}^{+0.3}$ & $0.399_{-0.069}^{+0.064}$ & $0.999_{-0.141}^{+0.016}$ & $1.20_{-0.12}^{+0.08}$ & $6.56_{-0.16}^{+0.20}$\\
$120$ & $-30$ & $312$ & $14.2_{-0.2}^{+20.5}$ & $0.998_{-0.316}^{+0.026}$ & $0.999_{-0.305}^{+0.029}$ & $1.19_{-0.10}^{+0.10}$ & $6.79_{-0.28}^{+0.65}$\\
$120$ & $30$ & $232$ & $37.9_{-23.8}^{+6.1}$ & $0.957_{-0.147}^{+0.015}$ & $0.999_{-0.087}^{+0.009}$ & $1.18_{-0.09}^{+0.11}$ & $7.34_{-0.42}^{+0.69}$\\
$150$ & $-90$ & $74$ & $39.7_{-11.5}^{+1.1}$ & $0.429_{-0.088}^{+0.080}$ & $0.993_{-0.233}^{+0.026}$ & $1.18_{-0.10}^{+0.10}$ & $6.85_{-0.38}^{+0.23}$\\
$150$ & $-30$ & $215$ & $38.4_{-25.4}^{+12.9}$ & $0.999_{-0.652}^{+0.077}$ & $0.996_{-0.109}^{+0.08}$ & $1.19_{-0.10}^{+0.10}$ & $7.67_{-0.56}^{+0.51}$\\
$150$ & $30$ & $277$ & $14.0_{-0.2}^{+8.7}$ & $0.997_{-0.199}^{+0.032}$ & $0.916_{-0.086}^{+0.051}$ & $1.18_{-0.10}^{+0.10}$ & $7.62_{-0.61}^{+0.57}$\\
$180$ & $-90$ & $161$ & $32.5_{-16.0}^{+3.7}$ & $0.997_{-0.118}^{+0.010}$ & $0.749_{-0.080}^{+0.183}$ & $1.22_{-0.13}^{+0.08}$ & $7.45_{-0.83}^{+0.75}$\\
$180$ & $-30$ & $318$ & $12.2_{-0.9}^{+8.5}$ & $0.994_{-0.434}^{+0.031}$ & $0.997_{-0.087}^{+0.004}$ & $1.22_{-0.11}^{+0.107}$ & $(9.30_{-0.56}^{+0.65})$\\
$180$ & $30$ & $377$ & $34.4_{-6.1}^{+0.8}$ & $(0.328_{-0.058}^{+0.034})$ & $0.998_{-0.060}^{+0.005}$ & $1.30_{-0.11}^{+0.09}$ & $(11.4_{-0.6}^{+0.8})$\\
$210$ & $-90$ & $98$ & $38.5_{-24.9}^{+6.2}$ & $0.6_{-0.2}^{+0.1}$ & $0.748_{-0.164}^{+0.070}$ & $1.22_{-0.12}^{+0.08}$ & $7.48_{-0.67}^{+0.10}$\\
$210$ & $-30$ & $387$ & $32.2_{-18.4}^{+3.4}$ & $0.997_{-0.297}^{+0.032}$ & $0.999_{-0.152}^{+0.016}$ & $1.16_{-0.10}^{+0.11}$ & $6.62_{-0.32}^{+0.41}$\\
$210$ & $30$ & $402$ & $31.8_{-9.2}^{+3.1}$ & $0.688_{-0.079}^{+0.067}$ & $0.999_{-0.040}^{+0.004}$ & $1.25_{-0.11}^{+0.08}$ & $9.81_{-0.46}^{+0.82}$\\
$240$ & $-30$ & $292$ & $36.7_{-22.4}^{+2.7}$ & $0.997_{-0.174}^{+0.014}$ & $0.996_{-0.162}^{+0.014}$ & $1.19_{-0.12}^{+0.08}$ & $6.98_{-0.32}^{+0.58}$\\
$240$ & $30$ & $476$ & $29.0_{-13.0}^{+3.57}$ & $0.786_{-0.061}^{+0.095}$ & $0.999_{-0.062}^{+0.006}$ & $1.18_{-0.10}^{+0.10}$ & $6.92_{-0.25}^{+0.68}$\\
$270$ & $-30$ & $20$ & $18.8_{-4.10}^{+4.52}$ & $0.992_{-0.390}^{+0.045}$ & $(0.544_{-0.261}^{+0.250})$ & $1.20_{-0.11}^{+0.08}$ & $7.10_{-1.02}^{+0.98}$\\
$270$ & $30$ & $521$ & $38.0_{-23.0}^{+3.49}$ & $0.939_{-0.062}^{+0.043}$ & $0.999_{-0.083}^{+0.009}$ & $1.17_{-0.08}^{+0.12}$ & $7.83_{-0.40}^{+0.60}$\\
$300$ & $-30$ & $2$ & $13.0_{-1.3}^{+9.8}$ & $0.984_{-0.722}^{+0.193}$ & $0.999_{-0.733}^{+0.197}$ & $1.20_{-0.10}^{+0.09}$ & $7.65_{-1.27}^{+0.74}$\\
$300$ & $30$ & $520$ & $26.0_{-11.6}^{+6.7}$ & $0.919_{-0.087}^{+0.058}$ & $0.999_{-0.019}^{+0.001}$ & $1.32_{-0.14}^{+0.09}$ & $(11.9_{-0.7}^{+0.9})$\\
$330$ & $-30$ & $102$ & $23.2_{-9.2}^{+9.9}$ & $0.988_{-0.248}^{+0.018}$ & $0.999_{-0.190}^{+0.021}$ & $1.19_{-0.07}^{+0.12}$ & $8.51_{-0.86}^{+0.65}$\\
$330$ & $30$ & $390$ & $27.6_{-7.71}^{+6.78}$ & $0.748_{-0.066}^{+0.090}$ & $0.999_{-0.029}^{+0.001}$ & $1.26_{-0.10}^{+0.12}$ & $10.3_{-0.5}^{+0.1}$\\
\enddata
\tablecomments{Best-fit values for the Einasto profile with $q(R_{\mathrm{gc}})$, when carried out on $\Delta l = 30 \arcdeg$, $\Delta b = 60 \arcdeg$ bins. The table give the bin limits from $(l,b)$ to $(l+\Delta l, b+\Delta b)$, the number of sources contained, as well as the best-fit model parameters. Only bins containing sources are listed. Values in brackets are unreliable for reasons mentioned in Sec. \ref{sec:LocalHaloProperties}.\newline
The $(l,b)$ distribution of these table values is depicted in Fig. \ref{fig:einasto_qr_60_heatmap}.}
\end{deluxetable*}   
\capstarttrue

\capstartfalse
\begin{deluxetable*}{ll}
\tablecolumns{2}
\tablecaption{Mean And Variance for Best-Fit Parameters on $\Delta l = 30 \arcdeg$, $\Delta b = 60 \arcdeg$ Bins\label{tab:mean_variance_60deg}}
\tablehead{
\colhead{density model} & \colhead{parameter mean and variance} 
}

\startdata
power law model & mean($n$) = 4.25, Var($n$) = 0.443,\\
 & mean($q$) = 0.840, Var($q$) =0.0333\\
\\
\tableline
Einasto profile & mean($r_{\mathrm{eff}}$) = 1.22, Var($r_{\mathrm{eff}}$) = 0.00252,  \\
  & mean($q$) = 0.873, Var($q$) = 0.0191 \\
  & mean($n$) = 3.85, Var($n$) = 8.76 \\
 \\
\tableline
power law model with $q(R_{\mathrm{gc}})$ &  mean($r_0$) = 36.0, Var($r_0$) = 36.3, \\
   & mean($q_0$) = 0.856, Var($q_0$) = 0.0443,\\
   & mean($q_{\infty}$) = 0.859, Var($q_{\infty}$) = 0.0559,\\
   & mean($n$) = 4.45, Var($n$) = 0.441\\
\\
\tableline
Einasto profile with $q(R_{\mathrm{gc}})$ & mean($r_0$) = 29.1, Var($r_0$) = 81.5,\\
&  mean($q_0$) =   0.790, Var($q_0$) = 0.058, \\
&  mean($q_{\infty}$) = 0.958, Var($q_{\infty}$) = 0.011, \\
&  mean($r_{\mathrm{eff}}$) = 1.21, Var($r_{\mathrm{eff}}$) = 0.001, \\
&  mean($n$) =7.961, Var($n$) = 2.01\\
\enddata
\tablecomments{
Mean and variance for the best-fit parameters of all four halo models on $\Delta l = 30 \arcdeg$, $\Delta b = 60 \arcdeg$ bins.
}
\end{deluxetable*}   
\capstarttrue

\capstartfalse
\begin{deluxetable*}{llll}
\tablecolumns{3}
\tablecaption{Model Parameters for selected Other Surveys\label{tab:compare_models_selected}}
\tablehead{
\colhead{paper} & \colhead{density model} & \colhead{parameter mean and variance} & \colhead{distance range} 
}

\startdata
\cite{Watkins2009} & BPL & $n_{\mathrm{inner}}=2.4$, $n_{\mathrm{inner}}=4.5$, & $5\; \mathrm{kpc}<{R_{\mathrm{gc}}}<115 \; \mathrm{kpc}$\\
& & $r_{\mathrm{break}}=23 \; \mathrm{kpc}$, $q=1$ (assumes no flattening)\\
\cite{Deason2014} & BPL with three segments &  $\alpha_1 = 2.5$, $\alpha_2 = 4.5$, $\alpha_{\mathrm{out}}=10$, & $10\; \mathrm{kpc}<{R_{\mathrm{gc}}}<100 \; \mathrm{kpc}$\\
& & $r_c=25 \; \mathrm{kpc}$, $r_{\mathrm{break}}=65 \; \mathrm{kpc}$, $q=1$ (assumes no flattening)\\
\cite{Cohen2015} & power law & $n = 3.8 \pm 0.3$, $q=1$ (assumes no flattening) & $50\; \mathrm{kpc}<{R_{\mathrm{gc}}}<115 \; \mathrm{kpc}$\\	
\cite{Xue2015} & power law with $q(R_{\mathrm{gc}})$ & $\alpha=4.2 \pm 0.1$, $q_0 = 0.2 \pm 0.1$, $q_{\infty} = 0.8 \pm 0.3$, $r_0=6 \pm 1 \; \mathrm{kpc}$& $10\; \mathrm{kpc}<{R_{\mathrm{gc}}}<80 \; \mathrm{kpc}$\\
\cite{Slater2016} & power law & $n=3.5 \pm 0.2$, $q=1$ (assumes no flattening) & $20 \; \mathrm{kpc}<{R_{\mathrm{gc}}}<80 \; \mathrm{kpc}$\\
\tableline
\enddata
\tablecomments{
Mean and, as far as available, variance for the best-fit parameters found in other work. The models are shown, together with our best-fit models, in Fig. \ref{fig:compare_models_selected}. Some authors call the exponent of their power law model $\alpha$ instead of $n$. We have kept their notation.
}
\end{deluxetable*}   
\capstarttrue

\end{document}